\documentclass[12pt,titlepage,a4paper]{book}
\usepackage{natbib, aas_macros, lscape}
\usepackage{amsmath, amssymb, booktabs}
\usepackage{bm}
\usepackage{color}
\usepackage{setspace, lscape}
\usepackage{epstopdf}
\usepackage{fancyhdr} 
\usepackage[T1]{fontenc}
\usepackage[dvipdfm,dvips,hiresbb]{graphicx}
\fancypagestyle{plain}{

\lhead[\leftmark]{\thepage}
\chead[]{}
\rhead[\thepage]{\rightmark}
\cfoot[]{}
}

\begin{document}

\newcommand\ion[2]{#1{\small\rmfamily{#2}}\relax}%


\setcounter{page}{1}
\pagenumbering{roman}
\chapter*{Abstract}
\markboth{Abstract}{}
\addcontentsline{toc}{chapter}{Abstract}
\thispagestyle{fancy}

The mechanism of magnetic flux maintenance on the solar surface in quiet regions on the Sun is observationally investigated.
There have been proposed two scenarios: One is by the local interactions only in the surface level and the other is by the frequent replacement 
of the fluxes among different atmospheric altitudes. 
In order to understand the energy transport and budget through the different levels of the solar atmosphere, 
it is important to quantitatively study the surface magnetic processes since they are one of the major drivers for the magnetic energy generation. 
The Solar Optical Telescope on board {\it Hinode} spacecraft makes, for the first time, such a study possible due to its 
stable and precise observation of the magnetic field.

We investigate surface processes of magnetic patches, namely merging, splitting, emergence, and cancellation, 
by using an auto-detection technique. 
We find that merging and splitting are locally predominant in the surface level, while the frequencies of the other two are less by one or two orders of
 magnitude.
The frequency dependences on flux content of surface processes are further investigated.
Based on these observations, we discuss a possible whole picture of the maintenance.
Our conclusion is that the photospheric magnetic field structure, especially its power-law nature, is maintained by the processes locally 
in the surface not by the interactions between different altitudes. 
We suggest a scenario of the flux maintenance as follows: 
The splitting and merging play a crucial role for the generation of the power-law distribution, not the emergence nor cancellation do. 
This power-law distribution results in another power-law one of the cancellation with an idea of the random convective transport. 
The cancellation and emergence have a common value for the power-law indices in their frequency distributions, 
which may suggest a ''recycle of fluxes by submergence and re-emergence''.

We summarize the previous studies and the purpose of the thesis in Chapter 1.
The description of data sets and method of the auto-detection code is explained in Chapter 2.
The main results and discussion of the thesis are described in Chapter 3$\--$5.

In Chapter 3, statistical properties of detected patches are investigated.
We use two data sets of magnetograms obtained by {\it Hinode}.
One has higher time resolution but an intermediate duration (data set 1).
The other has a long duration but lower time resolution (data set 2).
The frequency distributions of flux content are power-law distributions with indices of $-1.79\pm0.18$ (data set 1) 
and $-1.93\pm0.07$ (data set 2).
They are also consistent with the previous result by \cite{par2009}.
The average lifetimes over the whole magnetic patches are found to be $17.3$ minutes in data set 1 and 
$23.2$ minutes in data set 2. 
These different values show that the obtained lifetimes depend on the temporal resolution of the data sets, 
i.e. they become shorter as measured with a better resolution.
The lifetime of patches increases with the larger flux content and saturates at $\thicksim$10$^{18}$ Mx to the value of 60 min.
We investigate proper velocity of patches in data set 1.
The averaged proper velocity is $1.2$ km s$^{-1}$.
The flux dependence of proper velocity is found to be weak with a power-law index of $-0.23$.
The frequency of splitting has a flat distribution with a drop in the lower range of the flux content. 
It is interpreted that the actual frequency is independent of the flux content and that the obtained distribution is under the influence 
of the technical detection limit.

In Chapter 4, we investigate frequencies of surface processes.
It is found that occurrence rates of emergence and cancellation are much less than those of merging and splitting.
We found that probability distribution of merging has weak dependence on flux content, with a power-law index of $0.28$.
On the other hand, the frequency distribution of cancellation has a power-law distribution with an index of $-2.48$.

In Chapter 5, we present discussions of the results.
First we summarize magneto-chemistry equation \citep{sch1997b}, which describes number relationship between 
frequency distribution of flux content and those of magnetic processes.
Splitting has a time-independent solution of a power-law distribution with an index of $-2$.
Next, we also discuss the reason of steep slope of frequency distribution of cancellation.
Our discussion is based on assumptions that the patch motions are driven by convection which has random flow direction with constant 
velocity and that the power-law distribution of flux content is predominantly maintained. 
The steep slope is naturally obtained there. 
The derived power-law index is very similar to that of the emergence \citep{tho2011}.
It suggests that large parts are re-emergences of submerged loops through cancellations.

In Chapter 6, the conclusion of the thesis is given as follows:\\
1) Frequency distribution of the flux content is maintained to a power-law distribution by merging and splitting on the solar surface.\\
2) The frequency of cancellation can be interpreted as a result of collisions of patches under motions driven in random direction with 
constant velocities.\\
3) Most of emergences are interpreted as re-emergences of submerged loops recognized as cancellations.

\clearpage
\chapter*{Acknowledgement}
\markboth{Acknowledgement}{}
\addcontentsline{toc}{chapter}{Acknowledgement}
\thispagestyle{fancy}

First of all, I want to appreciate my supervisor, Dr. Takaaki Yokoyama, for his patient education and useful advices.
He supports all parts of the works done here.
Next, I show great appreciation to Dr. Mandy Hagenaar in Lockheed Martin Solar and Astronomical Laboratory.
Discussions with her at 3rd {\it Hinode} Science Conference gave a crucial moment to do the works done here.
The stay at LMSAL constitutes the greatest portion of the thesis. 
I also want to thank Global COE program From the Earth to $``$Earths$``$, which supports author's stay there.
The works in this thesis cannot be done without this support.

The seminar presentations helped me in a large part of the thesis.
I had fruitful discussions and comments at the seminar of National Astronomical Observatory in Japan, ISAS/JAXA, LMSAL, 
St. Andrews University, and University of Tokyo.
I thank to my colleagues in the room of university for their encouragements and patent treatment of me.

I also extend our appreciations to the proofreading/editing assistance from the GCOE program. 
Some works are supported by Grant-in-Aid for JSPS Fellows.
Hinode is a Japanese mission developed and launched by ISAS/JAXA, with NAOJ as domestic partner and NASA and STFC (UK) as
 international partners.
It is operated by these agencies in co-operation with ESA and NSC (Norway).
The highly qualified magnetograms obtained by {\it Hinode} enable us to investigate them automatically.
Finally, I want to appreciate my parents for their encouragement.

Without any of these people and projects, I cannot accomplish this work.
\vspace{1cm}
\begin{figure}[h]
\begin{flushright}
\includegraphics[bb=300 0 1800 500,width=0.35\textwidth]{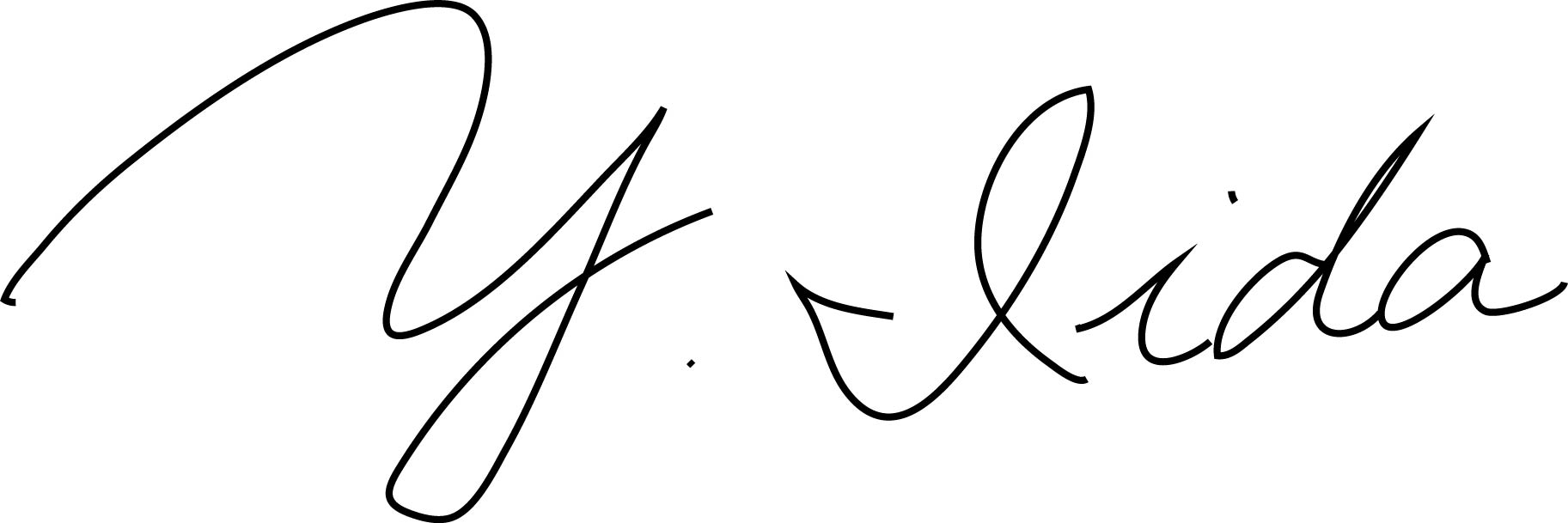}
\end{flushright}
\end{figure}

\vspace*{10cm}

\ifodd \arabic{page}
\else
  \thispagestyle{fancy}
  \mbox{}
  \newpage
  \clearpage
\fi

\pagestyle{fancy}
\addcontentsline{toc}{chapter}{Contents}
\addtocontents{toc}{\protect\thispagestyle{fancy}}
\tableofcontents


\ifodd \arabic{page}
\else
  \thispagestyle{fancy}
  \mbox{}
  \newpage
  \clearpage
\fi

\chapter[General Introduction]{General Introduction}
\setcounter{page}{1}
\pagenumbering{arabic}
\thispagestyle{fancy}

Brief reviews of previous studies and the purpose of this study are summarized in this chapter.
The main target of our thesis is magnetic field in quiet regions.
It is affected by convective motion of plasmas on the solar surface.
Magnetic structures on the solar surface are summarized in Section \ref{intro_mag}.
Magnetic processes i.e., merging, splitting, emergence, and cancellation, are
reviewed in Section \ref{intro_model}.
The purpose and strategy of this thesis are mentioned in Section \ref{intro_purpose}.

\pagestyle{fancy}
\section{Magnetic Structures on the Solar Surface}
\label{intro_mag}

\begin{figure}[p]
\centering
\includegraphics[bb=0 0 730 350,width=1.00\textwidth]{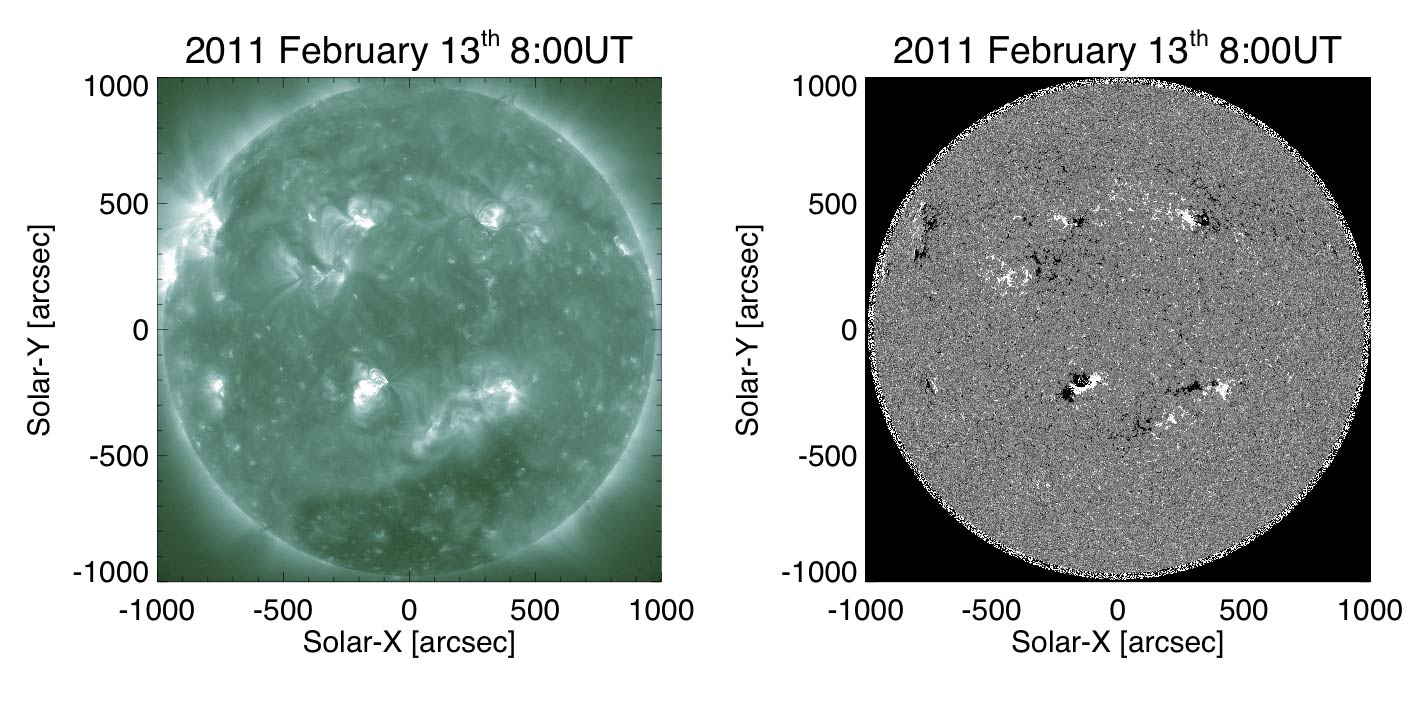}
\caption[EUV image and EUV image of the Sun]{
Extreme Ultra Violet (EUV) image (left) and corresponding line-of-sight magnetogram (right) of the whole Sun.
The EUV image is obtained by Atmospheric Image Assembly (AIA) on board Solar Dynamics Observatory (SDO).
The magnetogram is obtained by Helioseismic and Magnetic Imager (HMI) on board SDO.
The bright (hot) region coincides with the magnetized region.
}
\label{fig:magac}
\end{figure}

Magnetic field on the solar surface causes various energetic activities. 
Many researchers have been attracted in what properties it has and 
how it is structured (Figure \ref{fig:magac}).
It is important not only as the energy source of solar activities but also as the only magnetic field on the stellar surface that we
observe in the most detailed state.
 
\begin{figure}[p]
\centering
\includegraphics[bb=0 0 700 300,width=1.00\textwidth]{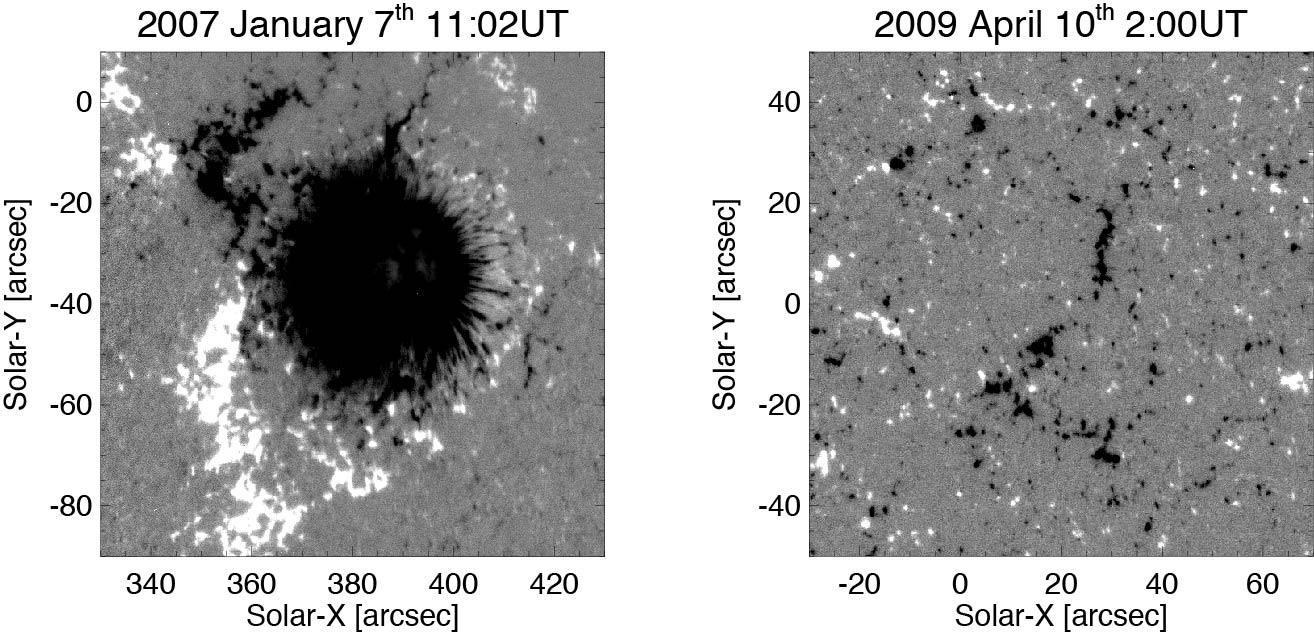}
\caption[Magnetogram of an active region and a quiet region.]{
Magnetograms of an active region (left) and a quiet region (right) obtained by Narrowband Filter Imager (NFI) on board 
{\it Hinode}.
Left:
Magnetogram of an active region.
There is a negative sunspot in the middle of the field of view.
Positive polarity has a less bundled structure than negative polarity.
Right:
Magnetogram of a quiet region.
Both polarities are fragmented and seem to be affected by convective motion.
There are some network structures ($\thicksim20'' \-- 40''$) in this image. 
}
\label{fig:magsun}
\end{figure}

Magnetic field structure on the solar surface is classified into two regions, i.e. active regions and quiet regions (Figure \ref{fig:magsun}).
An active region is a strongly magnetized region that often contains sunspots and plages.
Violent activities, solar flares, and ejections of plasmas (coronal mass ejections; CMEs) sometimes occur there.
On the other hand, a quiet region was thought to be a moderate region without such activities.

However, recent high-resolution observations change our picture of quiet regions.
One of the most noticeable results is that there is as large amount of magnetic flux content in quiet regions as in active regions.
Recent observations by ${\it Hinode}$/X-Ray Telescope (XRT) and ${\it Solar Dynamics Observatory}$ (SDO)/ Atmospheric Image Assembly (AIA)
 show that there are a plenty of brightenings which is thought to be events releasing magnetic energy there.
It is deduced here that a quiet region is not so moderate as we thought.
There is another important character in quiet regions.
The magnetic field is highly affected by convective motion on the solar surface, i.e. the plasma beta in quiet regions is higher than unity.
Right panel in Figure \ref{fig:magsun} shows line-of-sight magnetic field in a quiet region.
One can recognize that the field is organized as a collection of discrete units, here we call them magnetic patches.
They are swept to the edges of convective cells.

Magnetic field in a quiet region is divided into three categories, ephemeral region, network field, and 
internetwork field (same as intranetwork field in solar physics terminology).
Our target is network field.
Network field and internetwork field are ubiquitous structures in quiet regions.
Network field is a magnetic field swept to the edge of supergranular cells.
Convective motion is believed to have an important role in their formation and maintenance.
The field strength of network field is from a few hundred G to $\sim 1 \ {\rm kG}$.
Flux amount contained in each patch varies from $10^{18}$ Mx to $10^{19}$ Mx \citep{mar1988, wan1995}.
The lifetime varies from a few hours to a day \citep{tho2011}.
The observed flux content of internetwork field varies from $10^{16}$ Mx to $10^{18}$ Mx \citep{livi1975, zir1985, zir1987, wan1995}. 
On the other hand, internetwork field is in the middle of supergranular cells.
The field strength of internetwork field is much weaker than that of network field.
The smallest scale of internetwork field corresponds to the observational limit in the observation so far
\citep{livi1975, smi1975, liv1985, mar1988, wan1988, wan1995}.
Thus the frequency distribution of flux content is thought to continue below $10^{16}$ Mx 
\citep{soc2004, man2004, khom2005, domi2006, rez2007, sanc2007, har2007, oro2008, lit2008}.
A quiet region looks a mixed-polarity region from network field and internetwork field.
It is sometimes called magnetic carpet \citep{sch1998, par2001}. 
The other one, ephemeral region, is a bi-polar region \citep{har1973b, mar1988, hag2001}, 
which is thought as a remnant of a small flux emergence 
in a quiet region \citep{wan1988, har1993b}.
This prominent feature has a larger flux content than the other patches and larger proper velocity.
The typical flux amount varies from $(2-3) \times 10^{19}$ Mx to $10^{20}$ Mx \citep{har1973b, har1975, har1993b, cha2001b}. 
The typical lifetime is from a few days to 4.4 hr \citep{har1973b, har1975, har1993b}.
It significantly decreases with higher resolution.
At the end of the lifetime, the association of the opposite polarities in each ephemeral region is lost due to the random transport by the 
convective motion.

\begin{figure}[p]
\centering
\includegraphics[bb= 300 808 600 1188,width=0.60\textwidth]{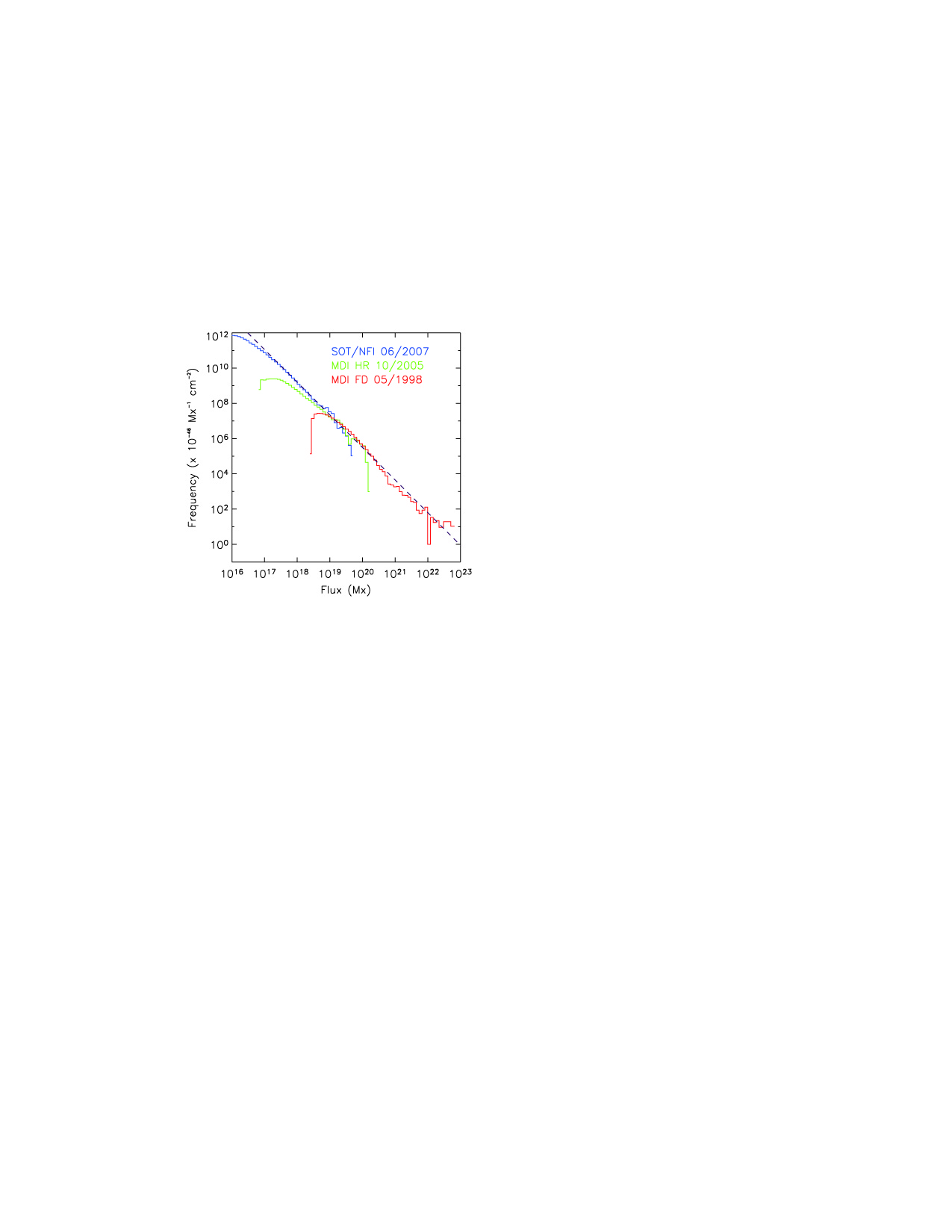}
\caption[Frequency distribution of flux content from Parnell et al.(2009)]{
Frequency distribution of flux content reported by Parnell et al.(2009).
Three different data sets are used for this plot.
They are different in solar cycle-phase and dynamic range of observation.
The obtained distribution is a single power-law distribution with an index of $-1.85 \pm 0.14$ from large active regions 
to granular-scale patches.
}
\label{fig:dist_par}
\end{figure}

One question arises.
What amount of magnetic flux is produced and transported?
The first step to answer this question is to investigate total flux amount in quiet regions.
It has been probed by the observations for a long time.
However, we have not determined the total flux amount because it increases as the spatial resolution of observation becomes better.
One approach for this question is to investigate a frequency distribution of magnetic flux content in discrete magnetic patch, namely
an investigation of scaling law.
We study whether patches with smaller or larger flux amount are dominant in total flux amount from the shape of the distribution.
Moreover, the distribution probably has the information for the flux maintenance mechanism.
There are several studies on this issue \citep{wan1995, sch1997b, hag1999, par2009, zha2010}.
But the distribution form is not settled for long time because it was different among above studies, such as an exponential distribution, 
a power-law distribution, and Weibull distribution.
\cite{par2009} gives one conclusion.
It is a power-law distribution with an index of $-1.85\pm0.14$ between $2 \times 10^{17}$ Mx and $10^{23}$ Mx (Figure \ref{fig:dist_par}).
There is a single power-law in the wide flux range spanning from large active regions to small patches in quiet network.
The collaborative usage of the whole Sun magnetograms obtained by {\it Solar and Heliospheric Observatory} ({\it SOHO})/MDI and 
high-resolution magnetogram obtained by {\it Hinode}/SOT enabled them to investigate in such a wide range of flux content.

The total flux amount, $\Phi_{\rm tot}$, is obtained as
\begin{equation}
\Phi_{\rm tot}=\int_{\phi_{\rm min}}^{\phi_{\rm max}}\phi n(\phi) d\phi
\end{equation}
where $n(\phi)$ is a frequency distribution of flux content, $\phi_{\rm min}$ and $\phi_{\rm max}$ are the 
minimum and maximum flux content respectively.
(Note that $\Phi$ is called the 'total flux amount' but actually it is spatially averaged one over the area, namely its unit is 
Mx cm$^{-2}$. In this thesis, we use this definition rule hereafter.)
When $n(\phi)=n_0 \left( \phi / \phi_0 \right)^{-\gamma}$, it becomes
\begin{equation}
\Phi_{\rm tot}=\frac{n_0 \phi_0^2}{-\gamma+2} \left[ \left( \frac{\phi_{\rm max}}{\phi_0} \right)^{-\gamma+2} -
\left( \frac{\phi_{\rm min}}{\phi_0} \right)^{-\gamma+2} \right].
\end{equation}
In a case of $\gamma < 2$, the frequency distribution is flat and patches with larger flux content is important for an estimation 
of the total flux amount.
On the contrary, the frequency distribution steeper than $\gamma > 2$ means that patches with smaller 
flux content is important for an estimation of the total flux amount.

The active region has a typical flux range between $10^{20}-10^{22}$ Mx.
The quiet region has a flux range below $10^{20}$ Mx.
From $n(\phi)=1.0\times10^{-33} \, (\phi/10^{16} {\rm Mx})^{-1.85}$ Mx$^{-1}$ cm$^{-2}$ from \cite{par2009}, one obtains
 $\Phi_{\rm tot}^{AR}=$ 1.99 Mx cm$^{-2}$ and $\Phi_{\rm tot}^{QS}=$ 1.50 Mx cm$^{-2}$.
The power-law index, $-1.85$, is not so different from $-2$, which cause an increase of total flux amount with higher resolution observations.

Next question arises at this point.
How the frequency distribution is achieved and sustained?
\cite{par2009} suggest two scenarios.
One is that frequency distribution on the solar surface represents distribution of flux tubes generated by dynamo mechanisms in the convective 
layer.
The other is that it is achieved and sustained by surface activities, namely emergence, merging, splitting, and cancellation of magnetic patches
on the photosphere.
The key investigation to distinguish them is to quantify the effect and amount of these magnetic activities.

\section[Surface Processes on the Solar Surface]{Surface Processes of Magnetic Patches on the Solar Surface}
\label{intro_model}
\begin{figure}[p]
\centering
\includegraphics[bb= 0 0 650 800,width=0.8\textwidth]{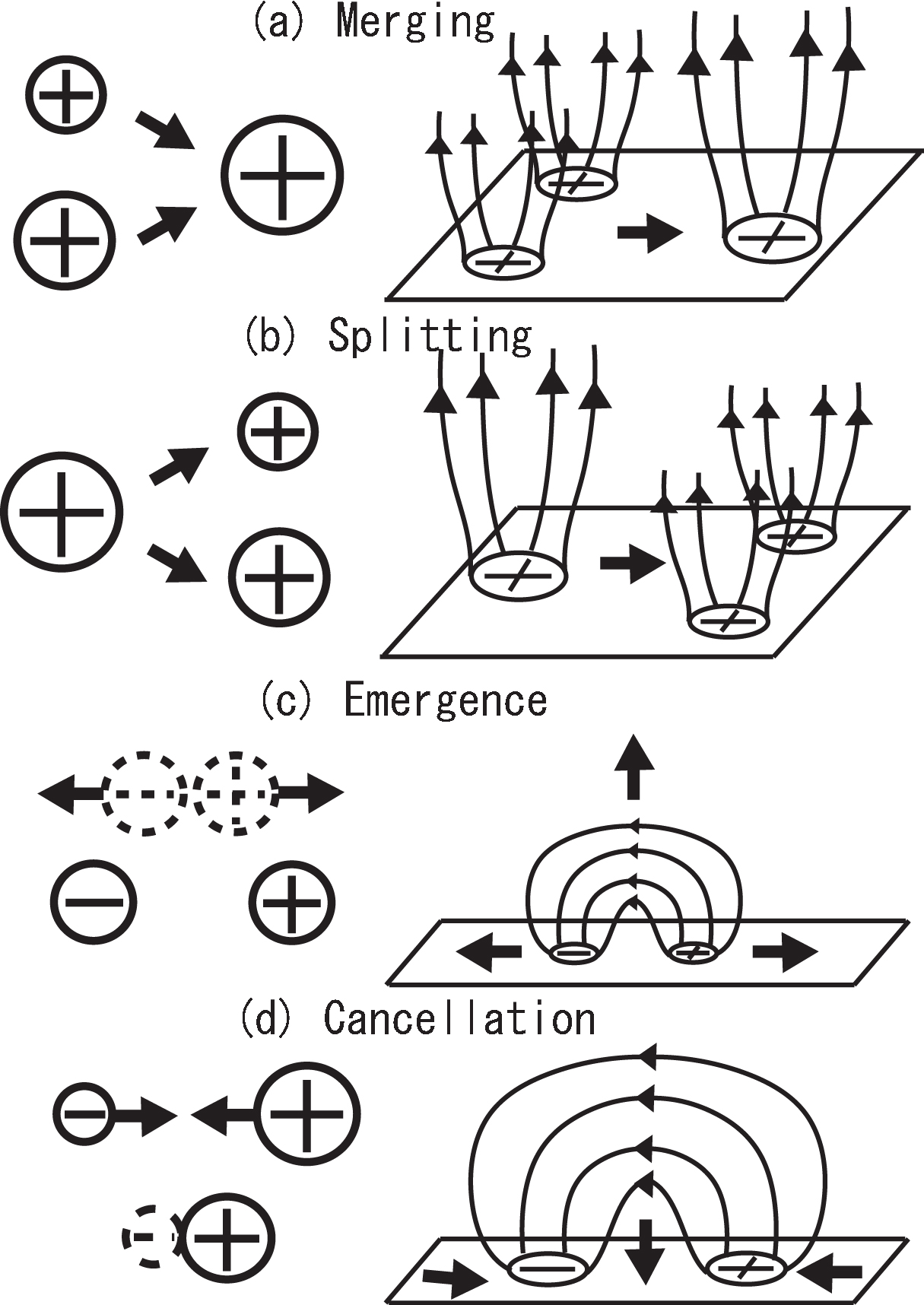}
\caption[Schematic pictures of elementary surface processes of magnetic patches.]
{Schematic pictures of magnetic activities between two patches (left column) and corresponding 3D 
pictures (right column).
(a) Merging. Two patches converge and coalesce into one patch.
(b) Splitting. One patch is divided into two patches.
(c) Emergence. Two opposite polarities with equal amount of flux content appear.
(d) Cancellation. Two opposite polarities converge with each other. One of them disappears.
 The other loses its flux content or disappears.}
\label{fig:pic_magac}
\end{figure}

The magnetic field near the surface is transported and entangled by the convective motion of the plasma.
The appearance of such modulation is observed as various activities in the line-of-sight magnetic component.
There are four surface magnetic processes to change a frequency distribution,
namely emergence, merging, splitting, and cancellation \citep{sch1997b, par2001}.
Left column of Figure \ref{fig:pic_magac} shows schematic pictures of four activities between two patches.
Note that there are diffusions and appearances of the unipolar patch in observational data, which do not correspond any of them.
\cite{lam2010} reports that an unipolar appearance in intermediate resolution data corresponds to merging of small patches in high resolution data.
It deduces that these unipolar process may be caused by the surface processes involving undetected patches.

Flux emergence is a divergence of opposite polarities in line-of-sight magnetogram.
The diverging velocity of polarities are reported by some authors \citep{bra1985,zwa1985, ots2007, ots2010}.
The separation speed is $\thicksim 4$ km s$^{-1}$ in the early phase and decreases to $\thicksim 1$ km s$^{-1}$.
It is much larger than that of convective motion on the solar surface, $\thicksim 1$ km s$^{-1}$.
The emergence contributes to the increase in total flux amount.
It is interpreted as a flux ascension from below the photosphere \citep{zir1972, zwa1987}.
Figure \ref{fig:pic_magac} (c) shows a schematic picture of it.
Downflow is observed at the foot point by Doppler velocity observation \citep{zir1972}.
It is interpreted as dropping material by the gravity of the Sun.
Upflow is also observed in the middle of the separating polarities \citep{bru1969,zir1972}.
These signatures support a picture of ascending flux tube from below the photosphere.

Flux cancellation is a convergence and a disappearance of magnetic fluxes of positive and negative polarities in 
line-of-sight magnetograms.
This phenomena is observed in both active regions \citep{zwa1978, mar1985, zir1985} and 
quiet regions \citep{har1976,liv1985,wan1988,har1985,har1993,har1996,sch1998}.
Two physical models, U-loop emergence and $\Omega$-loop submergence, are proposed \citep{zwa1987}.
Figure \ref{fig:pic_magac} (d) shows the $\Omega$-loop submergence model.
\cite{har1999} investigate timing difference of cancellation in different two layers: One 
should observe cancellation earlier in the lower layer than in the upper layer in U-loop emergence and vice versa in $\Omega$-loop submergence.
They obtained that most of cancellations (20 of 45 cancellations, $44\%$) are interpreted as an $\Omega$-loop submergence.
Some recent studies report the Doppler velocity in canceling region \citep{yur2001, cha2002, kub2010, iid2010, cha2010}.
\cite{cha2002} report downflow at the horizontal field connecting canceling patches near active region. 
\cite{iid2010} report time evolution of downflow around cancellation site in network field.
The lifetime of downflow is significantly larger than velocity fluctuation in surroundings.
These papers report same downward Doppler velocity, $1$ km s$^{-1}$.
As for more tiny cancellation in granular scale, significant signature of downflow has not been found so far \citep{kub2010}.
The other supporting material for $\Omega$-loop submergence is that cancellation in a quiet region often coincides an overlying bright point in 
EUV and X-ray image \citep{har1996}.
This suggests that the photospheric cancellation is a manifestation of the reconnection and corresponds a fieldline interaction 
in the upper atmosphere \citep{pri1994,par1994,par1994b,sch1998,litv1999}.
Some papers report result of numerical calculation including convective effect \citep{cheu2010}.
They found that cancellation takes place in a numerical box and it is an $\Omega$-loop submergence.
On the other hand, \cite{zha2009} report the blueshift motion around the cancellation sites from results of the spectropolarimetric observation.
There are no statistical investigations of the direct measurement of Doppler velocity around cancellation sites so far.
It is still an open question whether cancellation is an $\Omega$-loop submergence or a U-loop emergence.

Emergence and cancellation have direct relation with flux product and flux loss in line-of-sight magnetogram.
The flux replacement time scales by these processes are defined as 
\begin{equation}
\tau_{\rm emrg}=\Phi_{\rm tot} \, \left( \left. \frac{\partial \Phi_{\rm tot}}{\partial t} \right|_{\rm emrg} \right)^{-1}
\end{equation}
\begin{equation}
\tau_{\rm cnc}=\Phi_{\rm tot} \, \left( \left. \frac{\partial \Phi_{\rm tot}}{\partial t} \right|_{\rm cnc} \right)^{-1}
\end{equation}
where, $\partial \Phi_{\rm tot}/\partial t$ is flux change rate of each activity.
The previous papers show they drastically decrease from several days to several hours 
as spatial resolution becomes higher \citep{mar1985, sch1998, hag2001}.
It is not clear why they decrease with higher resolution.

Figure \ref{fig:pic_magac} (a) and (b) show schematic pictures of merging and splitting.
They do not change total flux amount on the solar surface but change the flux content of patches.
There are few reports for merging and splitting compared to those of emergence and cancellation.

Some analytical models are suggested. 
Magneto-chemistry (M-C) equation is useful in this point.
It is suggested by \cite{sch1997b}.
It describes a relationship between frequency distributions of surface processes and that of 
flux content. 
It is written as
\begin{align}
\displaystyle \frac{\partial n_{\rm \pm}(\phi)}{\partial t} & = S_{\pm}(\phi) \,  \notag\\
&+ \frac{1}{2}\int_{0}^{\infty}n_{\rm \pm}(x)n_{\pm}(\phi-x)l_{\pm}(x,\phi-x)dx \, - \,n_{\rm \pm}(\phi)\int_{0}^{\infty}n_{\rm \pm}(x)l_{\pm}(\phi,x)dx \, \notag\\
&+ 2\int_{0}^{\infty}n_{\rm \pm}(x)k_{\pm}(\phi,x-\phi)dx \, - \, n_{\rm \pm}(\phi)\int_{0}^{\phi}k_{\pm}(x,\phi-x) dx \notag\\
&+ \int_{0}^{\infty}n_{\rm \pm}(\phi+x)n_{\rm \mp}(x)m_{\pm}(\phi+x,x)dx \, - \,n_{\rm \pm}(\phi)\int_{0}^{\infty}n_{\rm \mp}(x)m_{\pm}(\phi,x)dx \notag \\
&\, 
\end{align}
where $n(\phi)$ is a frequency distribution of flux content, $S(\phi)$, $l(x,y)$, $k(x,y)$, and $m(x,y)$ are functions representing probability density distributions 
of emergence, merging, splitting, and cancellation respectively (See Section \ref{sec:magch} for more detailed explanation).
There are five unknown functions in this equation.
What we have obtained from observation is $n(\phi)$ for time independent case.
Some authors found sets of [$S(\phi)$,$l(x,y)$,$k(x,y)$,$m(x,y)$] to make a time-independent solution of $n(\phi)$ with some assumptions.
\cite{sch1997b} found one particular solution for an exponential distribution.
They assumed 1) re-appearance of submerged flux by cancellation, 2) $l(x,y)$=$m(x,y)$=constant, and 3) $k(x,y)$=constant.   
another set was found by \cite{par2002}.
They assume re-appearance of submerged flux through cancellation.
Probability density distributions are written as 
\begin{align}
k_{\pm}(x,y)&=\frac{k_0}{2A}f_{\pm}(x)f_{\pm}(y)g_{\pm}(x,y)\\
l_{\pm}(x,y)&=\frac{3k_0}{n_{\pm}}f_{\pm}(x+y)g_{\pm}(x,y)\\
m_{\pm}(x,y)&=\frac{k_0}{n_{\pm}}\frac{f_{\pm}(x-y)f_{\pm}(y)}{f_{\mp}(y)}g_{\pm}(x-y,y)
\end{align}
where $k_0$ is a constant, $n_{\pm}(\phi)=n_{\pm}/A \, f_{\pm}(\phi)$, $g(x,y)$ is an arbitrary function. 
Moreover, they assume 
\begin{equation}
g_{\pm}(x,y)=\frac{\exp(-(x+y)/\eta_{\pm})}{f_{\pm}(x)f_{\pm}(y)f_{\pm}(x+y)}.
\end{equation}
The deduced frequency distribution of emergence and cancellation becomes exponential in this solution.
These solutions satisfy the frequency distribution of flux content in their study.

We have not found a solution for a power-law distribution demanded from the observational result by \cite{par2009}.
It is difficult to obtain a proper solution because we need at least three assumptions to obtain all frequency distributions even in a 
time-independent solution.
One approach is to measure the frequency distribution of processes directly from observation.
The difficulty of this approach is that it needs a statistical investigation, namely that we need to look huge amount of data or 
develop an auto-detection code for surface processes of magnetic patches.
Frequency distribution of emerging flux has been investigated by several authors 
\citep{har1975, har1993, har1993b, tit2000, cha2001b, hag2001, hag2003, hag2008, tho2011}.
\cite{tho2011} gives one conclusion.
They found a power-law distribution with an index of $-2.69$ spanning from active region to inter-network field 
by using {\it Hinode}/SOT magnetogram and the previous result (Figure \ref{fig:dist_par}).
There are no reports of frequency distributions of other processes. 

\begin{figure}[p]
\centering
\includegraphics[bb=0 0 180 180,width=0.9\textwidth]{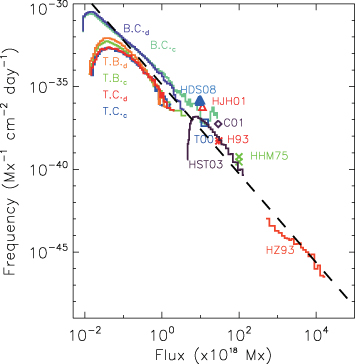}
\caption[Frequency distribution of flux emergence by Thornton $\&$ Parnell (2011)]{
Frequency distribution of flux emergence reported by Thornton $\&$ Parnell (2011).
They plot a frequency distribution of emergence on flux content obtained by their detection-code 
(indicated by B.C. and T.C. in the figure) 
and previous papers (indicated by the first character of authors and published year) together.
Their fitting results in a single power-law distribution with an index of $-2.69$. 
}
\label{fig:dist_tho}
\end{figure}

\section{Purpose of the Thesis}
\label{intro_purpose}

A quiet region is thought to have an important role in flux and energy transport on the solar surface 
because it has a plenty of flux budget and a large number of energetic activities.
\cite{par2009} reported that frequency distribution of flux content has a power-law form with an index of $-1.85\pm0.14$.

Our main purpose is to investigate what makes a power-law distribution.
The strategy of this study is that we investigate the frequencies of magnetic activities, namely emergence, splitting, merging, and cancellation, 
 based on observations and construct a model from the result. 
The auto-detection code is needed 
and is developed in this study to investigate frequencies of surface processes of magnetic patches.
Our target is network field. 
One reason is that most flux in quiet regions is contained in network field \citep{mar1990,sch1998}.
The other reason is that the lower limit of statistical study with recent high-resolution data, namely magnetograms obtained by {\it Hinode} 
spacecraft, is $\thicksim 10^{17}$ Mx that is good enough for network field while is insufficient for internetwork field.

The thesis has 6 chapters.
The detail of the data set and method of analysis are explained in Chapter 2. 
The description of data set and our auto-detection method (namely our definition of patches and activities) are explained there.
Detail description of M-C equation is summarized in Chapter 3.
The forms between frequencies of surface processes of magnetic patches and physical quantities are deduced there.
We introduce a detection limit for statistical study, above which we observe all phenomena in statistical sense.
Flux dependence of various properties of magnetic patches, size, flux content, proper velocity, and lifetime, are investigated in Chapter 4.
Frequencies of magnetic activities are investigated in Chapter 5.
We suggest one qualitative model in the discussion there from the observational results.
The summary of the thesis and future works are summarized in Chapter 6.

\ifodd \arabic{page}
\else
  \thispagestyle{fancy}
  \mbox{}
  \newpage
  \clearpage
\fi

\chapter{Data Description and Method of Analysis}
\thispagestyle{fancy}

We show data description used in the thesis and our method of detection of patches and surface processes of them in this chapter.
Two sets of magnetograms obtained by {\it Hinode} spacecraft are used in this study.
The instrumentation of {\it Hinode} spacecraft and data description are shown in Section \ref{sec:data_dscrpt}.
Our method of detection is explained in Section  \ref{sec:autodet}.

\pagestyle{fancy}
\section{Data Description}
\label{sec:data_dscrpt}

\subsection{\ion{Na}{I} D$_1$ Magnetogram Obtained by ${\it Hinode}$/NFI}
\label{sec:hinode}
Two sets of line-of-sight magnetograms obtained by the Narrowband Filter Imager (NFI) of the Solar Optical Telescope (SOT) on board 
${\it Hinode}$ spacecraft are used in this study.
The reason why we use {\it Hinode}/NFI magnetograms in this study is that the observed range of flux content is 
wide enough for an analysis of network field, which is not accomplished with the other instruments.
For example, the detection limit of MDI high-resolution data is $\thicksim 10^{18}$ Mx (see Figure \ref{fig:dist_par}).
We need wider range of flux content for a investigation of scale dependence of the network field ($<10^{19}$ Mx).
We summarize the description of magnetograms obtained by ${\it Hinode}$/NFI in this section.

\begin{figure}[thp]
\centering
\includegraphics[bb= 200 800 700 1350,width=0.50\textwidth]{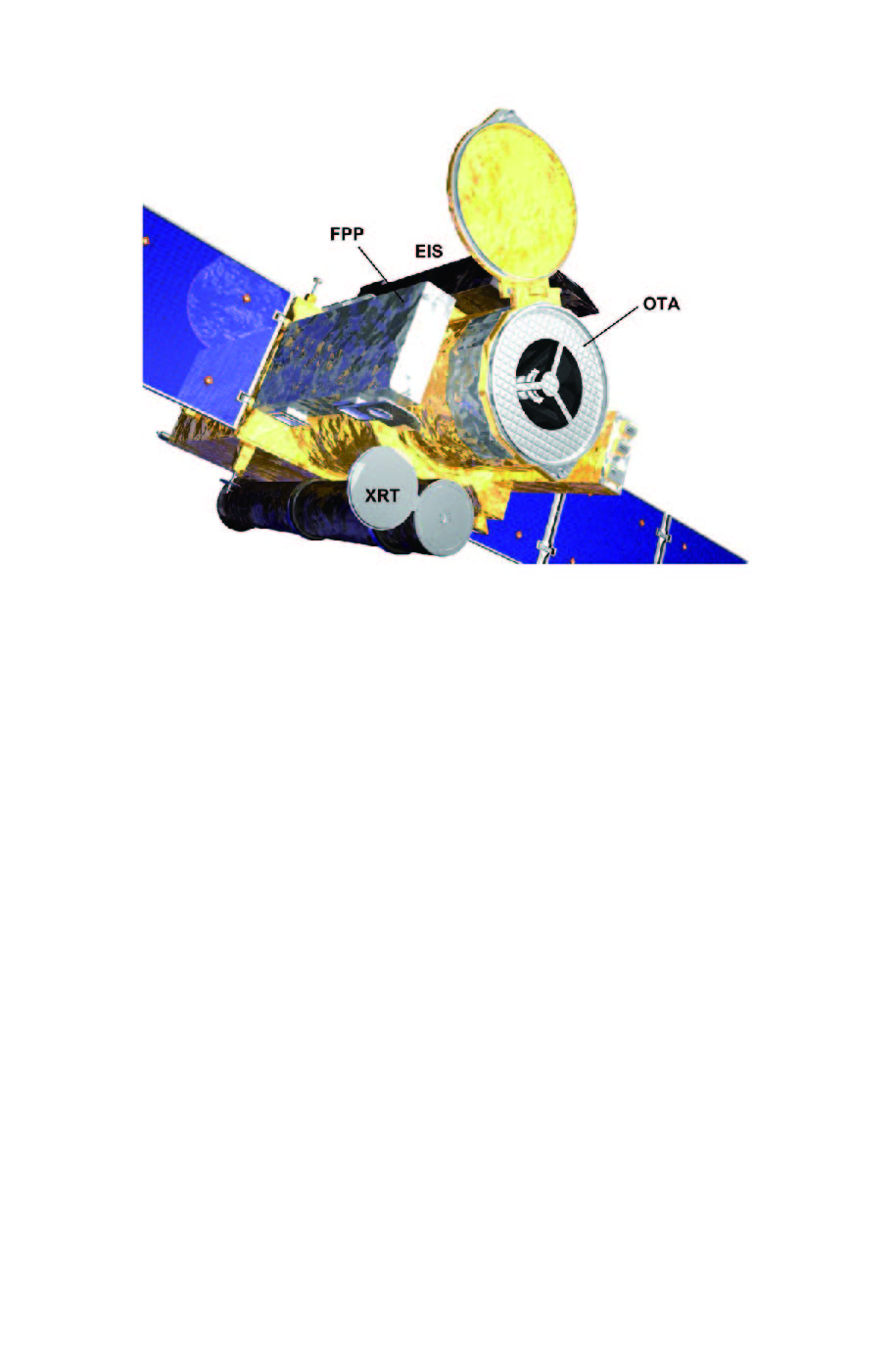}
\caption[Outlook of ${\it Hinode} spacecraft$.]{
Outlook of ${\it Hinode}$ spacecraft from \cite{tsu2008}.
}
\label{fig:hinode}
\end{figure}

${\it Hinode}$ is a Japanese mission of spacecraft observing the Sun \citep{kos2007}.
It is developed by ISAS/JAXA, with a collaboration of NAOJ, NASA, and STFC.
Figure \ref{fig:hinode} shows ${\it Hinode}$ spacecraft.
There are three instruments on board, Solar Optical Telescope \citep[SOT;][]{tsu2008, ich2008, shi2008, sue2008}, 
Extreme Ultra Violet Imaging Spectrometer \citep[EIS;][]{cul2007}, and X-Ray Telescope \citep[XRT;][]{golu2007}.
It is on a circular orbit with an altitude of $\thicksim 680$ km.
${\it Hinode}$ can continuously observe the Sun nine months a year.

SOT is the largest telescope which has a Gregorian optics with $50$-cm aperture.
Its diffraction limit is $0.2'' \-- 0.3''$ for the range of $3880$-$6302$ \AA .
The solar ray is divided to four paths in SOT, 
namely NFI, the Broadband Filter Imager (BFI), the Spectro Polarimeter (SP), and the Correlation Tracker (CT). 
NFI, which is used in this study, can record the filtergram image, dopplergram, and polarization image of some 
photospheric and chromospheric lines selected from \ion{Mg}{I} ($5172.7$\AA ), \ion{Fe}{I} ($5250.2$\AA , $5247.1$\AA , $5250.6$\AA , $5576.1$\AA ,
$6301.5$\AA , $6302.5$\AA), \ion{H}{I} ($6562.8$\AA ), and \ion{Na}{I} ($5895.9$\AA ). 
The full field of view of CCD camera is $328'' \times 164''$. 
The pixel size of CCD camera is $0.08''$.
The actual observation type is set to meet the demand and follow the restriction of data telemetry.

We use the circular polarization images of \ion{Na}{I} D$_1$ resonance line, whose center is $5896$\AA, in this study.
The formation height of this line is thought to be from the upper photosphere to the lower chromosphere.
The Lande factor of this line is 1.33. 
The circular polarization images shifted from the line center are taken by SOT/NFI in our data set.
They are converted to the magnetograms by considering the Zeeman effect.
We employ the weak field approximation in this conversion.
The detail of it is explained in the next chapter.  
The typical detection limit of longitudinal component of magnetic field strength is $\thicksim$ 11 G depending on signal-to-noise ratio.

\subsection{Description of Data Sets}
\label{sec:data}

We use two data sets of circular polarization images of \ion{Na}{I} D$_1$ resonance line.
One has a high time cadence but the observing period is relatively short. 
The other has a long period but the time cadence of it is relatively low.
From here, we call the former data set as data set 1 and the latter as data set 2
(Table 2.1).

\begin{figure}[tp]
\centering
\includegraphics[bb= 0 0 550 500,width=0.9\textwidth]{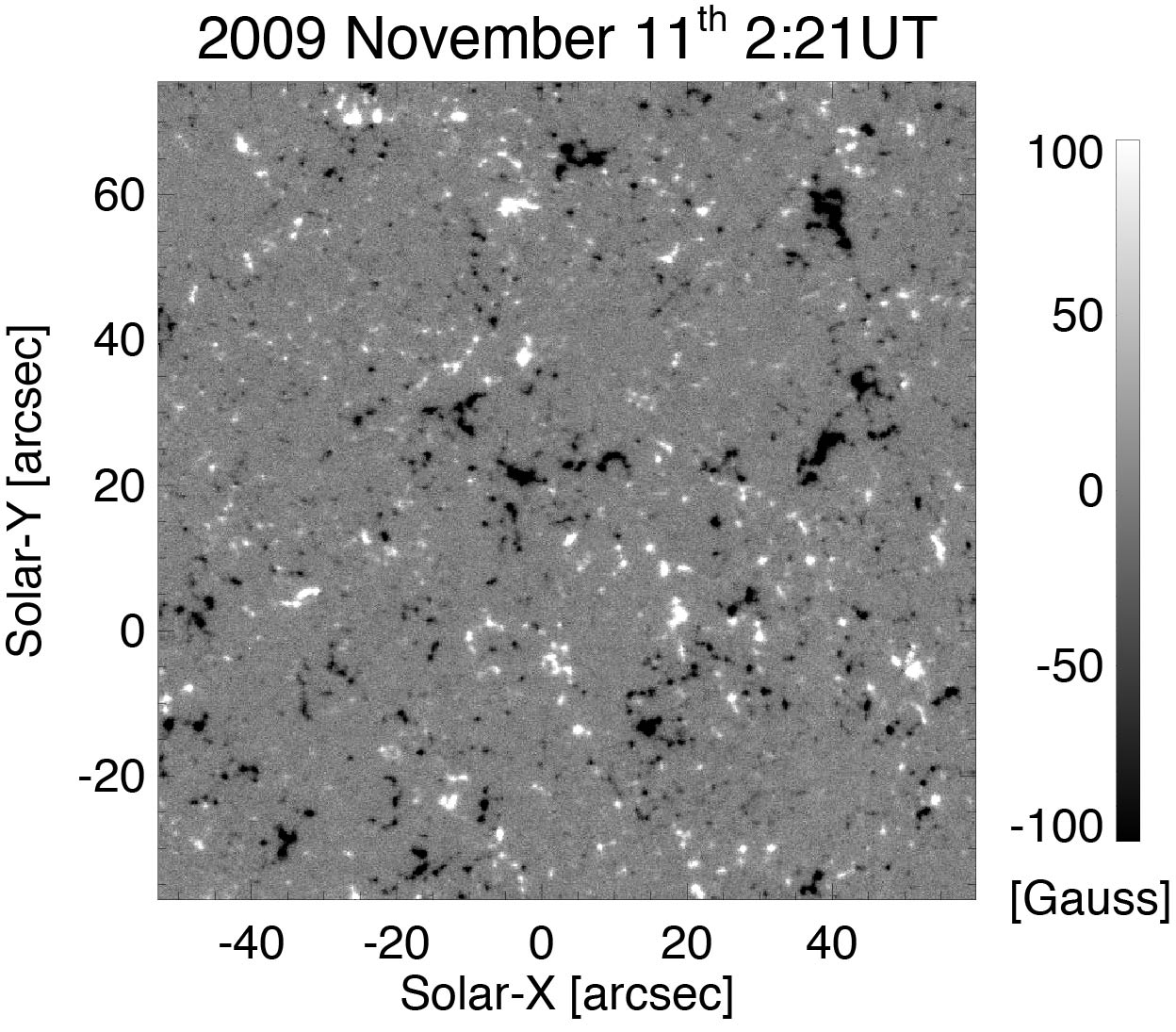}
\caption[Example of \ion{Na}{I} D$_1$ magnetograms in data set 1.]{
Example of \ion{Na}{I} D$_1$ magnetograms in data set 1 after the pre-process explained in section \ref{sec:auto_patch}.
This magnetogram is taken at 2:21UT on 2009 November 11, which is the middle of the observational period.}
\label{fig:sample_d1}
\end{figure}

The time period of data set 1 is from 0:33UT to 4:08UT on 2009 November 11$^{th}$.
The time duration is 3 hour 39 minutes. 
${\it Hinode}$ observes a quiet region near the disk center.
It obtains intensity signal (Stokes-I) and circular polarization signal (Stokes-V) 
at the two wavelength points shifted from the center of \ion{Na}{I} D$_1$ resonance line by $\pm$160m\AA.
The magnetogram and dopplergram, which are calculated from these signals on the spacecraft, are downlinked to the ground.
The center of the SOT view moves from $(-14.1'',19.0'')$ to $(20.2'',19.4'')$ in heliocentric coordinates during this period.
The data are summed by 2-pixels in both x and y directions.
The pixel size of images becomes $0.16''$.
Field of view in this data set is $113'' \times 113''$.
Some network cells, which have a typical size of $20'' \-- 40''$, are included in the field of view (Figure \ref{fig:sample_d1}).
The total number of images in this data set is 215 during the whole observational period.
There are sometimes data missing lines in some images, which are lost during data transferring.
We remove such images to obtain homogeneous images.
The total number of polarization images is 199 after the removal.
Time interval between consecuting magnetograms is $\thicksim 1$ minute.
It sometimes becomes 2 minutes due to our removal of data.

\begin{figure}[tp]
\centering
\includegraphics[bb=0 0 550 500,width=0.9\textwidth]{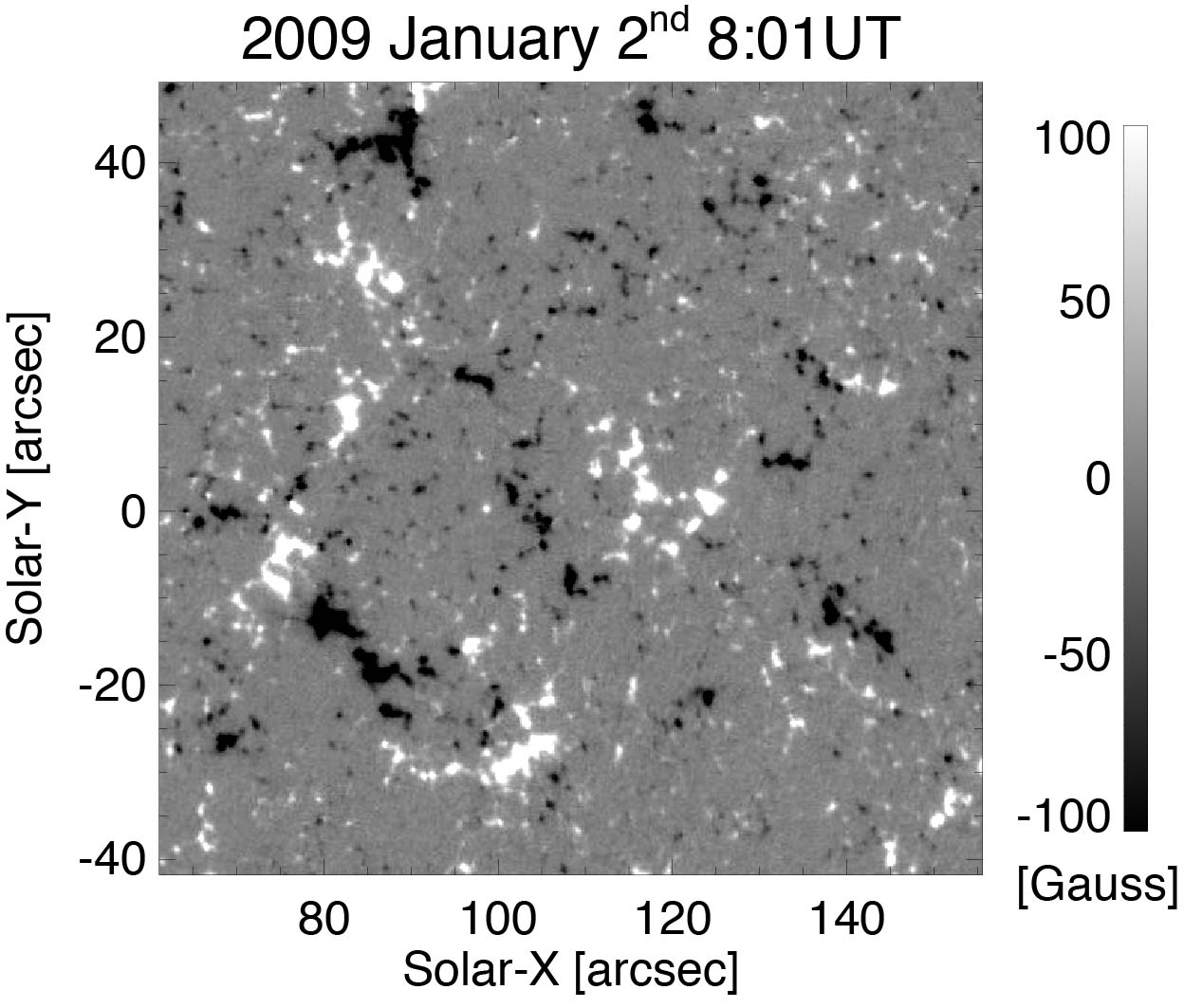}
\caption[Example of \ion{Na}{I} D$_1$ magnetograms in data set 2.]{
Example of \ion{Na}{I} D$_1$ magnetograms in data set 2 after the pre-process 
explained in the section \ref{sec:auto_patch}.
This magnetogram is taken at 8:01UT on 2009 January 2, which is the middle of observational period.}
\label{fig:sample_d2}
\end{figure}

The time period of data set 2 is from 10:24UT on 2008 December 30 to 5:37UT on 2009 January 5$^{th}$.
The time duration is 115 hour 13 minutes. 
${\it Hinode}$ tracks a quiet region near on the solar equator during this period.
SOT/NFI obtains circular polarization signal at only one wavelength point shifted from 
\ion{Na}{I} D$_1$ resonance line by 140m\AA.
The circular polarization degree is calculated on the spacecraft.
Note that the evaluation of magnetic flux density from this cirular polarization degree
contains an uncertainty with an influence of the Doppler shift which is not measured in this data set.
The center of SOT view moves from $(-520.9'',-9.3'')$ to $(695.9'',-6.9'')$ in solar coordinate during this period.
The data are summed by 2 pixels in both x and y directions. Then the pixel size becomes $0.16''$.
Full field of view in this data set is $113'' \times 113''$.
The studied field of view is reduced to $94.9'' \times 91.4''$ due to the removal of image edges for the correction of spacecraft jittering.
As same as data set 1, some network cells are included in the field of view (Figure \ref{fig:sample_d2}).
Total number of images in the data set is 1642 after removing the data which have missing lines.
Time interval between consecuting magnetograms is $\thicksim 5$ minutes and sometimes $\thicksim 10$ minutes.

\begin{table}
\begin{center}
\label{tab:sum_data}
\begin{tabular}{c|ccp{5zw}}
\hline\hline
&Data Set 1&Data Set 2\\
\hline \hline
Observing Line & \ion{Na}{I} D$_1$ $5896$ \AA  &\ion{Na}{I} D$_1$ $5896$\AA  \\
Pixel Size & $0.16''$ ($ 2 \times 2$ summed)& $0.16''$ ($2\times 2$ summed) \\
\hline
\hline
Observational Period& \ & \ \\
(Start)&2009 Nov. 11 0:30UT & 2008 Dec. 30 10:24UT\\
(End)&2009 Nov. 11 4:09UT & 2009 Jan. 5 5:37UT\\
\hline
Center of FoV&\ & \ \\
(Start)&($-14.1''$,$19.0''$)&($-520.9''$,$-9.3''$)\\
(End)&($20.2''$,$19.4''$)&($695.9''$,$-6.9''$)\\
\hline
Duration & 3hr 39min & 115hr 13min\\
Number of Images & 199 & 1642 \\
Time Cadence & $\thicksim$ 1 minute& $\thicksim$ 5 minutes\\
FoV&$113'' \times 113''$&$94.9'' \times 91.4''$\\
Obsevation Type&Stokes IV+DG&Stokes-V/I\\
Wavelength offset &+/-160m\AA&+140m\AA\\
\hline\hline
\end{tabular}
\end{center}
\caption[Summary of data sets.]{
Summary of data set description.
}
\end{table}

\section{Auto-Detection Algorithm}
\label{sec:autodet}

\begin{figure}[tp]
\centering
\includegraphics[bb=100 0 500 300,width=0.90\textwidth]{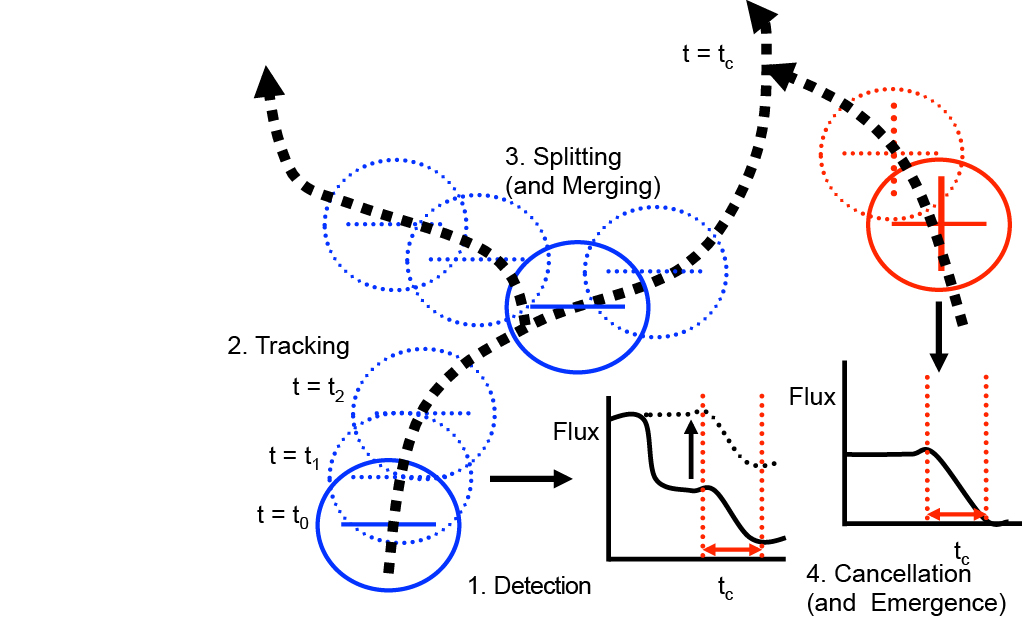}
\caption[Schematic picture of tracking of patches and detection of surface processes of magnetic patches.]{
Schematic picture of our patch tracking and detection f surface processes.
(1) Detection of patches with a clumping algorithm.
(2) Track of patches by examining overlaps.
(3) Detection of merging and splitting by examining overlaps.
(4) Detection of cancellation and emergence by making pairs of flux change events.
}
\label{fig:pic_autodet}
\end{figure}

We define four surface processes of magnetic patches and explain our method for identifications of them in this chapter.
Figure \ref{fig:pic_autodet} represents schematic pictures of it.
The basic chart is as follows:
1. Detection of patches with a clumping algorithm.
2. Track of patches by examining overlaps.
3. Detection of merging and splitting by examining overlaps.
4. Detection of cancellation and emergence by making pairs of flux change events.

\subsection{Preprocessing of Data Sets}
\label{sec:auto_pre}

\begin{figure}[tp]
\centering
\includegraphics[bb=0 0 520 600,width=0.90\textwidth]{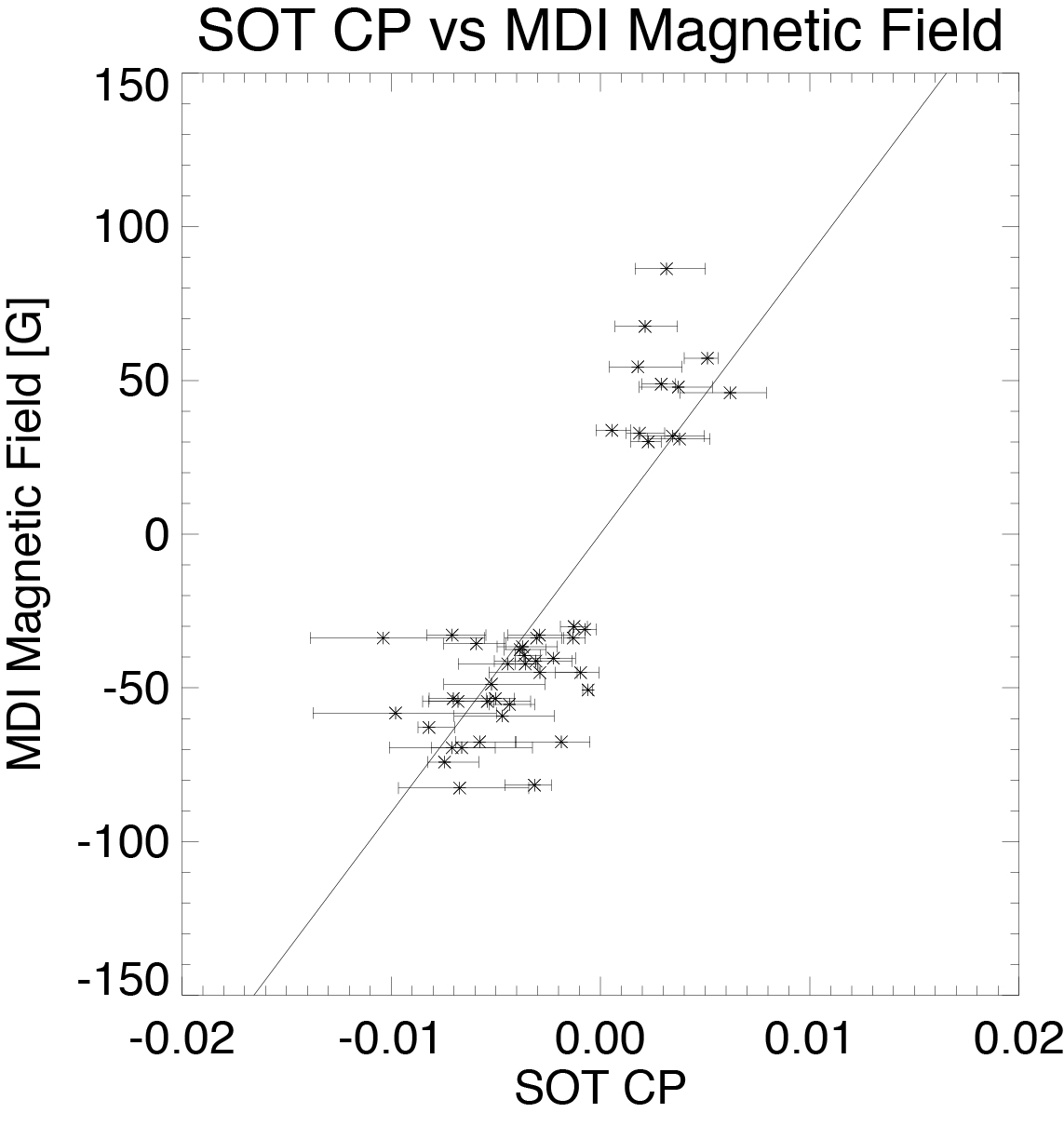}
\caption[Scatter plot of CP obtained by {\it Hinode}/SOT and the magnetic flux obtained by {\it SOHO}/MDI in data set 1.]{
Scatter plot of CP obtained by {\it Hinode}/SOT and the magnetic flux obtained by {\it SOHO}/MDI in data set 1.
Asterisks show the averaged SOT CP signal corresponding to 1 pixel of MDI. The horizontal bars indicate the minimum and maximum value of 
the SOT CP signal in each MDI pixel. 
The solid line shows a result of a linear fitting, whose slope is $9067.38$ G DN$^{-1}$.
The fitting range is from $30$ G to $100$ G in absolute magnetic field strength.
}
\label{fig:cpdn_d1}
\end{figure}

\begin{figure}[tp]
\centering
\includegraphics[bb=0 0 520 600,width=0.90\textwidth]{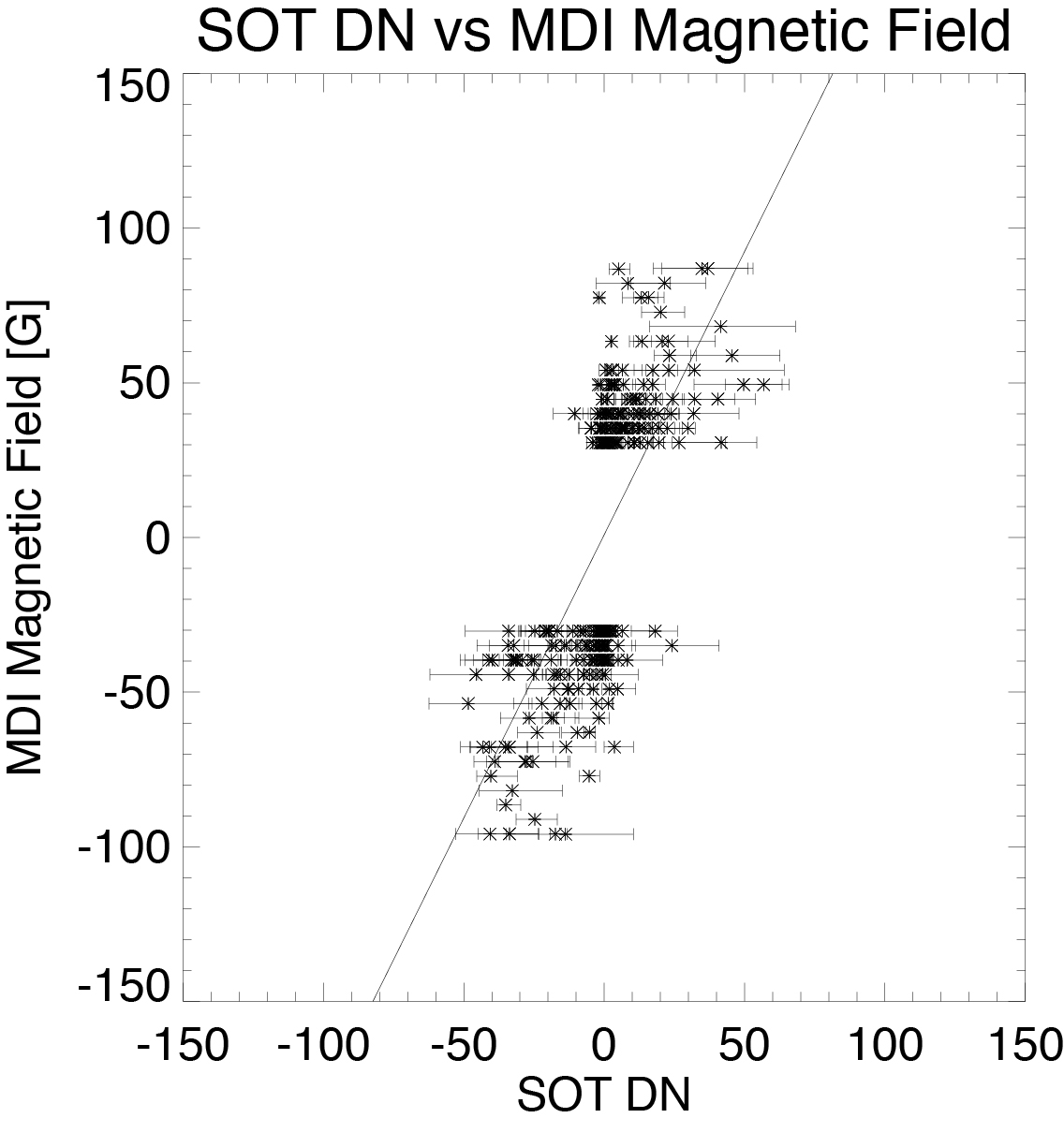}
\caption[Scatter plot of CP obtained by {\it Hinode}/SOT and magnetic flux obtained by {\it SOHO}/MDI in data set 2.]{
Same plot as Figure \ref{fig:cpdn_d1} but for data set 2.
The slope of fitting line is $1.83191$ G DN$^{-1}$.}
\label{fig:cpdn_d2}
\end{figure}

We have to convert polarization signal to magnetic field strength in data set 1.
The dark current and the flat field of the CCD camera are calibrated by using the procedure ${\textbf {\rm fg\_prep.pro}}$ included in 
the SolarSoftWare (SSW) package.
We apply the Zeeman weak-field approximation to obtain the line-of-sight magnetic field strength 
from the circular polarization (CP hereafter) signal and it is thought to be applicable to most of 
network field \citep{lan1973, lan2004, ase2011}.
The proportional relationship stands up in this approximation.
The condition for its approximation is that the effective wave shift of circular polarization by Zeeman effect 
is smaller that line broadening effect, which can be deduced as
\begin{equation}
\displaystyle B < \frac{4 \pi m c}{\overline{g} \lambda_0 e}\sqrt{\frac {2k_B T}{M} + v_{\rm mic}^2}
\end{equation}
where $m$ is the electron mass, $c$ is the speed of light, $\overline{g}$ is the effective Lande factor of the line,
$\lambda_0$ is the wavelength of the line center, $e$ is the electron charge, $k_B$ is the Boltzmann constant,
$T$ is temperature, $M$ is the mass of Na atoms, and $v_{\rm mic}$ is the microturbulence.
We substitute $\lambda_0=5896$ \AA , $\overline{g}=1.33$, $T=6000$ K, and $v_{\rm mic}=1$ km s$^{-1}$ for this case and 
obtain the condition, B< 1800 G.
This is satisfied in quiet regions.
The conversion coefficient is determined by comparing the {\it SOHO}/MDI magnetogram \citep{scher1995} and 
circular polarization in one image.
We spatially smear the SOT data to two times the MDI pixel size (Dr. R. A. Shine, private communication) 
and make a linear fitting between CP in the SOT and the magnetic field in the MDI.
Because the pixels in the SOT data, which are out of linear range, are too weak or too strong
, we make the fitting in the range from $30$ G to $100$ G as a magnetic field strength.
Figure \ref{fig:cpdn_d1} and Figure \ref{fig:cpdn_d2} shows scatter plots of signals obtained 
by SOT data numbers and magnetic field strength obtained by the MDI in data set 1 and 2.
We obtain $9067.98$ G DN$^{-1}$ and $1.83191$ G DN$^{-1}$ as a conversion coefficient respectively.

Some corrections are done before the detection.
Due to the change of the viewing angle into the solar surface in the long observing duration, 
line-of-sight magnetic field does not correspond to magnetic field perpendicular to the solar surface.
We project the observed line-of-sight magnetic field to perpendicular field to the solar surface 
assuming that the magnetic field is perpendicular to the solar surface.
There are sudden changes of field of view when the position of tip-tilt mirror hits the stroke limit due to the 
long observational period of data set 2.
We remove such change of field of view by making a correlation of magnetic field between consecutive images.
Then both data sets are rotated to the position at certain time, 2:03UT on November 11$^{th}$ for data set 1 
and 20:35UT on January 1$^{st}$ for data set 2,
 when the regions are near the disk center. 
There is sometimes a column wise offset of CP in NFI \citep{lam2010}, whose reason is not known.
We remove it by subtracting an average value of each line from all pixels in the same line.
Magnetograms are averaged over 3 continuous images and 3 pixels for smoothing.
Without these smoothings, the auto-detection does not give the stable results owing to the 
flipping noise around the threshold level.
Note that these smoothings deduce the time-resolution of data set, namely 3 minutes for data set 1 and 15 minutes for data set 2.

\subsection{Detection and Tracking of Magnetic Patches}
\label{sec:auto_patch}

\begin{figure}[tp]
\centering
\includegraphics[bb=0 0 900 750,width=0.90\textwidth]{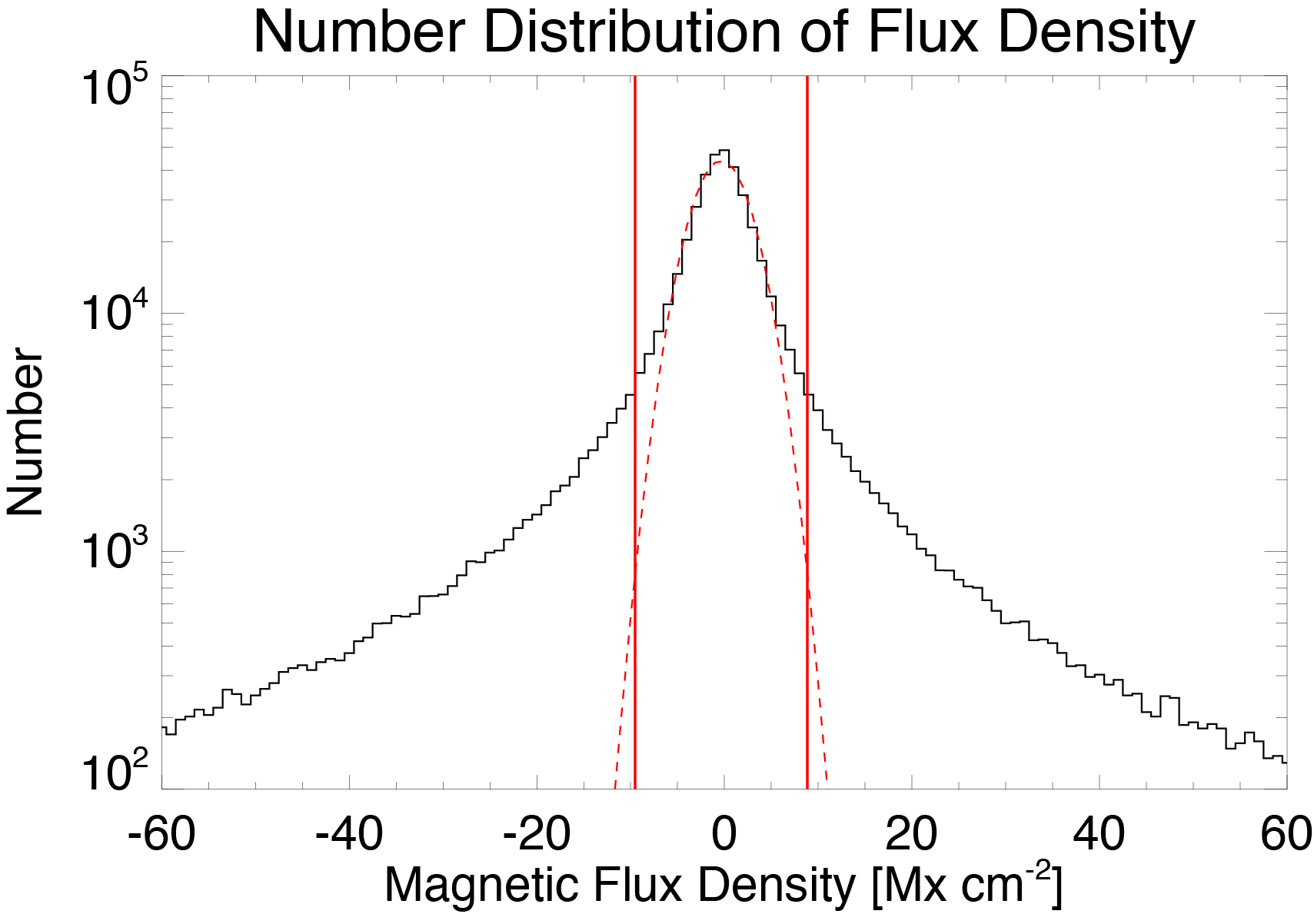}
\caption[Histogram of the magnetic flux density in one image of data set 1.]{
Histogram of the magnetic flux density in one magnetogram of data set 1.
The histogram represents the observational result and the red dotted curve represents fitting result.
Two vertical red lines indicate $2 \sigma$.}
\label{fig:hist_sgnl_d1}
\end{figure}

\begin{figure}[tp]
\centering
\includegraphics[bb=0 0 1200 1100,width=0.99\textwidth]{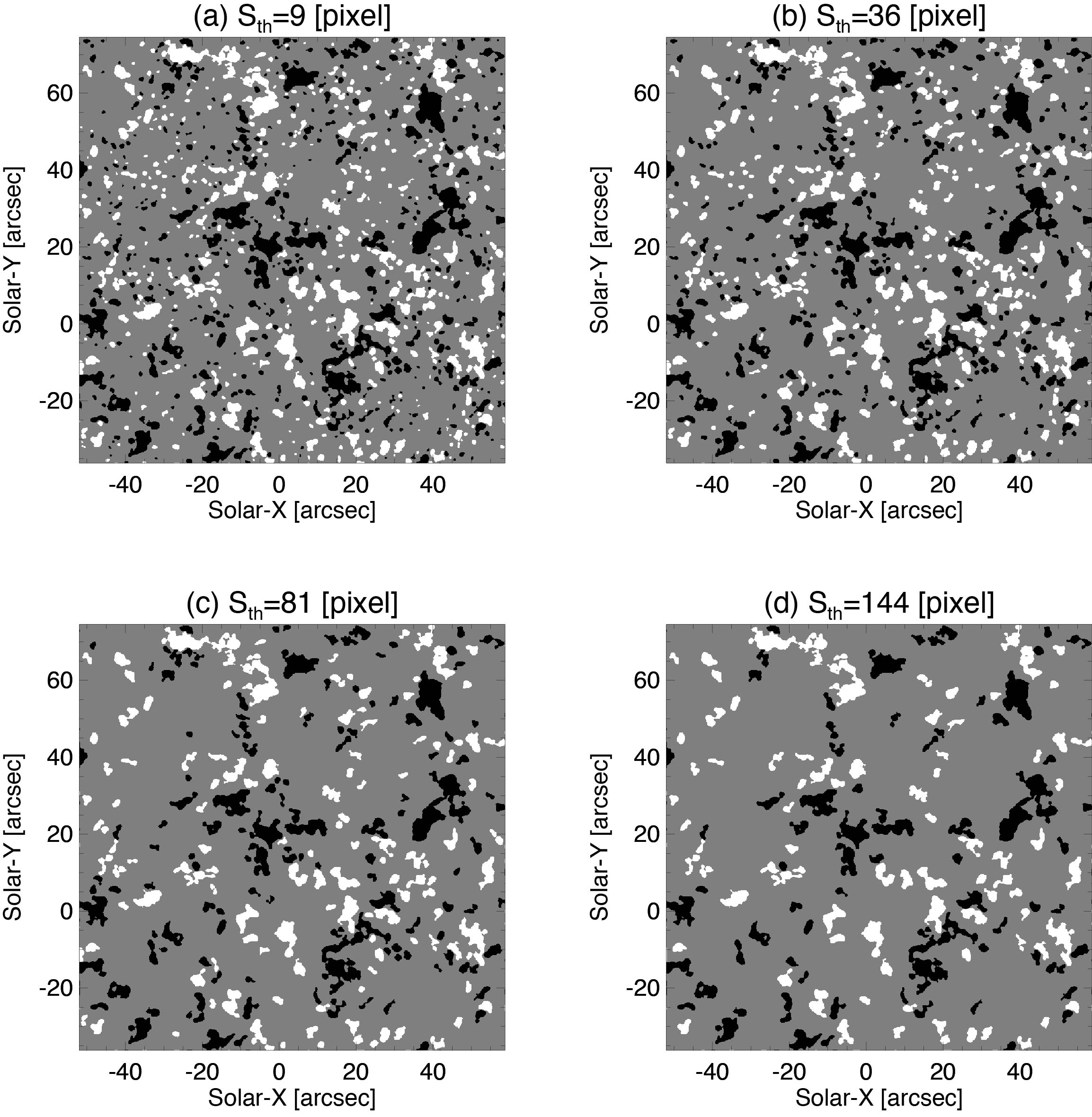}
\caption[Two-leveled magnetograms of data1 with different size thresholds.]{
Two-leveled magnetograms taken at 2:21 on 2009 November 11$^{th}$ with different size thresholds.
Size thresholds are set as (a) $9$ pixels, (b) $36$ pixels, (c) $81$ pixels, and (d) $144$ pixels.}
\label{fig:th_d1}
\end{figure}

Magnetic patches are detected in this step.
We adopt nearly same definition as \cite{par2002} in this study.
We use a clumping method for the detection of patches: 
each patch is picked up as a clump of marked pixels having magnetic strength beyond a given threshold.
We set a signal threshold in this method.
It is obtained by fitting the histogram of the signed magnetic field strength per pixel by Gaussian function in each magnetogram.
Figure \ref{fig:hist_sgnl_d1} shows an example of histogram and fitting result in one image of data set 1.
The horizontal axis indicates magnetic field strength and the vertical axis indicates number of patches per bin.
Histogram represents the observational result and dotted curve represents the fitting results.
Two vertical red lines indicate $2 \sigma$, which we use as the threshold.
The actual value of $\sigma$ is $\thicksim 5$ G pixel$\ \sim 6.8 \times 10^{14}$ Mx for each image.
This value is close to that of \cite{par2009}, who report $7.0 \times 10^{14}$ Mx.
We set a size threshold to pick up network field in addition to signal threshold.
One scale on the network field is granular size.
It has a typical scale of $10^3$ km ($\thicksim 9$ pixel size in SOT/NFI).
We pick up magnetic patches with sizes beyond 81 pixels for focusing our analysis on the network magnetic field. 
The validity of this choice is demonstrated in Figure \ref{fig:th_d1}, which shows two-leveled magnetograms with different size threshold.
As shown in this figure, one can pick up only the network fields without internetwork fields by adopting 81-pixels as a threshold.

The reasons why we adopt the clumping method are that it is simple and many of previous studies employ it.
One of the limitations of this method is that we cannot investigate the inner structures of the patches.
Note that \cite{def2007} reportd that there are differences of the obtained frequency distribution of flux content 
among the detection method below the range of $1.5 \times 10^{18}$ Mx with {\it SOHO}/MDI data.
We expect that there is no difference in the analysis range ($> 3.1 \times 10^{17}$) because the resolution of {\it Hinode}/SOT is 
better than that of {\it SOHO}/MDI by more than one-order of magnitude in the flux range

\begin{figure}[p]
\centering
\includegraphics[bb= 0 0 550 600,width=0.99\textwidth]{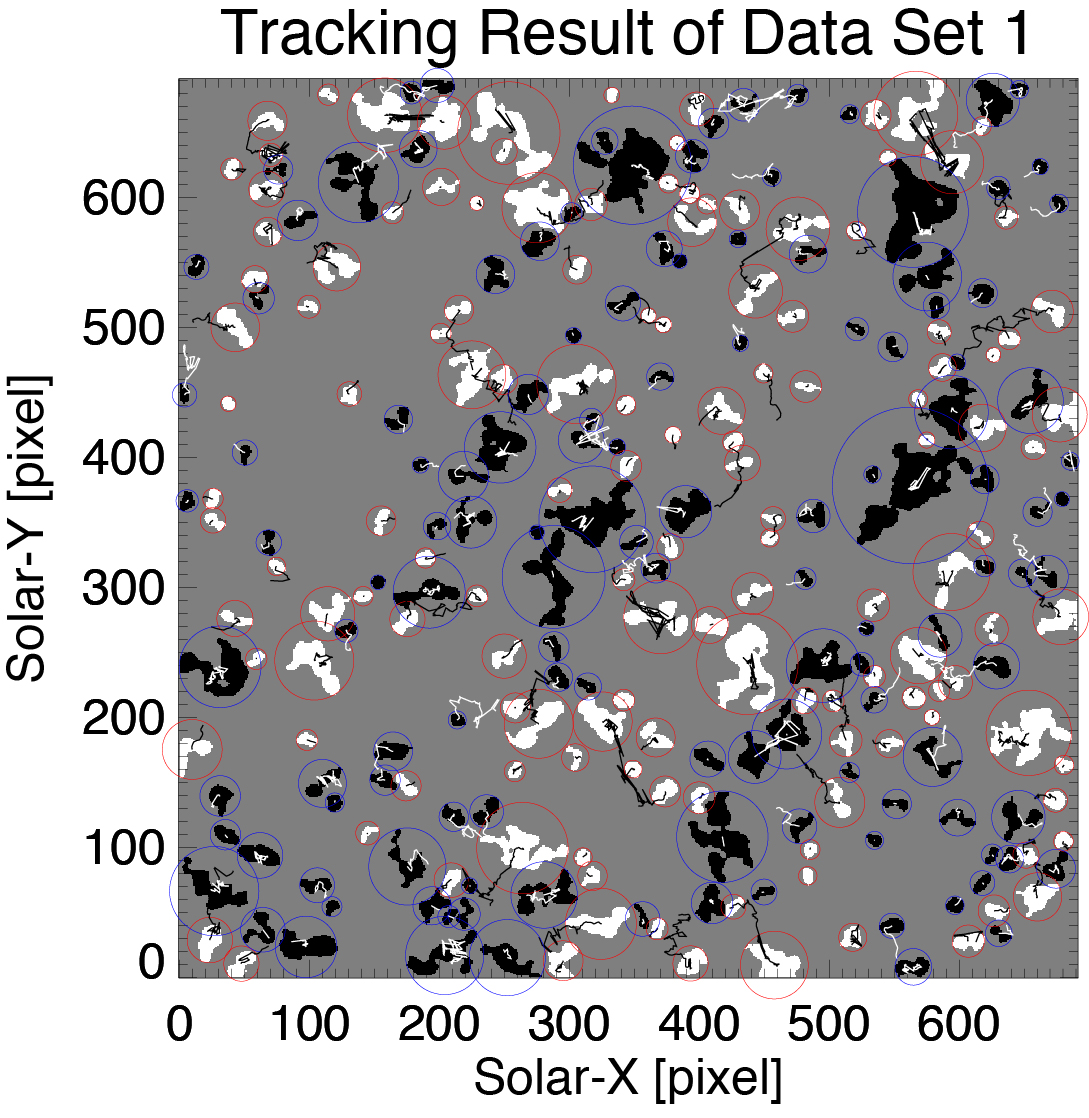}
\caption[Example of tracking result in data set 1.]{
Example of tracking result in data set 1.
The observational time is 0:30UT on 2009 November 11 
The red (blue) circles indicate positive (negative) patches detected in this image.
The black (white) lines show paths of center of detected patches.
}
\label{fig:track_d1}
\end{figure}

We track the motion of magnetic patches in consecutive images after the detection of them.
Patches are marked as identical when they have a spatial overlap in continuous images \citep{hag1999}.
The typical travel distance of magnetic patches in the data interval ($\sim 1 \ \rm{minute}$) is up to 
nearly 1 pixel size.
However, a typical travel distance of patches becomes more than $1$ pixel with a time interval more than $1$ minute.
We set margin to avoid this problem in data set 2.
The margin is set considering the typical velocity, which we set $2$ km s$^{-1}$ from twice of a typical patch motion 
(See Figure \ref{fig:dist_vel}), and pixel size.
It becomes $5$ pixels.
More than one patch in a previous image often have spatial overlaps with one patch 
in a consecutive image and vice versa in a recent high-resolution magnetogram.
To clear up this problem, we set two conditions when tracking patches.
First, we check spatial overlaps from a patch with larger flux content.
This is based on the concept that a smaller patch has a greater tendency to fall below the 
detection limit of the analysis by splitting and cancellation.
Second we select a patch with the most proximate flux content in case of overlaps of more than one patch.
Tracking paths of detected patches become unique with these conditions.
Figure \ref{fig:track_d1} show a tracking paths of center of detected patches in data set 1.

\subsection{Merging and Splitting}

\begin{figure}[p]
\centering
\includegraphics[bb= 0 0 1000 700,width=16.5cm,clip,angle=270]{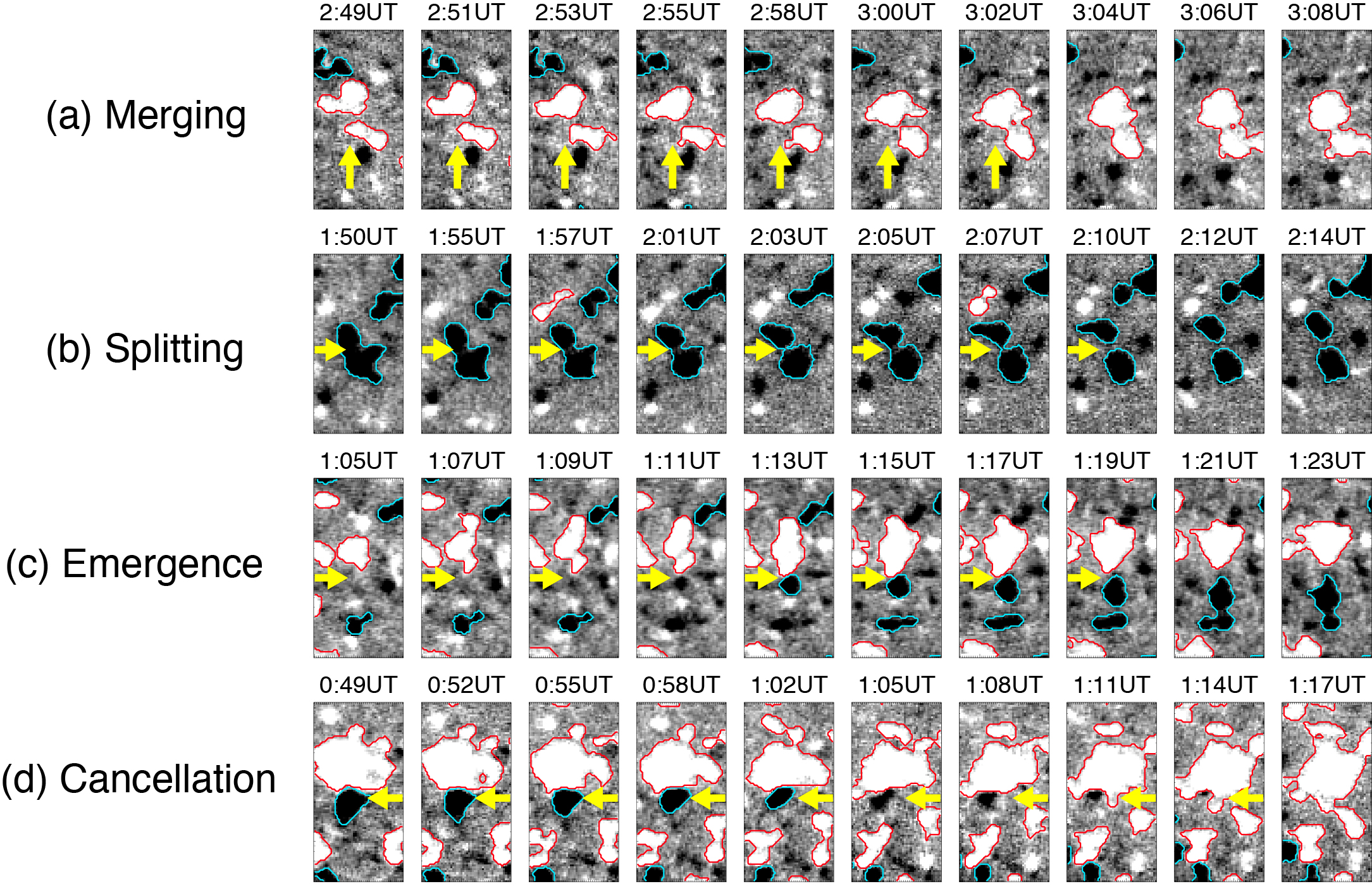}
\caption[Examples of detected surface processes of magnetic patches.]{
Examples of detected surface processes of magnetic patches, namely, 
(a) merging, (b) splitting, (c) emergence, and (d) cancellation.
The background shows the magnetic flux density obtained by ${\it Hinode}$/NFI.
The red (blue) contours indicate positive (negative) patches detected with our threshold.
The field of view is $7.2'' \times 14.4''$ for all the images for all images.
}
\label{fig:ex_magac_d1}
\end{figure}

Merging is a process where more than one patch of the same polarity converge and coalesce to one patch.
A merging event is defined as a feature that satisfies the following conditions: 
(1) that there are two or more parent patches in a previous magnetogram overlap one daughter patch in the consecutive magnetogram,
and (2) that more than one of the concerned patches in the previous magnetogram disappear in the time interval.
Figure \ref{fig:ex_magac_d1}(a) shows an example of the mergings.
There are two separated positive patches at 2:49UT.
They gradually converge and coalesce into one patch at 3:02UT.
They stayed to be one massif even at 3:08UT.

Splitting is a process where a single patch is divided into more than one patch.
A splitting event is defined as a feature that satisfies the two following conditions :(1) 
that there are one or more daughter patches in a magnetogram overlapping one parent patch in the previous magnetogram, and (2) 
that more than one of the concerned patches in the latter magnetogram appear in the time interval.
Figure \ref{fig:ex_magac_d1}(b) shows an example of the splittings.
There is one negative patch at 1:50UT.
A dip shape of the outline is formed between north-east and sourth-west parts of the patch at 2:06UT.
It grows at 1:55$\--$2:07UT and 
the patch finally splits into two negative patches at 2:10UT.
The distance between them continues to be larger at 2:10$\--$2:14UT.

\subsection{Emergence and Cancellation}

Emergence and cancellation are defined as a pair of flux increase and decrease events of different polarities respectively.
One can pick up a partial cancellation, where not all but only one patch of the involved pair disappears. 

\begin{figure}[tp]
\centering
\includegraphics[bb=0 0 850 550,width=0.99\textwidth]{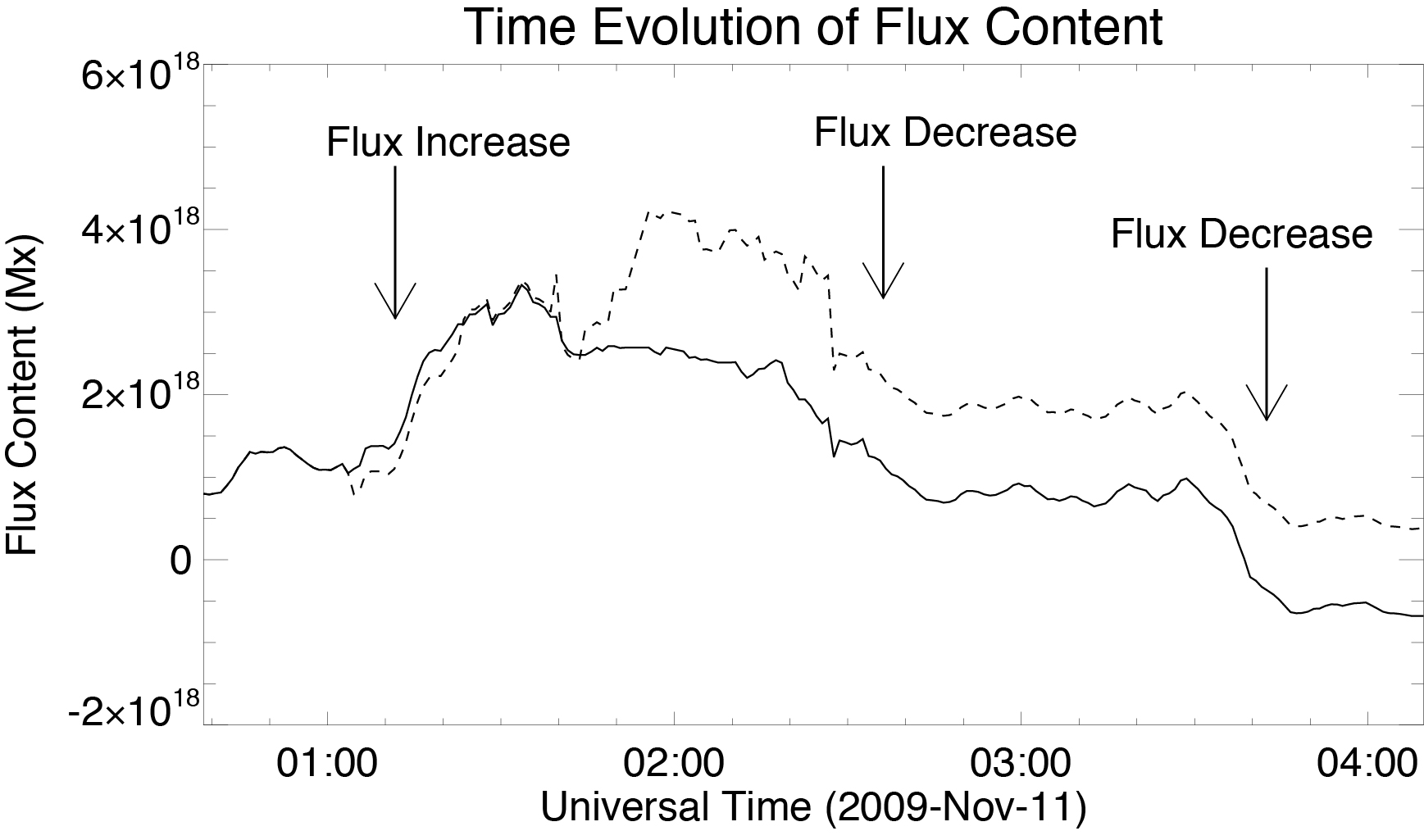}
\caption[An example of time evolution of flux content of one positive patch in data set 1.]{
An example of flux time series contained in one positive patch in data set 1 and flux change events there.
The dashed line indicates it.
The solid line indicates same plot but after removing flux change by mergings and splittings.
}
\label{fig:ex_teflux}
\end{figure}

First, we mark a flux change event in time series of flux content in each patch.
It is defined as a flux increase or decrease which has a duration and a flux change rate more than certain threshold.
Next, we exclude change of flux content by merging and splitting.
Cancellations in network field have a time scale of a few ten minutes \citep{cha2004b,iid2010}.
On the other, flux increase has a shorter time scale, less than $10$ minutes, in emergence event \citep{ots2007}. 
We want to set as short threshold for the duration as possible to pick up emergences.
We set the threshold for duration as 5 minutes for data set 1 and 10 minutes for data set 2.
The threshold for flux change rate is taken from a typical flux change rate of cancellation.
It is because a cancellation is thought to be a more moderate event in flux change rate than an emergence.
There are some reports of cancellations in network field \citep{cha2002, park2009}.
\cite{cha2002} reports a flux change rate of cancellation as $3.5 \times 10^{18}$ Mx hr$^{-1}$ from their analysis of magnetograms 
obtained by ${\it SOHO}$/MDI.
We use $10^{18}$ Mx hr$^{-1}$ as a threshold for flux change events.
The flux change events are marked with these thresholds after averaging magnetic flux content over three data point for smoothing.
Figure \ref{fig:ex_teflux} shows an example of time evolution of flux content in one positive patch in data set 1.
Dotted line presents time evolution of flux content in the patch.
Solid line is the same one but after removing flux changes by merging and splitting.
Three flux change events, which are indicated by arrows, are detected in this patch.

Then the flux change events in some distance threshold are paired.
We set the distance threshold for cancellation same as patch tracking, namely 1 pixel in data set 1 and 5 pixels in data set 2, because 
typical proper velocity of canceling patches is not so rapid.
On the other hand, there is a significant large velocity in emergence.
We employ 4 km s$^{-1}$, which corresponds to 4 pixel in data set 1 and 20 pixel in data set 2.

Our method is capable of detecting partial cancellations, i.e. those in which one or both of the involving patches do not disappear. 
It is also capable for the partial emergence. 
This method categorizes a hybrid event, in which the cancellation and emergence simultaneously occur, as a single cancellation, 
a single emergence or a mix of uni-polar processes. 
Thus, the detected number for the emergence and cancellation may be underestimated.

Figure \ref{fig:ex_magac_d1}(c) shows an example of the emergences.
There is a pre-existing positive patch in the north of the emerging region.
The negative polarity is small and unrecognized with our detection condition at 1:05$\--$1:07UT and
 grows large enough to be detected at 1:13UT.
The flux content of the positive polarity continues to increase at 1:13UT, which is defined as a flux increase in this duration.
The distance between patches continues to become larger at 1:19$\--$1:23UT.

Figure \ref{fig:ex_magac_d1}(d) shows an example of the cancellations.
One negative patch is converging to the larger positive patch that is located at the conjunction point of network.
They contact at 0:49UT and begin to cancel with each other at 0:52UT.
The negative patch continues to shrink and it becomes smaller than our size threshold at 1:05UT.
It totally disappears at 1:17UT.
On the other hand, positive patch remains after the cancellation because the flux content of the positive patch is larger than 
that of the negative patch. So this event is a partial cancellation.

\ifodd \arabic{page}
\else
  \thispagestyle{fancy}
  \mbox{}
  \pagebreak
\fi

\chapter[Magnetic Patch Characters]{Statistical Investigations of Magnetic Patch Characters}
\thispagestyle{fancy}
We investigate statistical properties of magnetic patches in this chapter.
Total number of patches and total flux amount are investigated in Section \ref{sec:pro_patch}.
The flux dependence of patch size is investigated to determine the statistical relationship 
between flux content and spatial scale in Section \ref{sec:size}.
Frequency distributions of flux content in both data sets are investigated in Section \ref{sec:dist_flux}.
We investigate proper velocity of patches, which is necessary to evaluate collision frequency of patches, in Section \ref{sec:vel}.
Lifetime i.e., a duration for which a patch has a flux content larger than our threshold, is investigated in Section \ref{sec:life}.
We will discuss these results and show our interpretations in Section 5.2 and 5.3.

\section{Total Number and Total Flux Amount of Patches}
\label{sec:pro_patch}

\begin{table}
\begin{center}
\begin{tabular}{c|ccp{5zw}}
\hline\hline
&Data Set 1&Data Set 2\\
\hline
Detected Patches&&\\
Positive& \shortstack{26745\\(134.4 per frame)}& \shortstack{97018\\(59.1 per frame)}\\
Negative& \shortstack{26459\\(133.0 per frame)}& \shortstack{92102\\(56.1 per frame)}\\
Total& \shortstack{53204\\(267.4 per frame)}& \shortstack{189120\\(115.2 per frame)}\\
\hline
Tracked Patches&&\\
Positive& 1636 & 21823\\
Negative& 1637 & 19544\\
Total& 3273 & 41367\\
\hline
Total Flux Amount& & \\
Positive & \shortstack{$1.63\times10^{20}$ Mx\\(2.53 Mx cm$^{-2}$)}& \shortstack{$1.70\times10^{20}$ Mx\\(3.76 Mx cm$^{-2}$)}\\
Negative & \shortstack{$2.32\times10^{20}$ Mx\\(3.60 Mx cm$^{-2}$)}& \shortstack{$1.54\times10^{20}$ Mx\\(3.42 Mx cm$^{-2}$)}\\
Total & \shortstack{$3.95\times10^{20}$ Mx\\(6.13 Mx cm$^{-2}$)}& \shortstack{$3.42\times10^{20}$ Mx\\(7.16 Mx cm$^{-2}$)}\\
\hline \hline
\end{tabular}
\end{center}
\label{tab:des_d1}
\caption[Number of patches and flux amount in each data set.]{
Number of patches and flux amount in the data sets.
}
\end{table}

Total number and total flux amount of patches are investigated.
Table 4.1 shows the results of both data sets.

26745 positive and 26459 negative are detected in data set 1.
They correspond to $\thicksim130$ patches of one polarity in one image. 
As for data set 2, 97018 positive and 92102 negative patches are found in data set 2.
These numbers are sufficient for statistical analysis.  
They correspond to $\thicksim60$ patches of one polarity in one image. 
1636 positive and 1637 negative patches are detected as one tracked patch in data set 1.
21823 positive and 19544 negative patches are detected as one tracked patch in data set 2.
The numbers of tracked patches may also be sufficient for statistical analysis.

\begin{figure}[p]
\centering
\includegraphics[bb=0 0 650 800,width=0.88\textwidth]{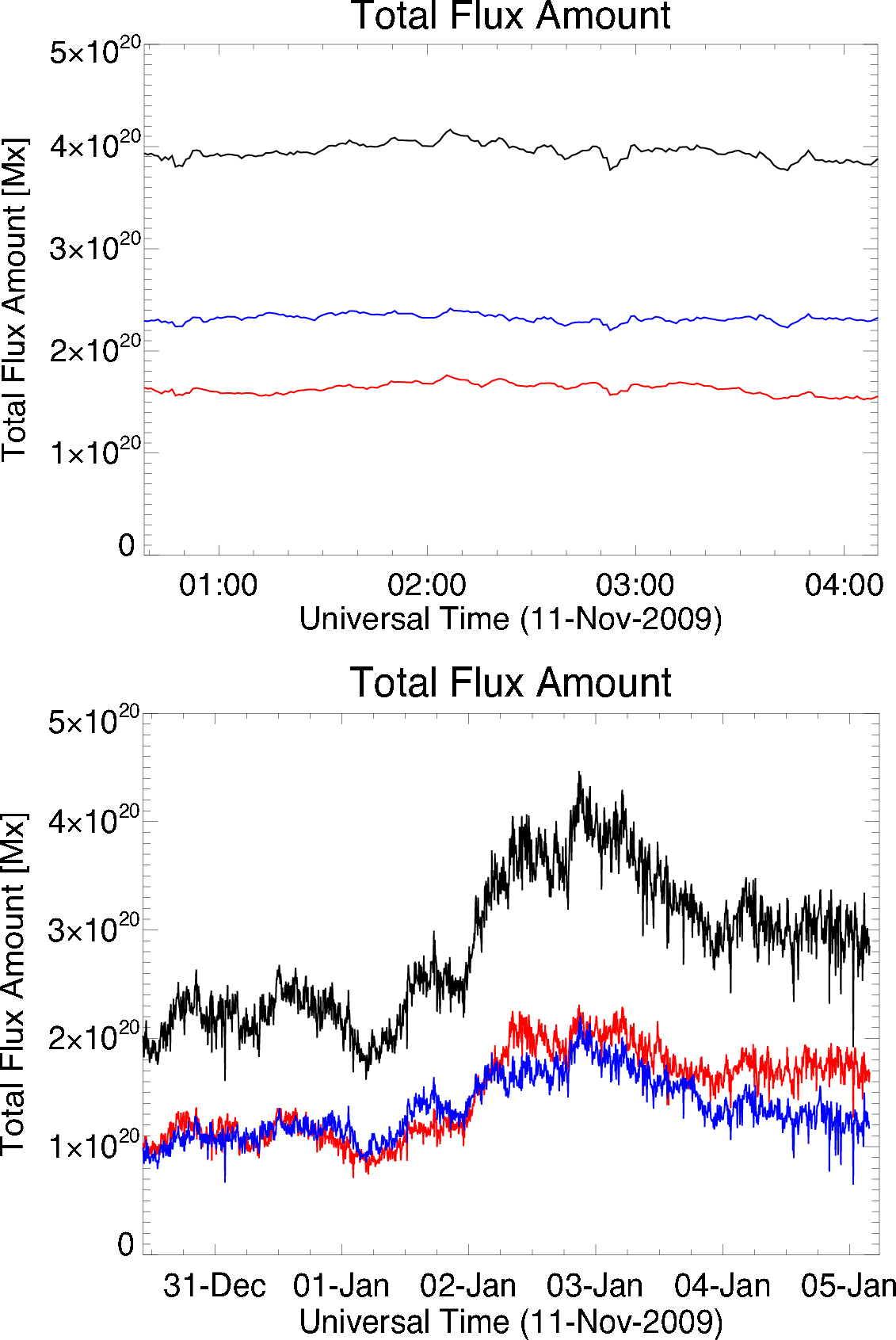}
\caption[Time series of the total flux amount contained in field of view during the observational period in each data.]{
Time series of the total flux amount contained in the field of view during the observational 
period in data set 1 (upper) and data set 2 (lower).
The red and blue line denote total flux amount of positive and negative polarities, respectively.
The black line denotes sum of them.}
\label{fig:flux_d1}
\end{figure}

Figure \ref{fig:flux_d1} shows time series of total flux amount in the observational period in the target regions.
The red lines denote total flux amount of positive polarity.
The blue ones denote those of negative polarity.
The black ones denote sum of them.
Time series of total flux amount in data set 1 is shown in the upper panel.
The averaged total flux in data set 1 is $1.63 \times 10^{20}$ Mx, $2.32 \times 10^{20}$ Mx and 
$3.95 \times 10^{20}$ Mx for positive, negative, and both polarities, respectively.
They correspond to $2.53$ Mx cm$^{-2}$, $3.60$ Mx cm$^{-2}$ and $6.13$ Mx cm$^{-2}$ as flux density averaged over whole area.
Negative flux content is larger than that of positive one by 42$\%$ in this region.
The maximum differences from average are $8.3\%$ and $5.1\%$ for positive and negative polarities.
Total flux amounts of both polarities are constant with an accuracy of $10\%$.
Same plot for data set 2 is shown in the lower panel.
There are many clear large flux emergences (but not so large as that forming active region) in data set 2.
The averaged total flux content is $1.70 \times 10^{20}$ Mx, $1.54 \times 10^{20}$ Mx and 
$3.42 \times 10^{20}$ Mx for positive, negative, and both polarities.
They correspond to $3.76$ Mx cm$^{-2}$, $3.42$ Mx cm$^{-2}$ and $7.16$ Mx cm$^{-2}$.
The difference between polarities is 9$\%$
The maximum differences from average are 45$\%$ and 52$\%$ for positive and negative polarities.

\section{Dependence of Patch Area on Flux Content}
\label{sec:size}

\begin{figure}[p]
\centering
\includegraphics[bb=0 0 750 700,width=0.99\textwidth]{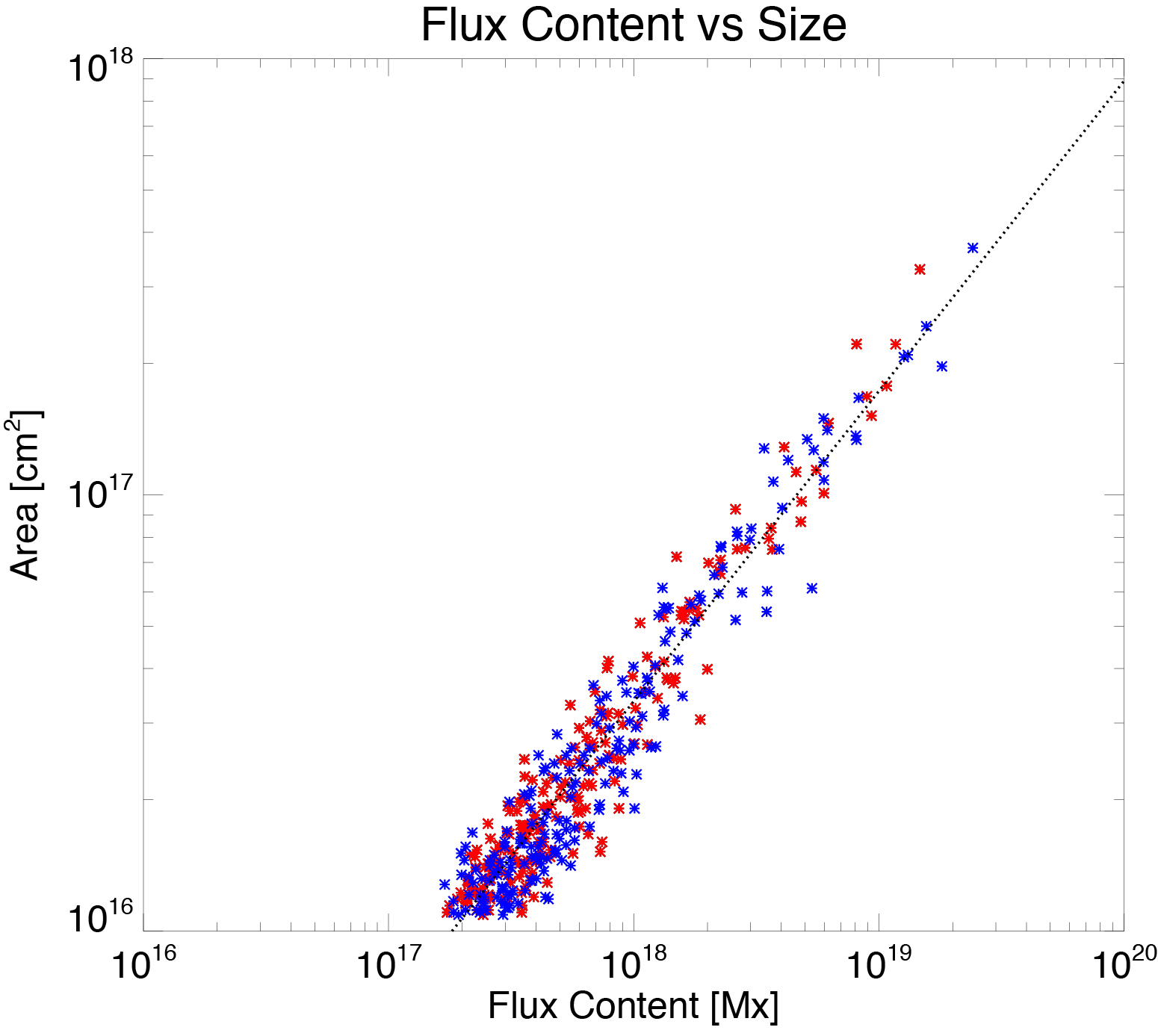}
\caption[Scatter plot of the flux content and the area of magnetic patches.]{
Scatter plot of the flux content and the area of magnetic patches in data set 1. 
The red (blue) asterisks denote those of positive (negative) polarity.
200 data points for each polarity are randomly selected for the plot. 
The solid line denotes the fitting result with an assumption of a power-law distribution.
}
\label{fig:sizeflux_d1}
\end{figure}

Figure \ref{fig:sizeflux_d1} shows a scatter plot of flux content and area of patches in data set 1.
There is a positive correlation between them.
We fit with a power-law function, namely
\begin{equation}
S(\phi)=S_0 \, \left( \frac{\phi}{\phi_0} \right)^{\alpha^{S}_{\phi}}.
\end{equation}
where $S_0$ is the reference area, $\phi_0$($=10^{18}$ Mx) 
is the reference magnetic flux content,
 and $\alpha^{S}_{\phi}$ is the power-law index.
We obtained $S_0=(3.33\pm0.12) \times 10^{16}$ cm$^2$ and 
$\alpha^{S}_{\phi}=0.711\pm0.001$ with the $1\sigma$-error of fitting.
Solid line in Figure \ref{fig:sizeflux_d1} denotes the fitting result.

\section{Frequency Distribution of Flux Content}
\label{sec:dist_flux}

\begin{figure}[p]
\centering
\includegraphics[bb=0 0 800 550,width=0.99\textwidth]{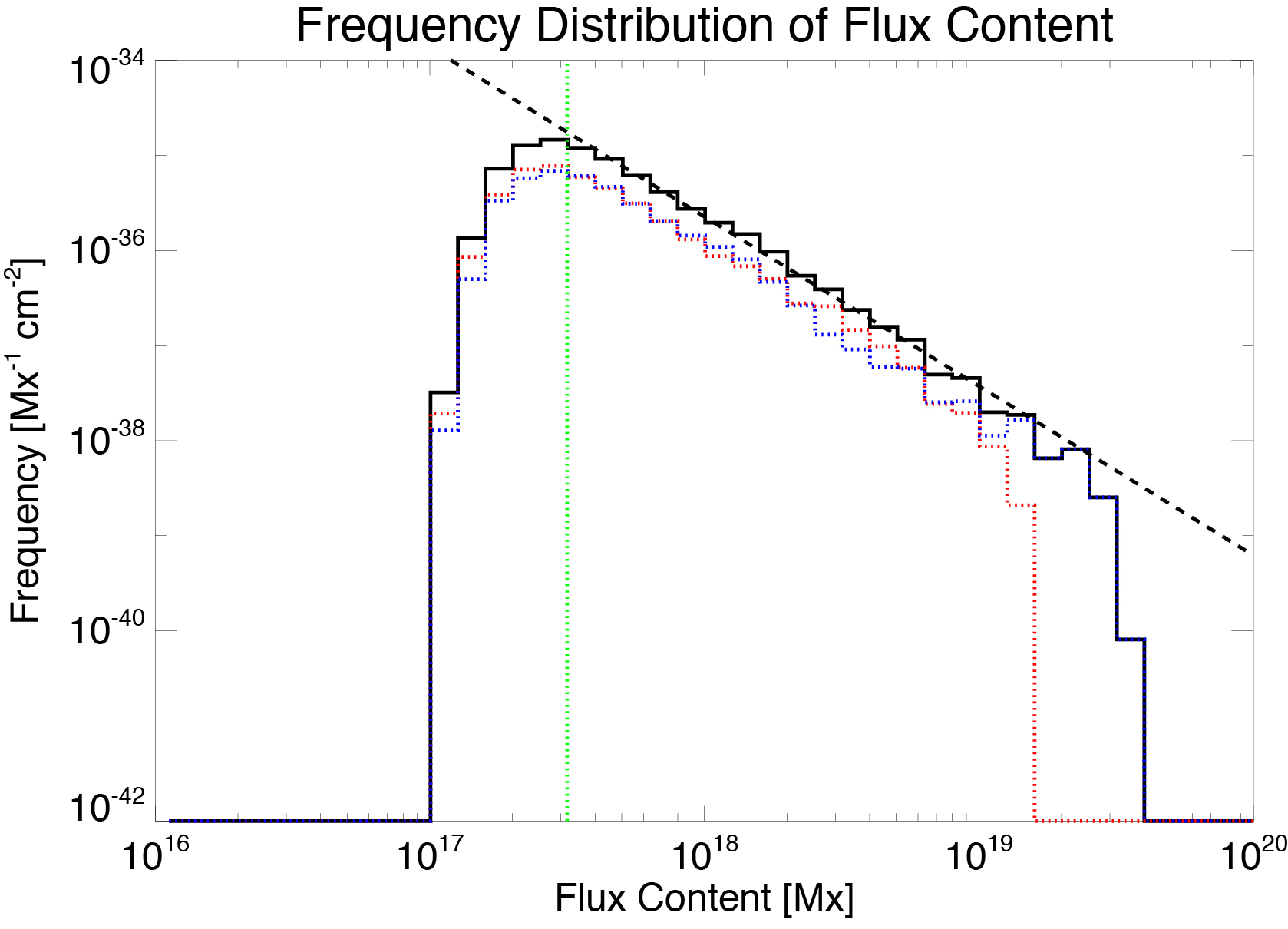}
\caption[Frequency distribution of the magnetic flux content in data set 1.]{
Frequency distribution of the magnetic flux content in data set 1.
The red (blue) dashed histogram denotes that of positive (negative) polarity.
The solid histogram denotes the sum of them.
The black dashed line indicates the fitting result of it.
The obtained power-law index is $-1.78$ with a fitting range between $10^{17.5}$ Mx and $10^{19}$ Mx.
The vertical green line denotes the detection limit, $\phi_{\rm th}=10^{17.5}$ Mx.}
\label{fig:dist_flux_d1}
\end{figure}

Figure \ref{fig:dist_flux_d1} shows a frequency distribution of the flux content in data set 1.
The red and blue dashed histograms denote frequency distributions of positive and negative polarities.
The solid histogram denotes the sum of them.
The distribution drops down below $\phi_{\rm{th}} = 10^{17.5} \ \rm{Mx}$, which is suggested as a detection limit defined in 
Section \ref{sec:magch_lmt}.
Only a limited number of the patches are detected below it. 
The vertical green line denotes it.
We set $\phi_{\rm th}=10^{17.5}$ Mx as a limit of statistical analysis in data set 1 from here.
We make a fitting of the power-law function between $10^{17.5} \ \rm{Mx}$ and $10^{19} \ \rm{Mx}$ with a 
least-squares fitting, namely
\begin{equation}
n(\phi)=n_0 \, \left(  \frac{\phi}{\phi_0}\right)^{-\gamma} \label{eq:dist}
\end{equation}
where $n_0$ is the reference frequency density, $\phi_0$($=10^{18}$ Mx) 
the typical scale of reference magnetic 
flux content, and $\gamma$ is negative value of a power-law index of the distribution.
The black dashed line indicates the fitting result.
We obtain $n_0 = 2.42 \times 10^{-36}$ Mx$^{-1}$ cm$^{-2}$ and $\gamma = 1.78 \pm  0.05$ 
with the $1 \sigma$-error of fitting.
\cite{par2009} reports that error of a power-law index from the selection of fitting range is larger than that of fitting itself.
We evaluate that error by changing the minimum value of fitting range. 
The index varies between $-1.78$ and $-1.91$ with the minimum value from $10^{17.5}$ Mx to $10^{18}$ Mx.
the total error of our fitting of the power-law index becomes $0.18$ with taking it into account.

\begin{figure}[p]
\centering
\includegraphics[bb=0 0 800 550,width=0.99\textwidth]{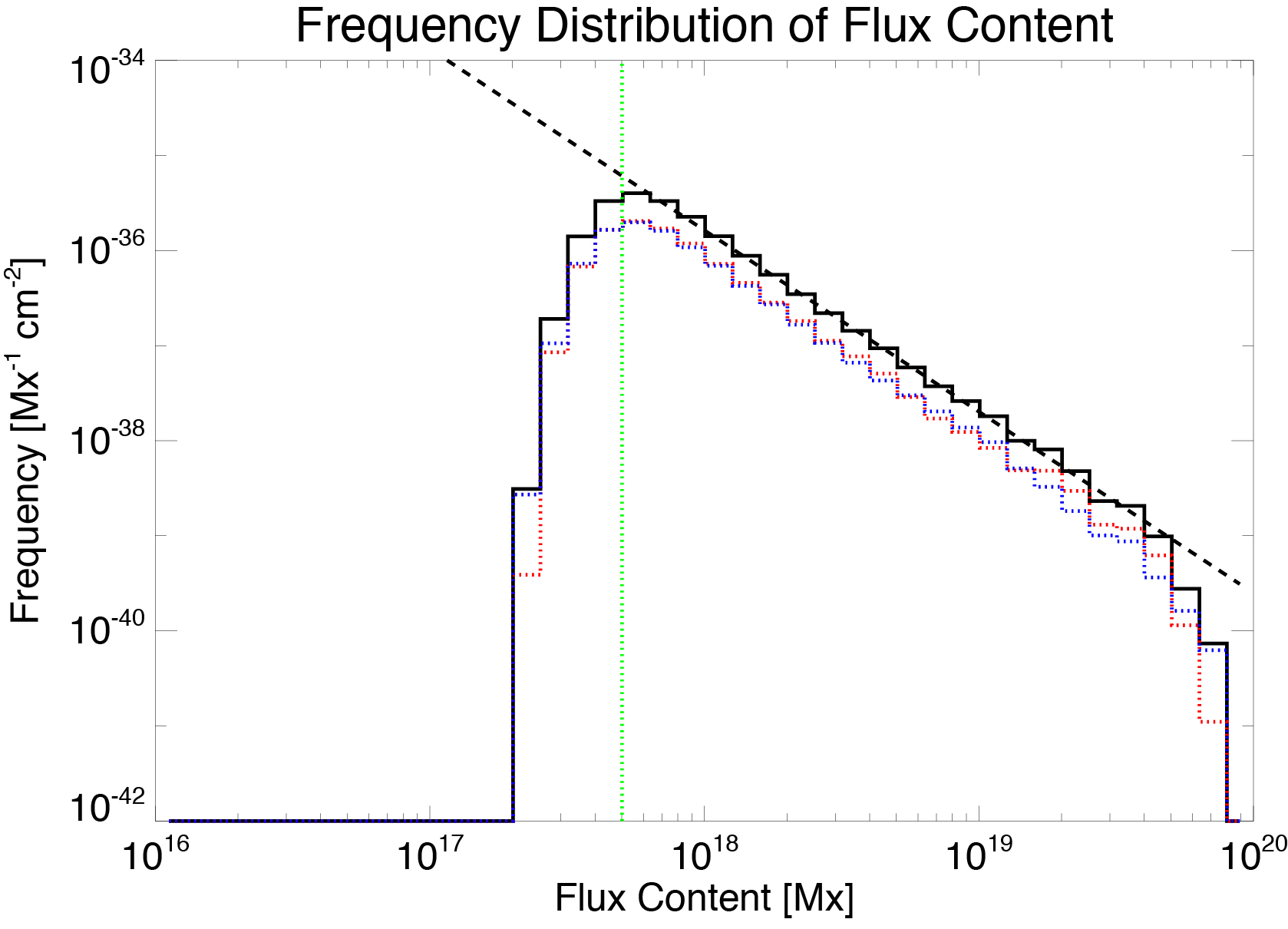}
\caption[Frequency distribution of the magnetic flux content in data set 2.]{
Same plot as Figure \ref{fig:dist_flux_d1} but for data set 2.
The power-law index is $-1.90$ with a fitting range between $10^{17.7}$ Mx and $10^{19}$ Mx.
The vertical green line denotes the detection limit, $\phi_{\rm th}=10^{17.7}$ Mx.}
\label{fig:dist_flux_d2}
\end{figure}

The same analysis is done for data set 2.
Figure \ref{fig:dist_flux_d2} shows a frequency distribution in data set 2.
The distribution is dropping down below $\phi_{\rm{th}} = 10^{17.5} \ \rm{Mx}$.
The difference of the detection limit may be caused by the fact that there are some emerging fluxes in this region during the 
observational period.
We obtain $n_0 = 1.63 \times 10^{-36}$ Mx$^{-1}$ cm$^{-2}$ and $\gamma = 1.93 \pm 0.03$ 
with the $1\sigma$-error of fitting.
The index varies between $-1.90$ and$-1.94$ with a minimum fitting range of $10^{17.7}$ Mx 
to $10^{18}$ Mx. The total error of our fitting of the power-law index becomes $0.07$.

These results are consistent with \cite{par2009}, who report $n_0=1.5\times10^{-37}$  Mx$^{-1}$ cm$^{-2}$ 
and $\gamma=-1.85\pm0.14$ with $\phi_0=10^{18}$ Mx for one polarity.

\section{Proper Velocity}
\label{sec:vel}
We investigate proper velocity of magnetic patches. 
The aim here is to investigate average velocity of patches in our data set.
The proper velocity is important to evaluate the collision frequency of patches \citep{sch1998}. 
Proper velocity is defined as a velocity of the center of a patch in this study.
The center of patches is determined by weighting pixel position with magnetic flux, namely
\begin{equation}
\displaystyle
\bm{\overline{x}}_{i}^j
=\frac{\sum_{k=1}^{N_i^j}\bm{B}^j(\bm{x}_{i,k}) \cdot \bm{x}_{i,k}^j}
{\sum_{k=1}^{N_i^j}\bm{x}_{i,k}^j}.
\end{equation}
Here $\bm{\overline{x}}_{i}^j$ is a position of center of $i$-th patch at $j$-th time step, $N_i^j$ is a pixel number of $i$-th patch at $j$-th time step, 
$\bm{B}^j(\bm{x})$ is a flux density at the position of $\bm{x}$ at $j$-th time step, and $\bm{x}_{i,k}^j$ is a position of $k$-th pixel 
of $i$-th patch at $j$-th time step. 
The proper velocity of patches at merging or splitting removed in the following statistical analysis since, 
in such timing, an artificially large displacement may occur.

\begin{figure}[p]
\centering
\includegraphics[bb=0 0 800 750,width=0.99\textwidth]{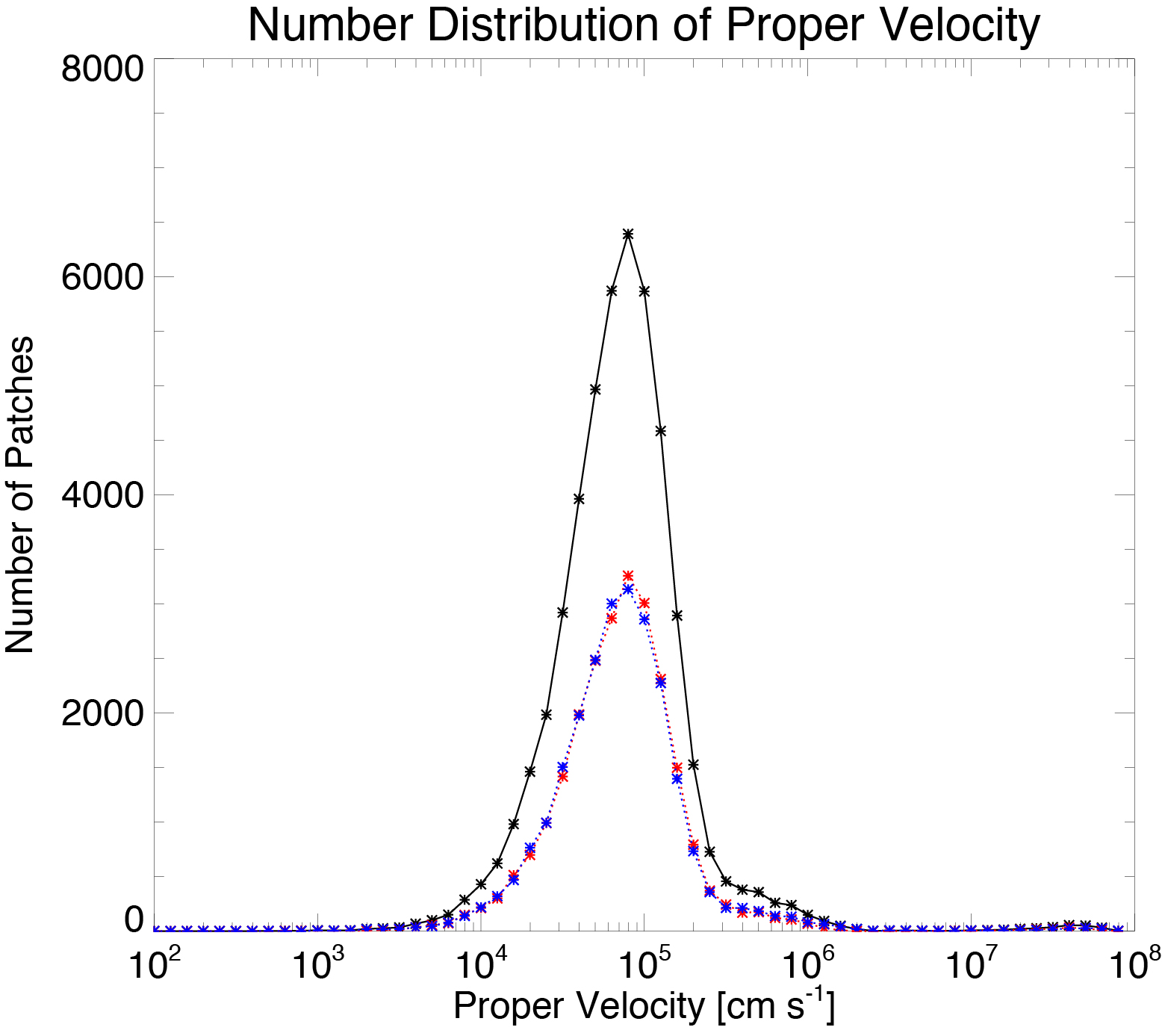}
\caption[Number distribution of the proper velocity.]{
Number distribution of the proper velocity.
The red (blue) dashed line indicates that of positive (negative) patches.
The black solid line indicates sum of them.
}
\label{fig:dist_vel}
\end{figure}

Figure \ref{fig:dist_vel} shows a number distribution of proper velocity.
The red dashed line indicates that of positive patches.
The blue one indicates that of negative patches.
The black solid one devotes sum of them.
They reach maximum value around $10^{5}$ cm s$^{-1}$.
There is an enhancement in the range of high velocity, namely from $10^7$ cm s$^{-1}$ and $10^8$ cm s$^{-1}$.
We found 123 and 124 points out of 24096 positive and 24068 negative points there.
They correspond to $\thicksim$0.5$\%$ of data points.  
Surface processes of patches involving undetected patches, which we cannot remove in this analysis, may cause them.  
Then we remove proper velocity larger than $10^7$ cm s$^{-1}$ and evaluate average proper velocity of patches. 
There are also enhancements between $3 \times 10^5$ and $2 \times 10^6$ cm s$^{-1}$.
Such relatively rapid motion of patches might corresponds to a peculiar events such as emergence driven by buoyancy, 
for which further investigation is necessary.
The obtained averaged proper velocities of positive, negative, and both patches are 
$1.10$ km s$^{-1}$, $1.19$ km s$^{-1}$, and $1.14$ km s$^{-1}$, respectively.
The medians of proper velocity are $0.81$ km s$^{-1}$, $0.79$ km s$^{-1}$, and $0.80$ km s$^{-1}$ 
for positive, negative, and both patches, respectively.
They are slightly smaller than the averages due to the peak offset in the distribution to the larger velocity side.

\begin{figure}[p]
\centering
\includegraphics[bb=0 0 800 750,width=0.99\textwidth]{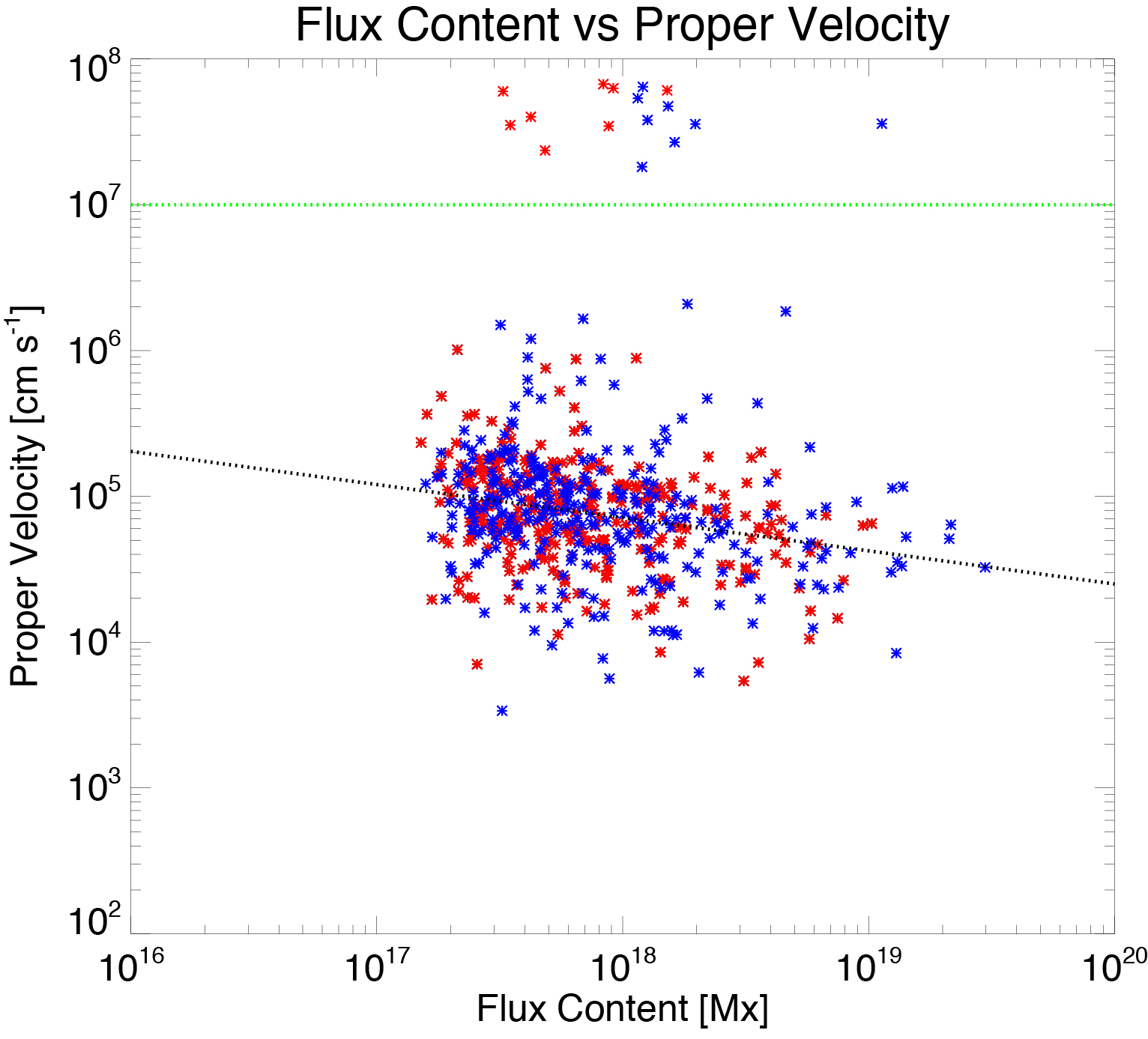}
\caption[Scatter plot of the flux content of the patches and their proper velocity.]{
Scatter plot of the flux content of the patches and their proper velocity in data set 1.
The red (blue) asterisks denote data point of positive (negative) patches.
The green horizontal line denotes the threshold to cut the large velocity enhancement.
The black dashed line indicates a fitting result with a power-law function.
}
\label{fig:sca_vel}
\end{figure}

Further a relationship between proper velocity and flux content is investigated.
Figure \ref{fig:sca_vel} shows a scatter plot between them.
The red and blue asterisks denote data point of positive and negative patches.
The green horizontal line denotes the threshold to cut the large velocity enhancement.
300 and 8 data points for each polarity are randomly selected for plot 
below $10^{7}$ cm s$^{-1}$ and above it, respectively.
We make a least-squares fitting with a power-law function. 
The fitting form is written as 
\begin{equation}
v(\phi)=v_0 \left( \frac{\phi}{\phi_0}\right)^{\alpha^{v}_{\phi}}
\end{equation}
where $v_0$ is the reference velocity, $\phi_0$ is the reference magnetic flux which we set $10^{18}$ Mx, and 
$\alpha^{v}_{\phi}$ is the power-law index of the dependence.
We obtain $v_0=(7.14\pm1.12) \times 10^4$ cm s$^{-1}$ and $\alpha^{v}_{\phi}=-0.227\pm0.003$.
The fitted result is shown by the black dashed line in Figure \ref{fig:sca_vel}.

\section{Frequency Distribution of Lifetime}
\label{sec:life}
\begin{figure}[tp]
\centering
\includegraphics[bb=0 0 800 750,width=0.99\textwidth]{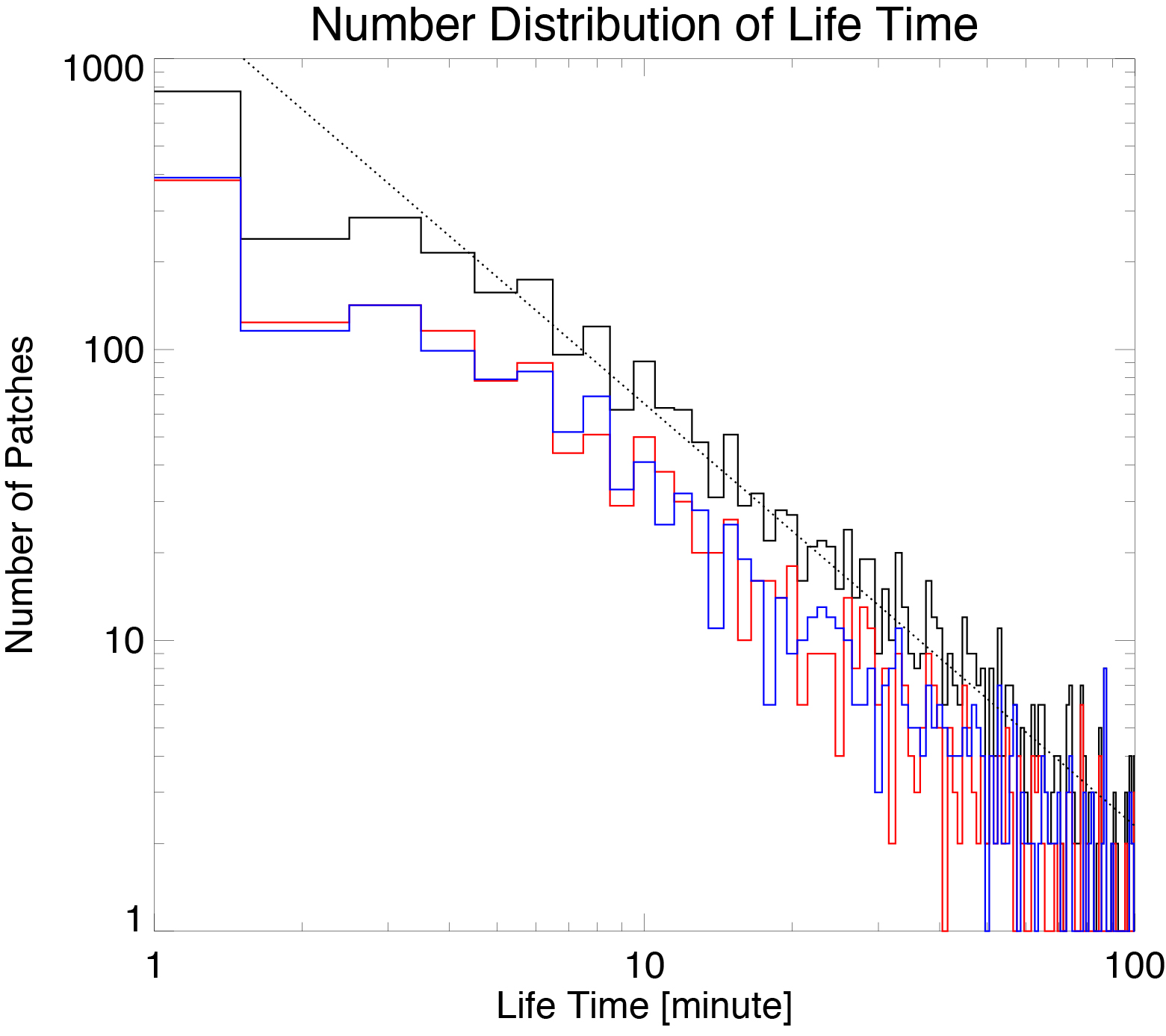}
\caption[Number distribution of the lifetime of the magnetic patches.]{
Number distribution of the lifetime of the detected patches in data set 1 in log-log plot.
The red/blue/black line denotes that of positive/negative/both polarity.
The dashed line indicates a fitting result of both polarities.
It has a power-law index of $-1.45$.}
\label{fig:dist_lt_d1}
\end{figure}

We investigate the frequency distribution of lifetime of patches in this section.
We define lifetime as a duration for which a patch is recognized based on our detection, namely from the appearance to the disappearance in two-leveled 
images.
In data set 1, the averaged lifetime of positive, negative and both polarities are $17.4$ minutes, $17.2$ minutes, and $17.3$ minutes, respectively.
Figure \ref{fig:dist_lt_d1} denotes a number distribution of lifetime of tracked patches.
The red, blue, and black lines denote those of positive, negative, and both polarities.
We make a least-squares fitting of a power-law function in the range between $6$ minutes and $30$ minutes.
The fitting form is  
\begin{equation}
  N(\tau)=N_0 \, \left( \frac{\tau}{\tau_0}\right)^{\alpha^{N}_{\tau}}.
\end{equation}
where $N_0$ is the reference number, $\tau_0$ is the reference lifetime that we set $10$ minutes, and 
$\alpha^{N}_{\tau}$ is the power-law index of the dependence.
We obtained $N_0=(6.51\pm0.05)\times10$ and $\alpha^{N}_{\tau}=-1.45\pm0.13$ with a $1\sigma$-error of fitting.
The dashed line in Figure \ref{fig:dist_lt_d1} represents the fitting result for the distribution of both polarities.
The power-law index steeper than $-1$ in number distribution means that the lifetime of patches becomes shorter as their smaller flux content.

\begin{figure}[tp]
\centering
\includegraphics[bb=0 0 800 750,width=0.99\textwidth]{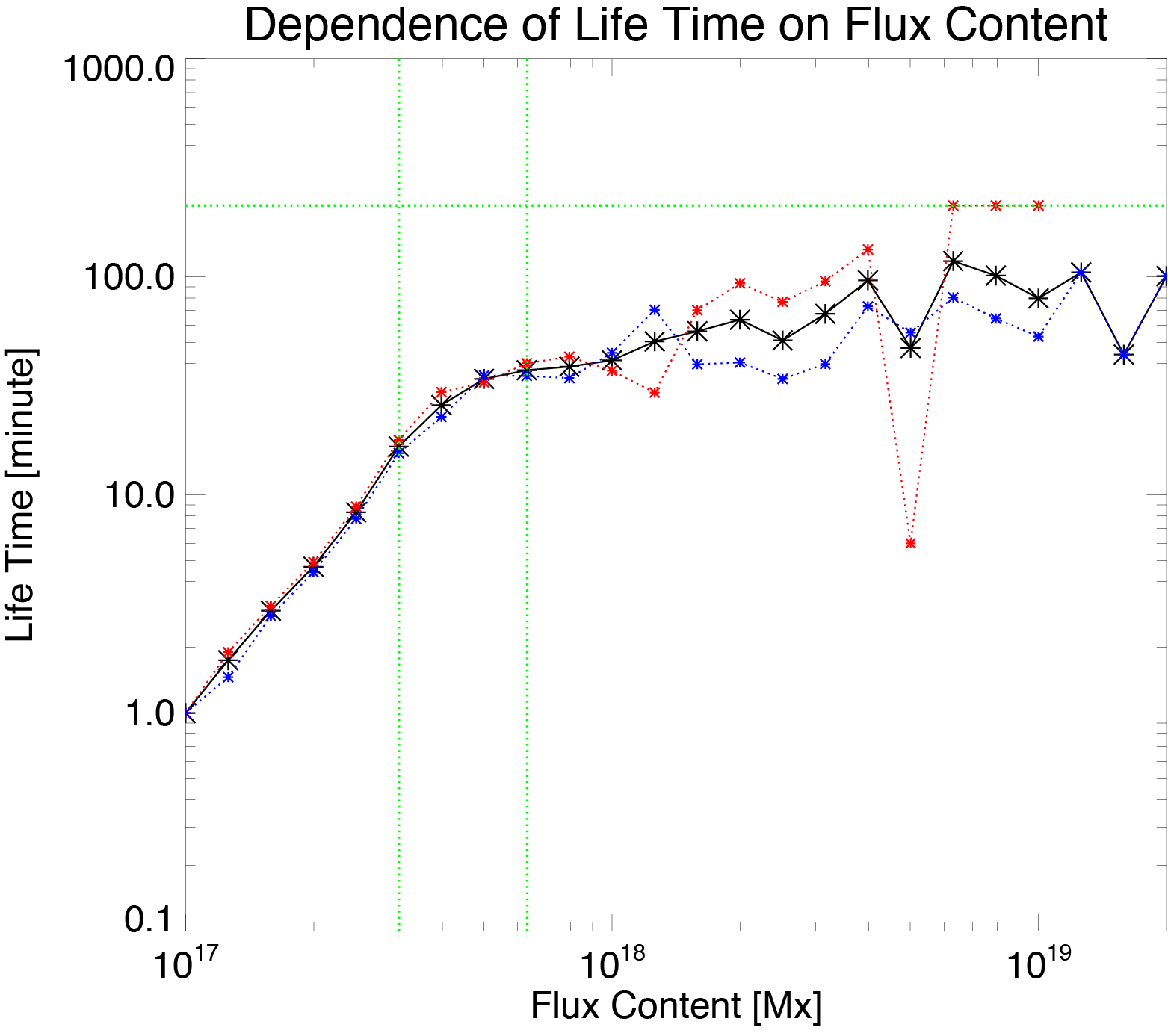}
\caption[Dependence of the averaged lifetime on the magnetic flux content in data set 1.]{
Dependence of the averaged lifetime on the magnetic flux content in data set 1.
The red and blue lines indicate those of positive and negative polarities respectively.
The solid line indicates that of both polarities.
The horizontal dashed line indicates the observational period of this data set.
The vertical dashed lines denote $\phi_{\rm th}$ and $2\phi_{\rm th}$ in data set 1. 
}
\label{fig:fllt_d1}
\end{figure}

Further dependence of lifetime on flux content is investigated.
We define flux content of tracked patches by averaging flux content over whole their lifetime.
Figure \ref{fig:fllt_d1} shows the result of it.
Red and blue dashed lines indicate those of positive and negative polarities, respectively.
Solid line indicates that of both polarities and horizontal dashed line indicates the observational period of this data set.
Vertical dashed lines denote detection limit in this study, namely $\phi_{\rm th}=10^{17.5}$ Mx, and twice of it.

\begin{figure}[tp]
\centering
\includegraphics[bb=0 0 800 750,width=0.99\textwidth]{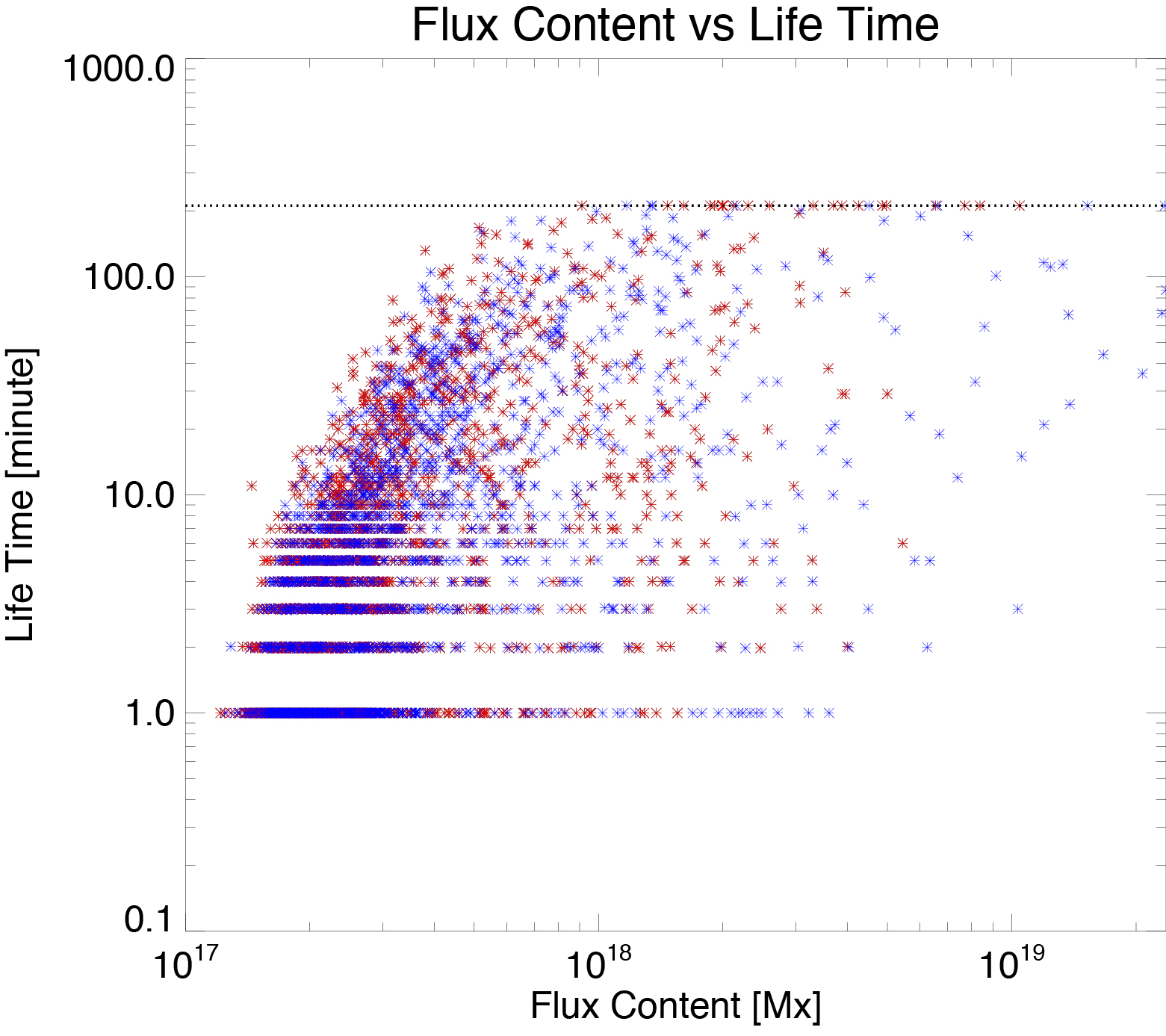}
\caption[Scatter plot of the magnetic flux content and the lifetime in data set 1.]{
Scatter plot of the magnetic flux content and the lifetime in data set 1.
The red (blue) asterisks indicate data points of positive (negative) polarities.
The horizontal dashed line indicates the whole observational duration of this data set.
}
\label{fig:fltau_d1}
\end{figure}

There are two domains from the discussion in Section \ref{sec:magch_lmt}.
One is below $2\phi_{\rm th}$, where we cannot distinct the apparent disappearance from the actual disappearance, 
and the other is above it, where one apparent disappearance cannot remove the patch.
There seems to be a change of dependence around it.
However, the lifetime hits near the observational duration around $10^{18}$ Mx.
To investigate this effect, we investigate a scatter plot of magnetic flux content of tracked patches and their lifetime.
Figure \ref{fig:fltau_d1} shows the result.
Red and blue asterisks indicate data points of positive and negative polarity.
Total numbers of positive and negative patches are 1636 and 1637 respectively.
The horizontal dashed line presents the whole observational period of this data set.
Some of data points hit the observational duration around $10^{18}$ Mx. 
It is difficult to distinct the effect of it and physical change of dependence. 

\begin{figure}[p]
\centering
\includegraphics[bb=0 0 800 750,width=0.99\textwidth]{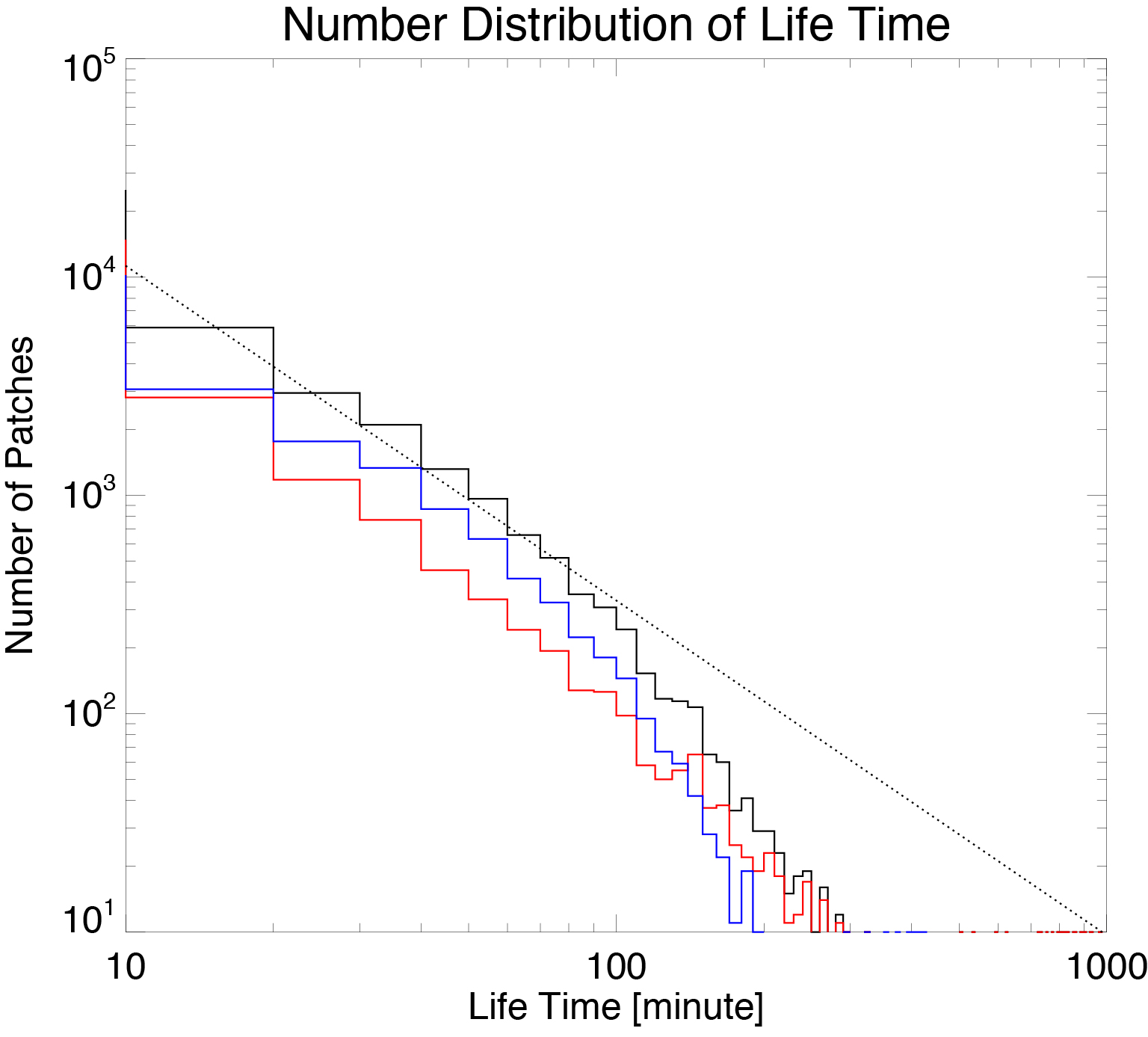}
\caption[Number distribution of lifetime of magnetic patches.]{
Same plot as Figure \ref{fig:dist_lt_d1} but for data set 2. 
The power-law index of the fitting line is $-1.53$.}
\label{fig:dist_lt_d2}
\end{figure}

\begin{figure}[p]
\centering
\includegraphics[bb=0 0 800 750,width=0.99\textwidth]{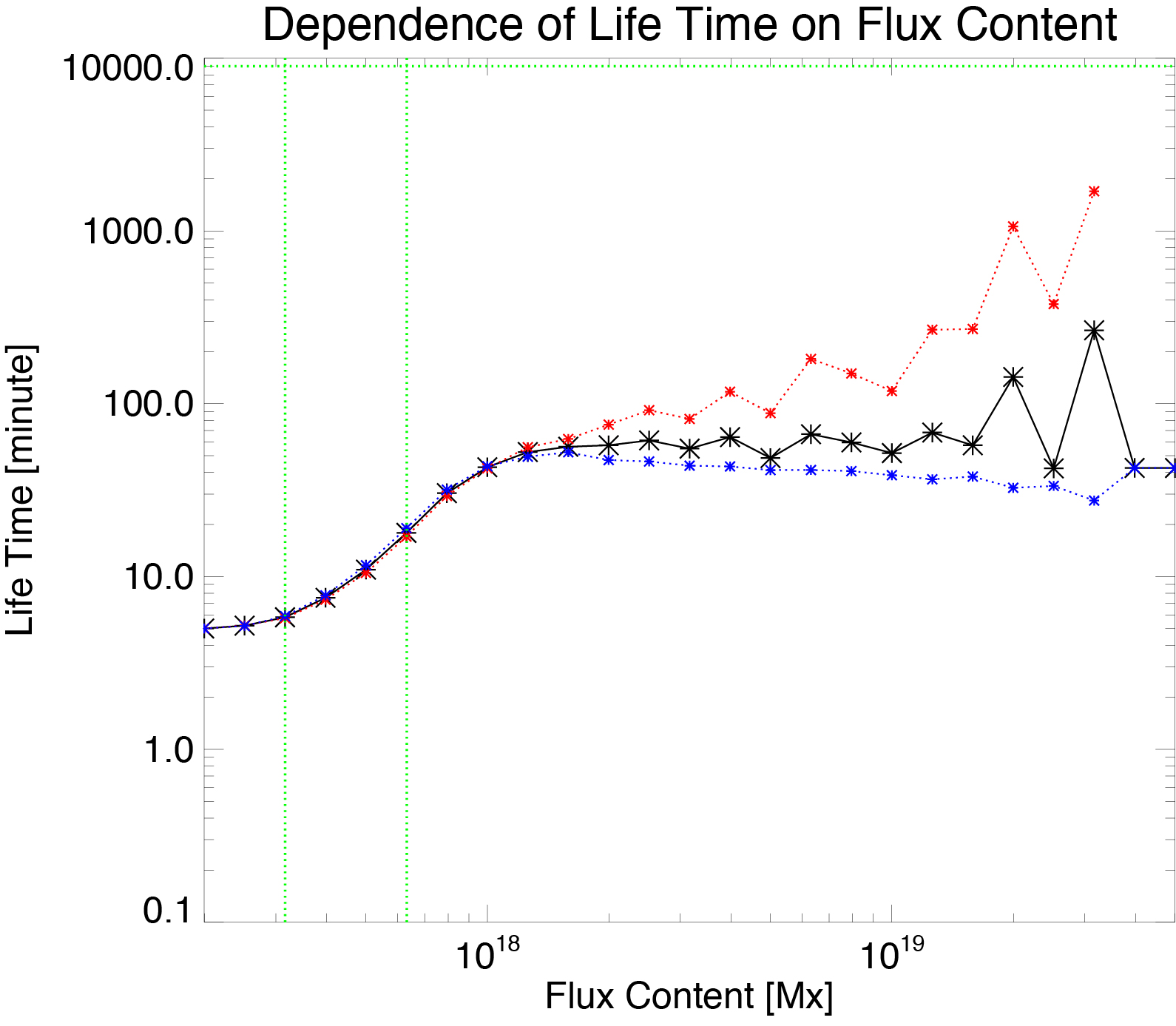}
\caption[Dependence of averaged lifetime on magnetic flux content in data set 2.]{
Same plot as Figure \ref{fig:fllt_d1} but for data set 2.
}
\label{fig:fllt_d2}
\end{figure}

\begin{figure}[p]
\centering
\includegraphics[bb=0 0 900 800,width=0.99\textwidth]{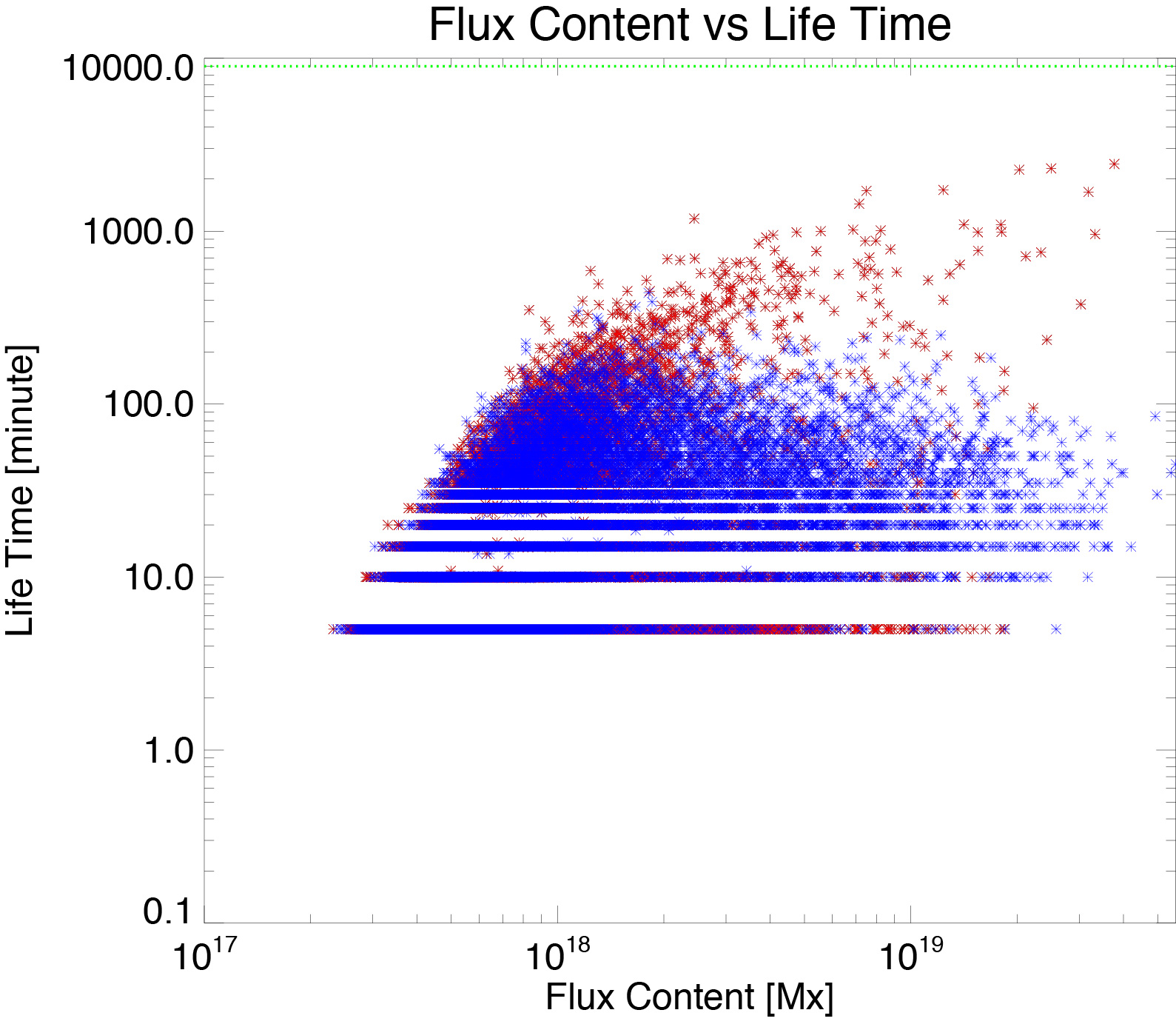}
\caption[Scatter plot of magnetic flux content and lifetime in data set 2.]{
Same plot as Figure \ref{fig:fltau_d1} but for data set 2.
}
\label{fig:fltau_d2}
\end{figure}

Same analysis is done for data set 2.
The observational period is much larger than that of data set 1.
The averaged lifetime is $22.5$ minutes, $23.9$ minutes, and $23.2$ minutes for positive, 
negative, and both polarities.
Figure \ref{fig:dist_lt_d2} shows a number distribution of lifetime. 
It seems a power-law distribution in the range below 70 minutes. 
We make a least-squares fitting in the range between $30$ $\--$ $60$ minutes.
$N_0=(1.34\pm0.21)\times10^3$ and $\alpha^{N}_{\tau}=-1.53\pm0.06$ are obtained with $\tau_0=40$ minutes.
There is a rapid drop above it, which is different from data set 1.
Figure \ref{fig:fltau_d2} shows a scatter plot of flux content and lifetime.
The horizontal dashed line shows the observational duration of this data set.
We see that the longest lifetime is shorter than the observational period.
Figure \ref{fig:fllt_d2} shows a dependence of averaged lifetime on flux content.
There is a change of slope around $2 \phi_{\rm th}$.
The lifetime decreases again below it.
However, the lifetime becomes nearly independent of flux content above it.
The value of constant lifetime is $\thicksim 60$ minutes, which is similar value of upper limit of power-law distribution 
in Figure \ref{fig:dist_lt_d2}.

\ifodd \arabic{page}
\else
  \thispagestyle{fancy}
  \mbox{}
  \newpage
  \clearpage
\fi

\chapter[Freqeuncies of surface processes]{Frequencies and Flux Amounts of Surface Processes of Magnetic Patches}
\thispagestyle{fancy}
Frequencies and flux amounts of surface processes are investigated in this chapter.
Total frequencies and flux change rates of processes in both data sets are summarized in Section \ref{sec:dscrpt_magpro}.
The purpose of this investigation is to determine what processes is dominant in surface processes.
Further, flux dependences of merging, splitting, and cancellation are investigated in Section \ref{sec:obs_distpro}.
We will construct a physical picture of flux maintenance in quiet regions from these results in Section \ref{sec:discuss_obs}

\section{Total Frequencies and Flux Amounts of Surface Processes}
\label{sec:dscrpt_magpro}

We investigate changing rates of total flux amount by four surface processes of magnetic patches.
Along with the total numbers of patches, their flux density averaged over the field of 
view and the observational period
(in Mx cm$^{-2}$), the frequencies of events in the magnetic flux density 
(in Mx cm$^{-2}$ s$^{-1}$) are summarized in Table \ref{tab:sum_magpro}.

The total numbers of merging for positive and negative patches are $493$ and $482$ in data set 1.
They are $5905$ and $4252$ in data set 2.
The total number of splitting has a same order of magnitude.
Those of splitting for positive and negative patches are $536$ and $535$ in data set 1.
Those in data set 2 are $5764$ and $4143$.
The total numbers of emergence in data set 1 and in data set 2 are $3$ and $27$ respectively.
They are much smaller than those of merging and splitting in both data sets. 
The total numbers of cancellation are $86$ and $775$, which are again smaller than those of merging and splitting.

To determine what processes dominate a flux balance in the region, we investigate change rates of total flux 
amount by each process.
The occurrence time is obtained by dividing the total flux by this rate.
Note that these are averages over the flux content in patches.
The flux dependence of the occurrence times is discussed later in Section \ref{sec:obs_distpro}.
The results are shown in the parenthesis in Table 4.1.
The change rates of flux density involving merging process are defined as sum of flux content of patches before the events.
Those in data set 1 are $1.65 \times 10^{-3}$ Mx cm$^{-2}$ s$^{-1}$ and $3.53 \times 10^{-3}$ Mx cm$^{-2}$ s$^{-1}$ for positive and 
negative patches, respectively.
Time scales of those processes, $\Phi_{\rm tot}/ ( \left. \partial \Phi_{\rm tot}/\partial t \right|_{\rm mrg})$, are evaluated as 
$1.53 \times 10^3$ s and $1.02 \times 10^3$ s from these values.
We obtain $3.36 \times 10^{-3}$ Mx cm$^{-2}$ s$^{-1}$ and $1.77 \times 10^{-3}$ Mx cm$^{-2}$ s$^{-1}$ 
for positive and negative patches in data set 2.
They correspond to the time scales of $1.21 \times 10^3$ s and $1.93 \times 10^3$ s, respectively.
The time scales of these processes are much shorter than the estimated time scales by cancellation and emergence reported in previous studies 
\citep{liv1985, sch1998, hag2001, tho2011}. 
The change rates of flux density involving splitting process are defined as sum of flux content patches before the events. 
We obtain $1.48 \times 10^{-3}$ Mx cm$^{-2}$ s$^{-1}$ and $3.03 \times 10^{-3}$ Mx cm$^{-2}$ s$^{-1}$ for positive and negative patches 
in data set 1, which
corresponds $1.71 \times 10^3$ s and $1.19 \times 10^3$ s, respectively.
In data set 2, we obtain $2.38 \times 10^{-3}$ Mx cm$^{-2}$ s$^{-1}$ and $1.72 \times 10^{-3}$ Mx cm$^{-2}$ s$^{-1}$ 
for positive and negative patches.
The time scales of positive and negative patches are $1.58 \times 10^3$ s and $1.99 \times 10^3$ s.
They are same order of those of merging in both data set and much shorter than those of emergence and cancellation in the previous 
studies again.
The timescale of cancellation is investigated in data set 2.
We detect $610$ cancellations.
Change rate of flux amount is $2.09 \times 10^{-5}$ Mx cm$^{-2}$ s$^{-1}$, which corresponds $1.71\times10^5$ s as a time scale 
of cancellation.
Time scales of emergence and cancellation are in the same order with the previous studies.
It is much longer than those of merging and splitting.
These results deduce that merging and splitting are dominant surface processes in quiet regions.

We should note that there are also many unipolar increases and decreases in both data set.
Total numbers of the detected unipolar processes are shown in the last two lines in Table 5.1.
These events are interpreted as the surface processes with the patches below the detection limit.
We show some examples of these processes and discussions in Appendix B to support the interpretation.

\begin{table}
\begin{center}
\label{tab:sum_magpro}
\begin{tabular}{c|cc|cc}
\hline\hline
&\multicolumn{2}{|c|}{Data Set 1}&\multicolumn{2}{|c}{Data Set 2}\\
&Positive&Negative&Positive&Negative\\
\hline
Tracked Patch&1636&1637&21823&19544\\
&(2.53)&(3.60)&(3.76)&(3.42)\\
Merging&536&535&5905&4252\\
&(1.65 $\times$ 10$^{-3}$)&(3.53 $\times$ 10$^{-3}$)&(3.36 $\times$ 10$^{-3}$)&(1.77 $\times$ 10$^{-3}$)\\
Splitting&493&482&5764&4143\\
&(1.48 $\times$ 10$^{-3}$)&(3.03 $\times$ 10$^{-3}$)&(2.38 $\times$ 10$^{-3}$)&(1.72 $\times$ 10$^{-3}$)\\
\hline
Emergence&\multicolumn{2}{|c|}{3}&\multicolumn{2}{|c}{27} \\
&\multicolumn{2}{|c|}{}&\multicolumn{2}{|c}{}\\
Cancellation&\multicolumn{2}{|c|}{86}&\multicolumn{2}{|c}{775} \\
&\multicolumn{2}{|c|}{}&\multicolumn{2}{|c}{(2.09 $\times$ 10$^{-5}$)}\\
\hline
Unipolar Events&&&&\\
&&&&\\
\shortstack{Increase$\&$ \\ Appearance}&503&556&6185&7633\\
&&&&\\
\shortstack{Decrease$\&$ \\ Disappearance}&381&398&5440&5400\\
\hline
\hline
\end{tabular}
\end{center}
\caption[Total numbers and flux amounts of the magnetic patches and activities in data 1.]{
Summary of total numbers of the patches and surface processes in our data sets.
Flux density and change rates of the flux amount are shown in the parentheses 
with units of Mx cm$^{-2}$ and Mx cm$^{-2}$ s$^{-1}$, respectively.}
\end{table}

\section[Frequency Distributions of Surface Processes]{Frequency Distributions of Interactions of Magnetic Patches}
\label{sec:obs_distpro}

\subsection{Merging}
\label{sec:mrg}

We investigate flux dependences of occurrence of mergings in data set 1 namely, the frequency distribution 
and the probability distribution.
The reason using data set 1 is that the time interval is short enough compared with the occurrence time ($\thicksim$17 minutes).

\begin{figure}[p]
\centering
\includegraphics[bb=0 0 800 780,width=0.9\textwidth]{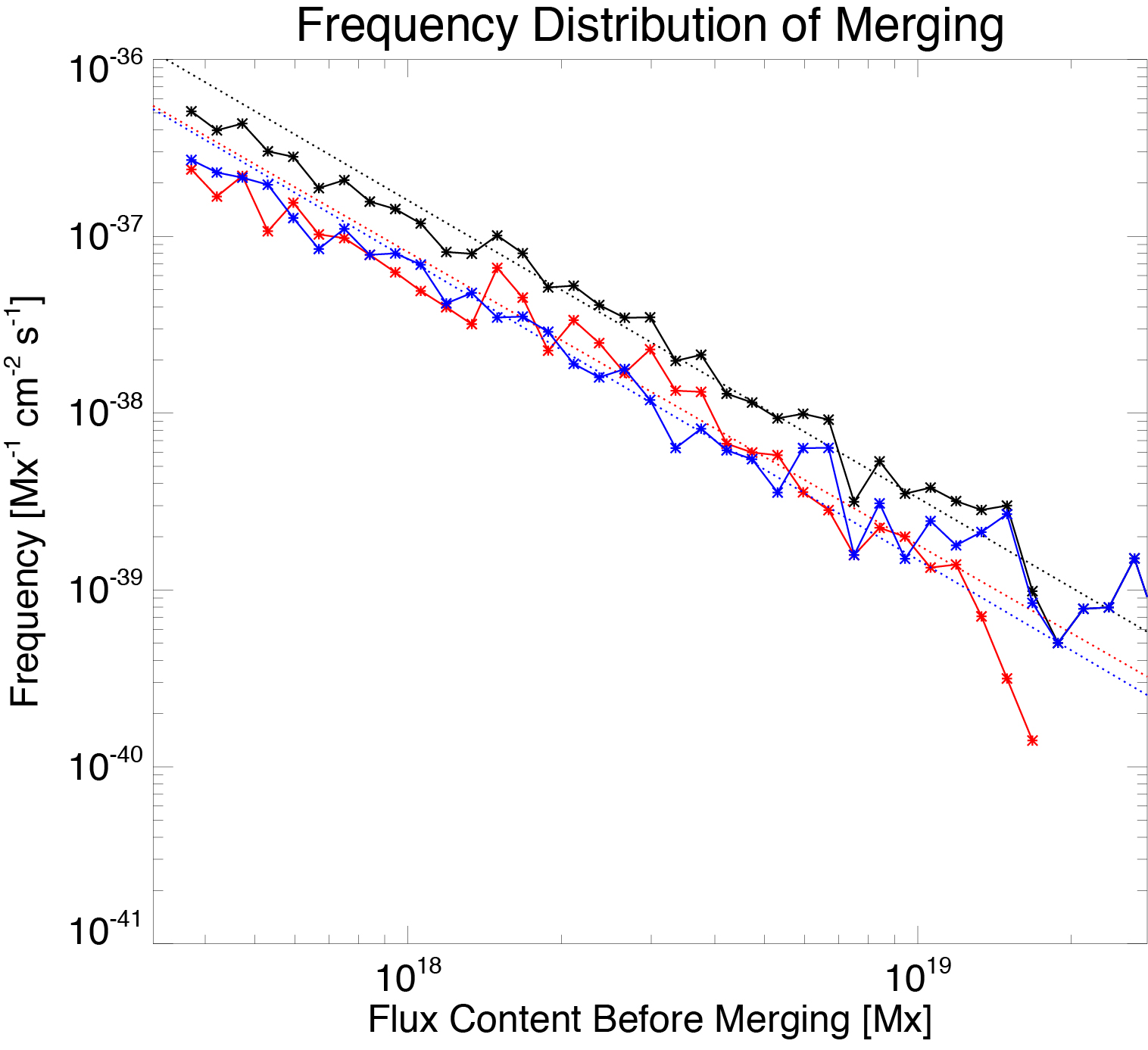}
\caption[Apparent frequency distribution of mergings in data set 1.]{
Apparent frequency distribution of mergings with $\phi_{\rm th}=10^{17.5}$ Mx in data set 1.
It is made from merging where the parent patches have a flux content of more than $\phi_{\rm th}$. 
The red/blue/black solid lines indicate observational results for positive/negative/both patches.
The red/blue/black dashed lines indicate fitting results with a range of $10^{17.5} \-- 10^{19}$ Mx. 
The power-law indexes of the fitting lines are $-1.66.$, $-1.70$, and $-1.68$. 
}
\label{fig:freqmrg_d1}
\end{figure}

First, we investigate the frequency distribution of mergings.
The freqeuncy distribution of merging is deduced by
\begin{equation}
\frac{\partial n_{\rm mrg}^{\rm APP}}{\partial t} = \frac{N_{\rm mrg}^{\rm tot}}{t^{\rm tot} \, S \, \Delta \phi},
\end{equation}
where $N_{\rm mrg}^{\rm tot}$ denotes the total occurrence of mergings throughout the data, 
$t^{\rm tot}$ the observational period, $S$ the observational area, and $\Delta \phi$ the bin size in the flux content.
The result is shown in Figure \ref{fig:freqmrg_d1}.
The distribution seems to have a power-law dependence on flux content and can be a least-squares fitted in the forms as,
\begin{equation}
\frac{\partial n_{\rm mrg}^{\rm APP}}{\partial t}=n_{0,{\rm mrg}} \left( \frac{\phi}{\phi_0} \right)^{-\alpha_{\rm mrg}}.
\end{equation}
$n_{0,{\rm mrg}}$ is the reference frequency at $\phi_0$, $\phi_0(=10^{18}$ Mx$)$ is the reference flux content, and 
$\alpha_{\rm mrg}$ is the power-law index of the probability distribution of merging.
We obtained $n_{0,{\rm mrg}}=(8.14\pm0.86) \times 10^{-38}$ Mx$^{-1}$ cm$^{-2}$ s$^{-1}$ and $\alpha_{\rm mrg}=1.66\pm0.07$ 
for the positive patches, 
$n_{0,{\rm mrg}}=(7.41.\pm0.72) \times 10^{-38}$ Mx$^{-1}$ cm$^{-2}$ s$^{-1}$ and $\alpha_{\rm mrg}=1.70\pm0.06$ for the negative patches, 
and $n_{0,{\rm mrg}}=(1.60\pm0.77) \times 10^{-37}$ Mx$^{-1}$ cm$^{-2}$ s$^{-1}$ and $\alpha_{\rm mrg}=1.68\pm0.05$ for both patches.
The errors mean a $1 \sigma$-error of the least-squares fitting.

\begin{figure}[p]
\centering
\includegraphics[bb=0 0 800 750,width=0.9\textwidth]{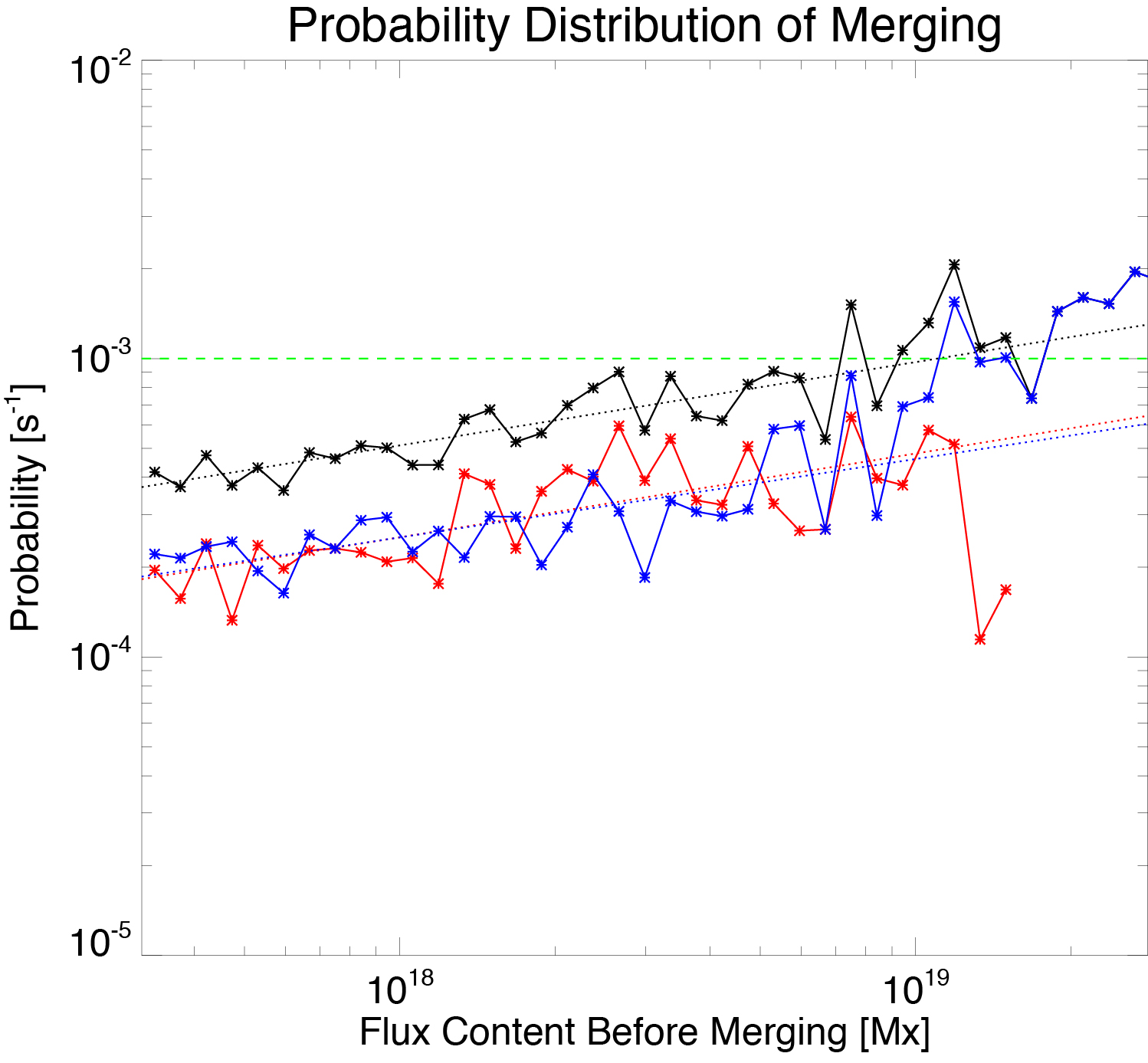}
\caption[Apparent probability distribution of mergings in data set 1.]{
Apparent probability distribution of mergings with $\phi_{\rm th}=10^{17.5}$ Mx in data set 1.
It is made from merging where the parent patches have a flux content of more than $\phi_{\rm th}$. 
The red/blue/black solid lines indicate observational results for positive/negative/both patches.
The red/blue/black dashed lines indicate fitting results with a range of $10^{17.5}$ $\--$ $10^{19}$ Mx. 
The power-law indexes of the fitting lines are $0.28$, $0.26$, and $0.28$. 
The horizontal and vertical dashed lines indicate time scale of $33$ minutes and the detection limit, respectively.}
\label{fig:distmrg_d1}
\end{figure}

Next, we investigate the probability distribution of merging.
The apparent probability distribution of merging is obtained by
\begin{equation}
\frac{\partial P_{\rm mrg}^{\rm APP}}{\partial t} = \frac{1}{n(\phi)} \frac{\partial n_{\rm mrg}^{\rm APP}}{\partial t}.
\end{equation}
Figure \ref{fig:distmrg_d1} shows the result.
We make a least-squares fitting in a range of $10^{17.5}$ $\--$ $10^{19}$ Mx, 
where the number of the merging event is enough for the fitting.
The fitting form is 
\begin{equation}
\frac{\partial P_{\rm mrg}^{\rm APP}}{\partial t}=P_{0,{\rm mrg}} \left( \frac{\phi}{\phi_0} \right)^{\beta_{\rm mrg}}
\end{equation}
where $P_{0,{\rm mrg}}$ is the reference probability at $\phi_0$, $\phi_0$($=10^{18}$ Mx) 
the reference flux content, and 
$\beta_{\rm mrg}$ the power-law index.
We obtained $P_{0,{\rm mrg}}=(2.52\pm0.08) \times 10^{-4}$ s$^{-1}$ and $\beta_{\rm mrg}=0.28\pm0.05$ for the positive patches, 
$P_{0,{\rm mrg}}=(2.52\pm0.08) \times 10^{-4}$ s$^{-1}$ and $\beta_{\rm mrg}=0.26\pm0.05$ for the negative patches, 
and $P_{0,{\rm mrg}}=(5.12\pm0.11) \times 10^{-4}$ s$^{-1}$ and $\beta_{\rm mrg}=0.28\pm0.04$ for both patches with $\phi_0=10^{18}$ Mx.
The errors mean $1 \sigma$-errors of least-squares fitting.

\subsection{Splitting}
\label{sec:splt}

We investigate the flux dependence of splitting occurrence in the same manner as that for mergings in the previous section.

\begin{figure}[p]
\centering
\includegraphics[bb= 0 0 800 750,width=0.9\textwidth]{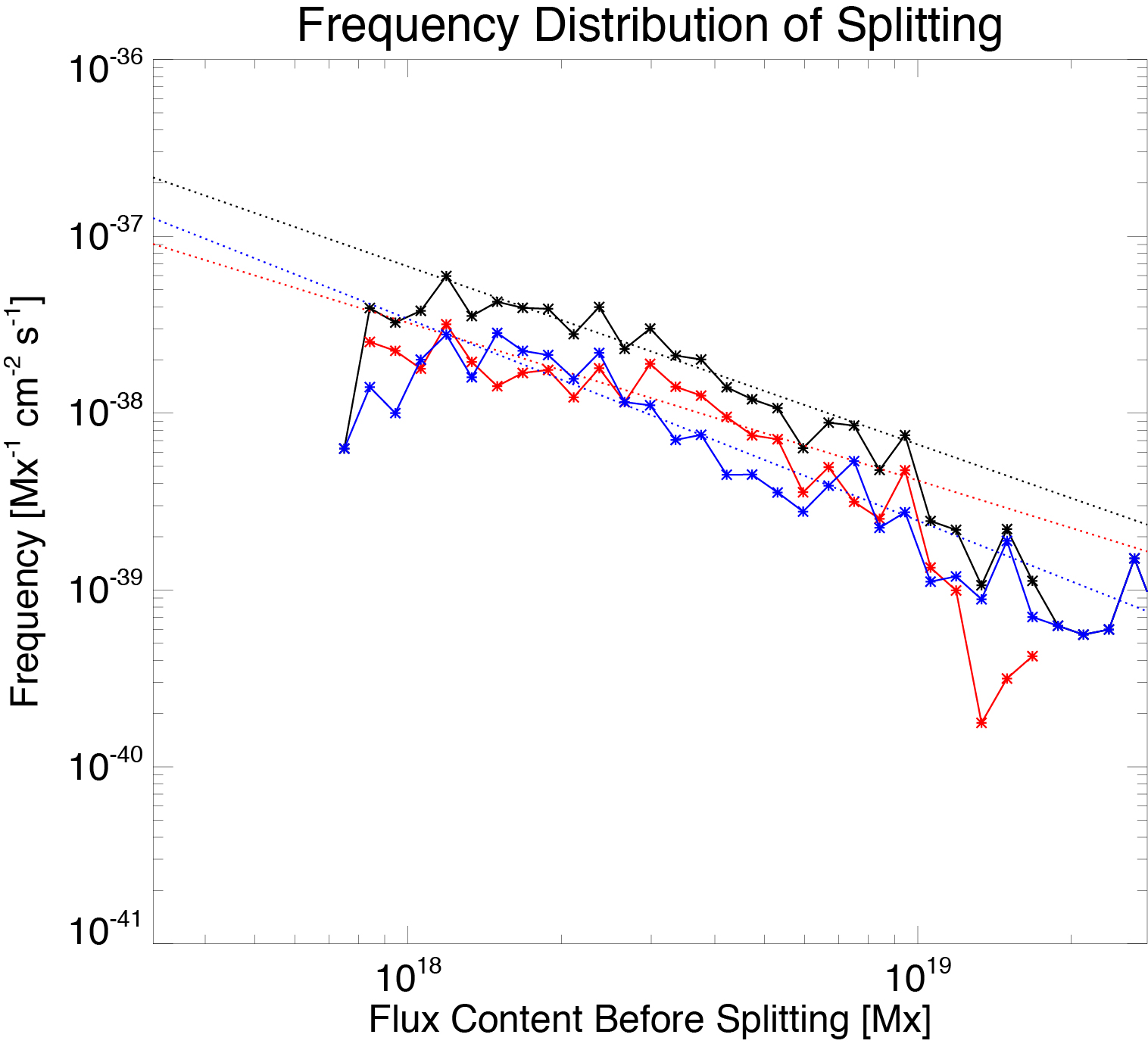}
\caption[Apparent frequency distribution of splittings in data set 1.]{
Apparent frequency distribution of splittings with $\phi_{\rm th}=10^{17.5}$ Mx in data set 1.
It is made from splitting where the parent patches have a flux content of more than $\phi_{\rm th}$. 
The red/blue/black solid lines indicate observational results for positive/negative/both patches.
The red/blue/black dashed lines indicate fitting results with a range of $10^{17.5} \-- 10^{19}$ Mx. 
The power-law indexes of the fitting lines are $-0.89$, $-1.14$, and $-1.01$. 
}
\label{fig:freqsplt_d1}
\end{figure}

First, we investigate the frequency distribution of splittings.
It is
\begin{equation}
\frac{\partial n_{\rm splt}^{\rm APP}}{\partial t} = \frac{N_{\rm splt}^{\rm tot}}{t^{\rm tot} \, S \, \Delta \phi}.
\end{equation}
The result is shown in Figure \ref{fig:freqsplt_d1}.
The distribution seems to be a power-law dependence on flux content again and we make a least-squares fitting in the forms as,
\begin{equation}
\frac{\partial n_{\rm splt}^{\rm APP}}{\partial t}=n_{0,{\rm splt}} \left( \frac{\phi}{\phi_0} \right)^{-\alpha_{\rm splt}}.
\end{equation}
We obtained $n_{0,{\rm mrg}}=(3.24\pm0.62) \times 10^{-38}$ Mx$^{-1}$ cm$^{-2}$ s$^{-1}$ and $\alpha_{\rm mrg}=1.01\pm0.07$ 
for the positive patches, 
$n_{0,{\rm mrg}}=(3.41\pm0.55) \times 10^{-38}$ Mx$^{-1}$ cm$^{-2}$ s$^{-1}$ and $\alpha_{\rm mrg}=1.14\pm0.08$ for the negative patches, 
and $n_{0,{\rm mrg}}=(6.75\pm0.95) \times 10^{-37}$ Mx$^{-1}$ cm$^{-2}$ s$^{-1}$ and $\alpha_{\rm mrg}=1.01\pm0.07$ for both patches.

\begin{figure}[p]
\centering
\includegraphics[bb=0 0 800 750,width=0.9\textwidth]{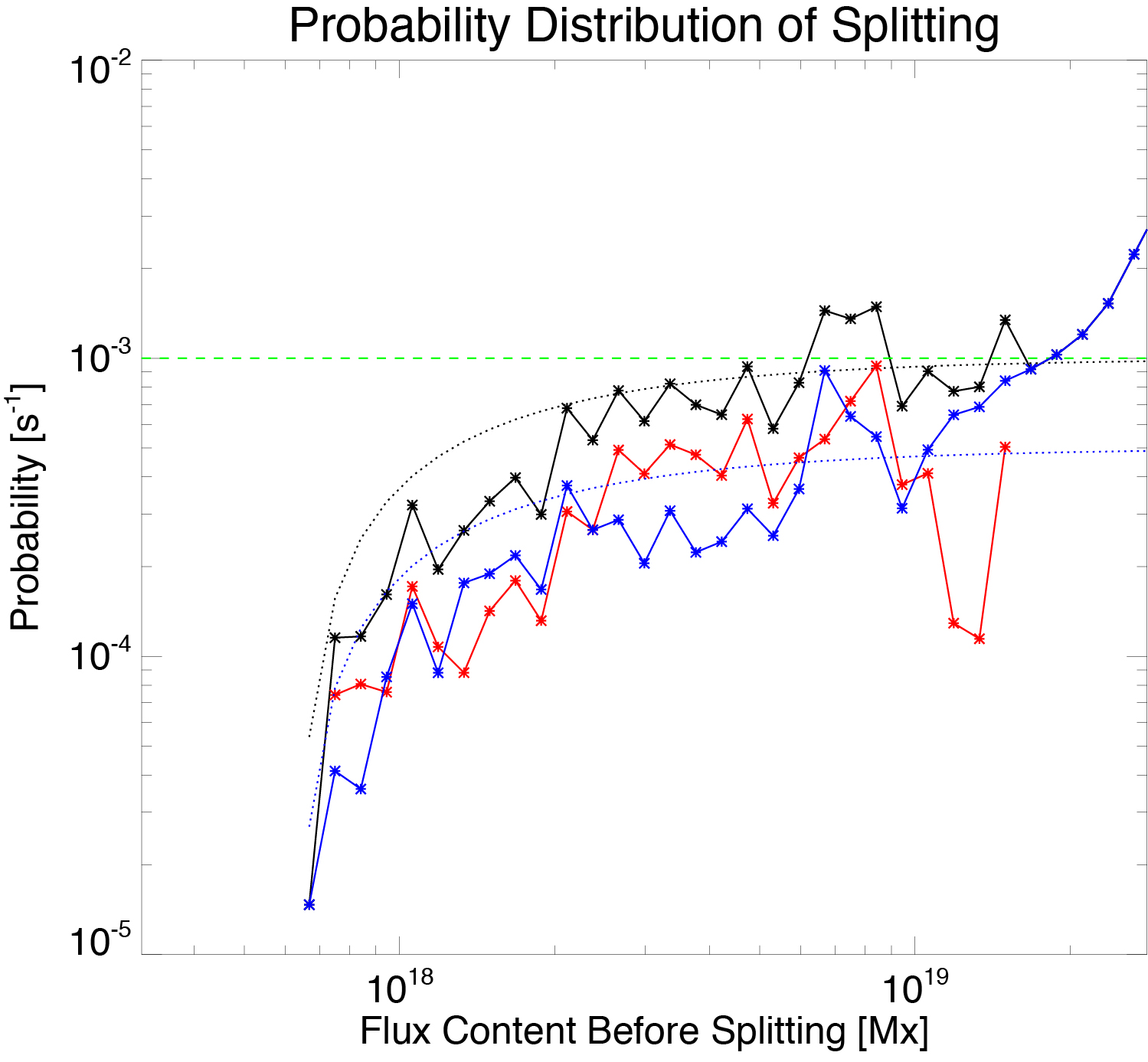}
\caption[Apparent probability distribution of splittings in data 1.]{
Apparent probability distribution of splittings with $\phi_{\rm th}=10^{17.5}$ Mx in data set 1.
The red/blue/black solid lines indicate observational results for positive/negative/both patches.
The blue and black dashed lines indicate analytical curves with $k_0=5.0\times10^{-4}$ s$^{-1}$ and $1.0\times10^{-3}$ s$^{-1}$, 
which are discussed in Section \ref{sec:discuss_form}. 
The horizontal dashed line indicates a time scale of $33$ minutes.
}
\label{fig:distsplt_d1}
\end{figure}

Next, we investigate an apparent probability distribution of splitting in data set 1.
The apparent probability distribution of splitting is obtained in the same manner as that of merging.
The apparent probability distribution of splitting is induced as
\begin{equation}
\frac{\partial P_{\rm splt}^{\rm APP}}{\partial t} = \frac{1}{n(\phi)} \frac{\partial n_{\rm splt}^{\rm APP}}{\partial t}.
\end{equation}
Figure \ref{fig:distsplt_d1} shows the result.
The strong increase in the range larger than $10^{19}$ Mx is probably caused by the lack of a patch number in the analysis.
On the other hand, there is a drop in the flux range near $\phi_{\rm{th}}$ where the number of patches is enough for a statistical study. 
We interpret this dropping as an effect of splitting into the area below $\phi_{\rm{th}}$. 
We quantify this effect in the discussion in Section \ref{sec:discuss_form}. 
We see that the probability of splitting is almost constant as $1.0 \times 10^{-3} \  \rm{s}^{-1}$, 
which means a time scale of $33$ minutes, 
in the range enough above $\phi_{\rm{th}}$, $3.0 \times 10^{18} - 1.0 \times 10^{19}$ Mx.
It means that frequency of splitting is independent of the parents' flux content.

\subsection{Cancellation}
\label{sec:cnc}

\begin{figure}[p]
\centering
\includegraphics[bb=0 0 800 750,width=0.9\textwidth]{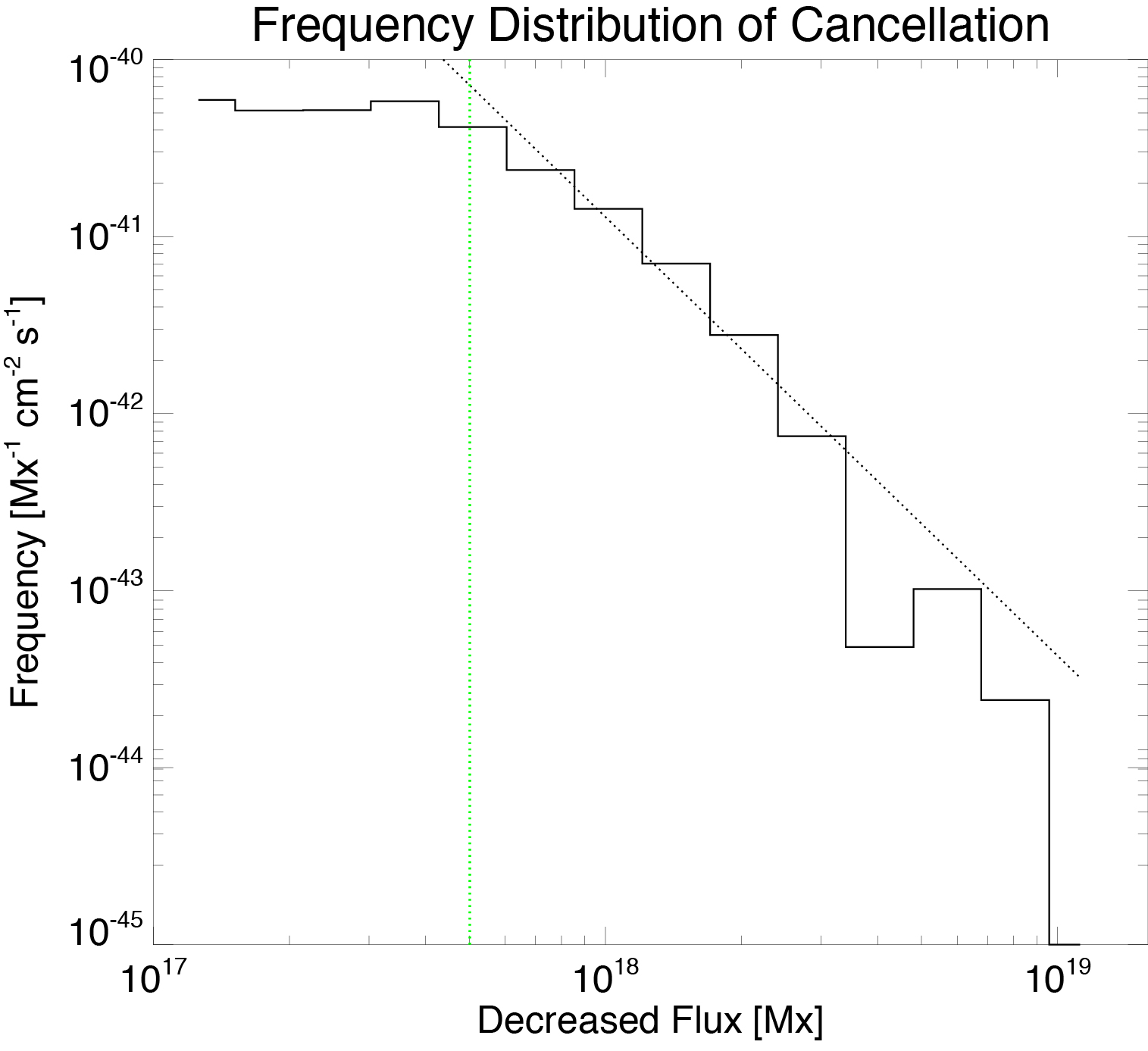}
\caption[Frequency distribution of cancellations in data 2.]{
Frequency distribution of cancellations on decreased flux in data 2. 
The histogram and dashed line indicate the observational result and 
the fitting result with a power-law distribution, respectively.
The vertical green line shows detection limit, $\phi_{\rm th}=10^{17.7}$ Mx.
The fitted power-law index is $-2.48$.}
\label{fig:distcnc_d2}
\end{figure}

We investigate an apparent frequency distribution of cancellations in this section.
Because the total number of cancellations in data set 1 is too small, we use data set 2.
The frequency distribution of cancellation is obtained in the same manner as the merging and splitting, namely
\begin{equation}
\frac{\partial n_{\rm cnc}^{\rm APP}}{\partial t} = \frac{N_{\rm cnc}^{\rm tot}}{t^{\rm tot} \, S \, \Delta \phi}.
\end{equation}
Figure \ref{fig:distcnc_d2} shows the result.
We make a least-squares fitting with a form of 
\begin{equation}
\frac{\partial n_{\rm cnc}^{\rm APP}}{\partial t}=\frac{\partial n_{0,{\rm cnc}}}{\partial t} \left( \frac{\phi}{\phi_0} \right)^{-\gamma_{\rm cnc}}.
\end{equation}
We obtained $\partial n_{0,{\rm cnc}} / \partial t=(1.29\pm0.12) \times 10^{-41}$ Mx$^{-1}$ cm$^{-2}$ s$^{-1}$ and 
$\gamma_{\rm cnc}=2.48\pm0.26$ and the fitting 
range from $10^{17.7}$ Mx to $10^{18.4}$ Mx.
The errors indicate $1 \sigma$-errors of least-squares fitting.

\begin{figure}[p]
\centering
\includegraphics[bb=0 0 550 600,width=0.9\textwidth]{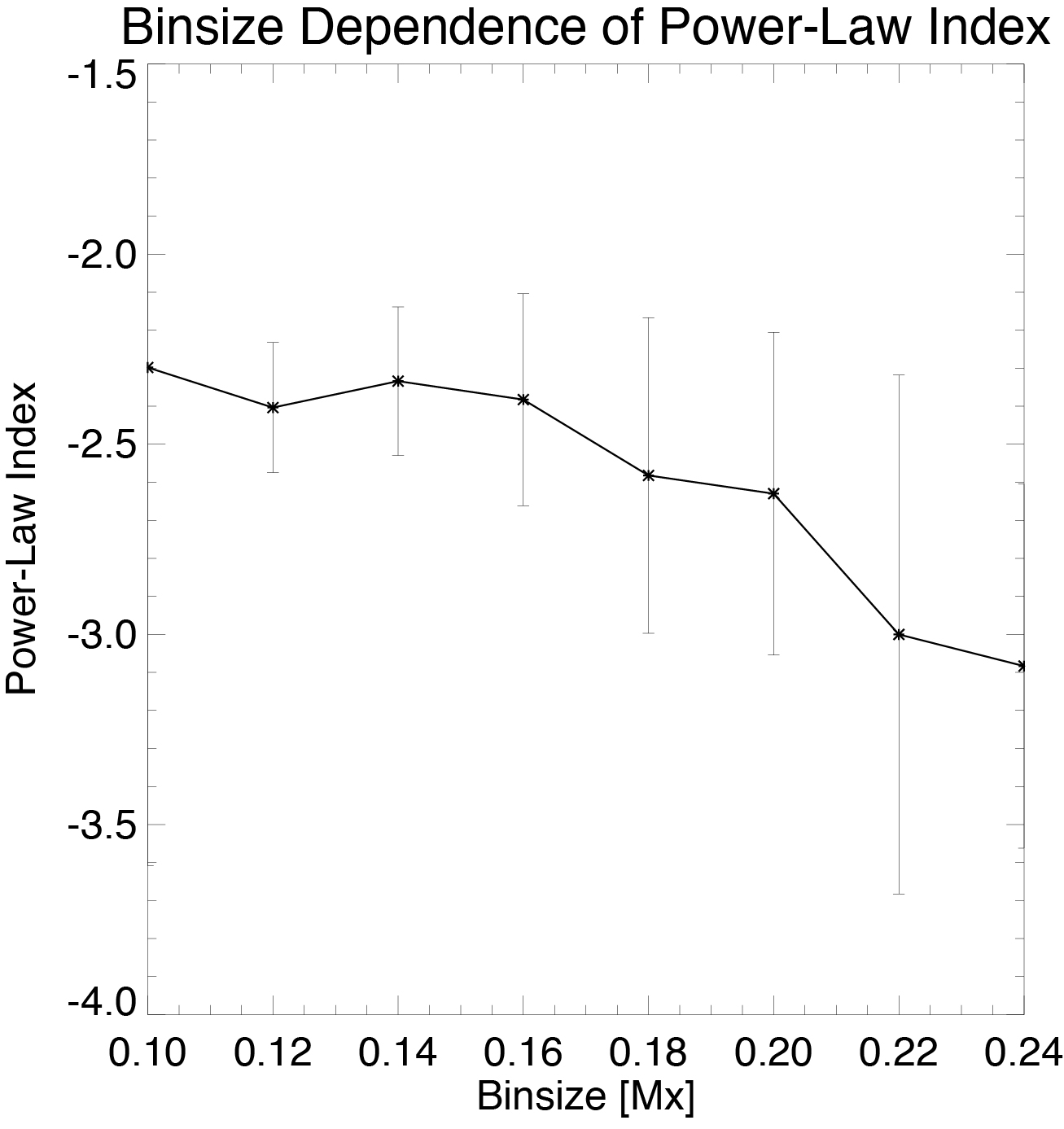}
\caption[Dependence on fitting bin size of the power-law index of cancellation in data 2.]{
Dependence on fitting bin size of the power-law index of cancellation in data 2.
The error bars show 1$\sigma$-error of the fitting.}
\label{fig:bin_index_cnc}
\end{figure}

We investigate the dependence of the power-law index on the bin size for the least-squares fitting because the total number of 
cancellation is not so large in data set 2.
The result is shown in Figure \ref{fig:bin_index_cnc}.
The error bars indicate $1 \sigma$-errors of the least-squares fitting in each case.
We see that the error has the tendency to increase with the larger bin size, which is caused by the decrease of the fitting points.
The proper power-law index is roughly from $-2.3$ to $-2.6$.

\begin{figure}[p]
\centering
\includegraphics[bb=0 0 800 750,width=0.9\textwidth]{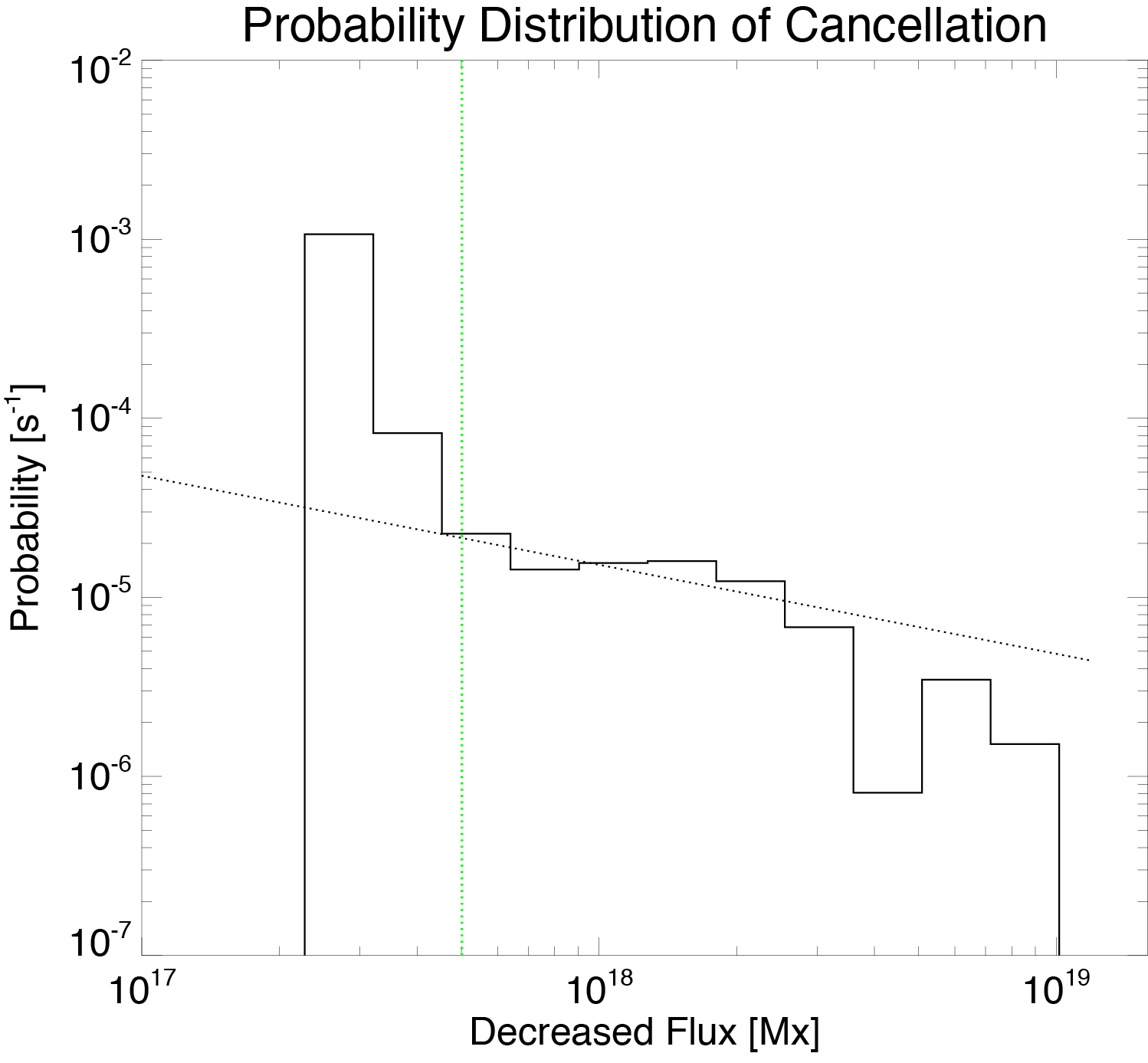}
\caption[Probability distribution of cancellation in data set 2.]{
Probability distribution of cancellation in data set 2.
The histogram and dashed line indicate the observational result and 
the fitting result with a power-law distribution, respectively.
The vertical green line shows detection limit, $\phi_{\rm th}=10^{17.7}$ Mx.
The fitted power-law index is $-0.50$.}
\label{fig:prob_cnc}
\end{figure}

Then, we investigate an apparent probability distribution of cancellation.
It is given as
\begin{equation}
\frac{\partial P_{\rm cnc}^{\rm APP}}{\partial t} = \frac{1}{n(\phi)} \frac{\partial n_{\rm cnc}^{\rm APP}}{\partial t}.
\end{equation}
Figure \ref{fig:prob_cnc} shows the result.
We make a least-squares fitting in a range of $10^{17.7}$ $\--$ $10^{18.4}$ Mx in the same manner as that for 
merging and splitting. The fitting form is 
\begin{equation}
\frac{\partial P_{\rm cnc}^{\rm APP}}{\partial t}=P_{0,{\rm cnc}} \left( \frac{\phi}{\phi_0} \right)^{\beta_{\rm cnc}}
\end{equation}
We obtained $P_{0,{\rm cnc}}=(1.52\pm0.15) \times 10^{-5}$ s$^{-1}$ and $\beta_{\rm cnc}=-0.50\pm0.24$.
The error bars mean $1 \sigma$-errors of the least-squares fitting.

\ifodd \arabic{page}
\else
  \thispagestyle{fancy}
  \mbox{}
  \newpage
  \clearpage
\fi

\chapter{Discussions and Interpretations}
\thispagestyle{fancy}
We discuss the results of Chapter 3 and Chapter 4 in this chapter.
First we summarize and expand magneto-chemistry equation (M-C equation hereafter), 
which is suggested by \cite{sch1998} and useful to evaluate the observational results, in Section \ref{sec:magch}.
Some forms connecting observational results and elementary functions in M-C equation are deduced.
Discussions of characteristics of patches and frequencies of surface processes are done in Section 
\ref{sec:discuss_patch} and \ref{sec:discuss_obs}.
We suggest a physical picture of flux maintenance on the solar surface from these discussions in Section \ref{sec:phys_pict}.

\pagestyle{fancy}
\section{Magneto-Chemistry Equation}
\label{sec:magch}

The detailed description of magneto-chemistry equation is summarized and expanded in this chapter.
We deduce some forms connecting frequency distributions of surface processes with observables, namely total number of patches, 
total flux amount, total numbers of surface processes, time scales by surface processes, and averaged lifetime.    
A general description of magneto-chemistry equation is summarized in section \ref{sec:gen_magch}.
We deduce general forms in section \ref{sec:magch_cmp}. 
In section \ref{sec:magch_lmt}, we take the detection limit of analysis into the account.
We stand on one interpretation of unipolar processes that they are surface processes between magnetic patches below the detection limit.

\subsection{General Description}
\label{sec:gen_magch}

The magneto-chemistry equation was proposed by \cite{sch1997b}, 
which describes a relation among frequency distribution of flux content, $n(\phi)$, and those of four surface processes. 
This equation treats magnetic structures as isolated patches.
This treatment is valid with continuity and no dissipation of the field, here $\bold{B}$.
The remained condition, no dissipation of magnetic field, is probably valid because of the large magnetic Raynolds number, $R_m=L_0V_0/\eta$, on the soar surface.
We obtain $R_m\thicksim 10^3-10^7$ in the solar photosphere, assuming partial ionization and collision dominated plasma \citep{spi1962,pri1984}.
Then magneto-chemistry equation is written as 
\begin{equation}
\frac{\partial n_{\pm}(\phi)}{\partial t} = 
 S_{\pm}(\phi) + L_{\pm}(\phi) + K_{\pm}(\phi) + M_{\pm}(\phi)
\end{equation}
where $n_{\pm}(\phi)$ denotes a frequency distribution of positive (negative) patches with a flux content $\phi$, 
$S_{\pm}(\phi)$, $L_{\pm}(\phi)$, $K_{\pm}(\phi)$, and $M_{\pm}(\phi)$ denotes 
changes of number density of positive (negative) patches with a flux content of $\phi$ 
by emergence, merging, splitting, and cancellation, respectively.
We describe them with probability density distributions of activities, $l(x,y)$, $k(x,y)$, and $m(x,y)$.
Here, $l(x,y)$ is defined as a probability density of mergings where patches with flux contents of $x$ and $y$ merge 
into a patch with a flux content 
of $x+y$,  $k(x,y)$ is defined as a probability density of splittings where a patch with flux content $x+y$ splits 
into patches with flux contents 
of $x$ and $y$,  and $m(x,y)$ is defined as a probability density of cancellations where patches with flux contents 
of $x$ and $y$ cancel 
and results in a patch with a flux content of $|x+y|$.
By this definition, one obtain 
\begin{eqnarray}
\displaystyle L_{\pm}&=& \frac{1}{2} \int_0^{\phi}n_{\pm}(x)n_{\pm}(\phi-x)l(x,\phi-x)dx \notag \\
& \,& \ \ \ \ \ \ \ \ \ \ \ \ \ \ \ \ \ \ \ \ \ \ \ \ \ \ \ \ \ \ - n_{\pm}(\phi) \int_0^{\infty}n_{\pm}(x)l(\phi,x)dx \label{eq:mrg_gen}\\
\displaystyle K_{\pm}&=&  2 \int_0^{\infty}n_{\pm}(x)k(\phi,x-\phi)dx - n_{\pm}(\phi) \int_{0}^{\phi}k(x,\phi-x)dx
\label{eq:splt_gen}
\end{eqnarray}
\begin{eqnarray}
\displaystyle M_{\pm}&=& \int_{0}^{\infty}n_{\pm}(\phi+x)n_{\mp}(x)m(\phi+x,x)dx \notag \\
& \,& \ \ \ \ \ \ \ \ \ \ \ \ \ \ \ \ \ \ \ \ \ \ \ \ \ \ \ \ \ \ - n_{\pm}(\phi)\int_{0}^{\infty}n_{\mp}(x)m(\phi,x)dx.
\label{eq:cnc_gen}
\end{eqnarray}
The first and second terms in equations (\ref{eq:mrg_gen})-(\ref{eq:cnc_gen}) are production and loss terms respectively.
We remove difference between polarities as
\begin{eqnarray}
n_{+}(\phi)&=&n_{-}(\phi)=n(\phi) \label{eq:dst_flx_eq}\\
S_{+}(\phi)&=&S_{-}(\phi)=S(\phi) \label{eq:dst_emg_eq}\\
l_{+}(\phi)&=&l_{-}(\phi)=l(\phi) \label{eq:dst_mrg_eq}\\
k_{+}(\phi)&=&k_{-}(\phi)=k(\phi) \label{eq:dst_splt_eq}\\
m_{+}(\phi)&=&m_{-}(\phi)=m(\phi)\label{eq:dst_cnc_eq}.
\end{eqnarray}
Equation (\ref{eq:dst_flx_eq}) is applicable only in well mixed-polarity and flux balanced region, which is 
probably applicable in quiet regions.

\subsection{Relationships with Observables}
\label{sec:magch_cmp}

We show specific forms connecting probability density distributions, namely $l(x,y)$, $k(x,y)$, and $m(x,y)$, and observables in this section.
These forms are used in the discussion of chapter 4 and 5.

First, we summarize the forms of total number of patches, $N_{\rm tot}$, and total flux amount, $\Phi_{\rm tot}$, in the system from \cite{sch1997b}.
We assume the homogeneity of a frequency distribution in this study.
Then  we obtain
\begin{eqnarray}
\displaystyle N_{\rm tot}&=& \int_0^{\infty}n(\phi) d\phi\\
\displaystyle \Phi_{\rm tot}&=& \int_0^{\infty}\phi \, n(\phi) d\phi \label{eq:flux}.
\end{eqnarray}
These variables are more directly comparable with the observables.  

Next we deduce forms representing probability density distributions, total occurrence rates, and total numbers of surface processes.
Probability density distribution means probability density of one patch.
It would represent the physical character more clearly. 
Frequency distributions of merging and splitting in one polarity equal to loss terms in Equation (\ref{eq:mrg_gen}) 
and Equation (\ref{eq:splt_gen}).  
On the other hand, that for cancellation equals twice of loss term because we have to sum up occurrence of cancellation in both polarity. 
We obtain them as
\begin{eqnarray}
\displaystyle \frac{\partial n_{\rm mrg}}{\partial t}(\phi) &=&  n(\phi)\int_0^{\infty}n(x)l(\phi,x)dx \label{eq:n_mrg}\\
\displaystyle \frac{\partial n_{\rm splt}}{\partial t}(\phi) &=&  n(\phi)\int_0^{\phi}k(x,\phi-x)dx \label{eq:n_splt}\\
\displaystyle \frac{\partial n_{\rm cnc}}{\partial t}(\phi) &=&  2n(\phi)\int_0^{\infty}n(x)m(\phi,x)dx \label{eq:n_cnc}
\end{eqnarray}
where $\partial n_{\rm mrg} / \partial t$, $\partial n_{\rm splt}/\partial t$ and $\partial n_{\rm cnc}/\partial t$ are change rates of 
frequency distribution 
by merging, splitting, and cancellation respectively.
Frequency distribution of emergence, $\partial n_{\rm emrg} / \partial t$, equals to a source term itself from their definitions, namely 
\begin{equation}
\displaystyle \frac{\partial n_{\rm emrg}}{\partial t}(\phi) = 2 n(\phi) \int_0^{\infty}n(\phi+x)m(\phi,\phi+x)dx \label{eq:n_emrg}.
\end{equation}
We obtain probability density distributions simply by dividing frequency distributions by frequency distribution of flux content as
\begin{eqnarray}
\displaystyle \frac{\partial P_{\rm emrg}}{\partial t}(\phi) &=& \frac{1}{n(\phi)} \frac{\partial n_{\rm emrg}}{\partial t}
=  2 \int_0^{\infty}n(x)n(\phi+x)m(\phi,\phi+x)dx \label{eq:p_emrg}\\
\displaystyle \frac{\partial P_{\rm mrg}}{\partial t}(\phi) &=& \frac{1}{n(\phi)} \frac{\partial n_{\rm mrg}}{\partial t}
=  \int_0^{\infty}n(x)l(\phi,x)dx \label{eq:p_mrg}\\
\displaystyle \frac{\partial P_{\rm splt}}{\partial t}(\phi) &=& \frac{1}{n(\phi)} \frac{\partial n_{\rm splt}}{\partial t}
= \int_0^{\phi}k(x,\phi-x)dx \label{eq:p_splt}\\
\displaystyle \frac{\partial P_{\rm cnc}}{\partial t}(\phi) &=& \frac{1}{n(\phi)} \frac{\partial n_{\rm cnc}}{\partial t}
=  2 \int_0^{\infty}n(x)m(\phi,x)dx \label{eq:p_cnc}
\end{eqnarray}
where $\partial P_{\rm emrg} / \partial t$, $\partial P_{\rm mrg} / \partial t$, $\partial P_{\rm splt} / \partial t$ and 
$\partial P_{\rm cnc} / \partial t$ are
probability density distributions of emergence, merging, splitting and cancellation respectively.
We evaluate total occurrence rates of them.
By integrating Equation (\ref{eq:n_mrg})-(\ref{eq:n_emrg}) on $\phi$ from $0$ to $\infty$, we obtain
\begin{eqnarray}
\displaystyle \left. \frac{\partial N_{\rm tot}}{\partial t} \right|_{\rm emrg} 
&=& 2 \int_0^{\infty} \left[ n(\phi) \int_0^{\infty}n(\phi+x)m(\phi,\phi+x)dx \right] d\phi \label{eq:N_emrg} \\
\displaystyle \left. \frac{\partial N_{\rm tot}}{\partial t} \right|_{\rm mrg} 
&=& \int_0^{\infty} \left[ n(\phi) \int_0^{\infty}n(x)l(\phi,x)dx \right] d\phi \label{eq:N_mrg} \\
\displaystyle \left. \frac{\partial N_{\rm tot}}{\partial t} \right|_{\rm splt} 
&=& A\int_0^{\infty} \left[ n(\phi) \int_0^{\phi}k(x,\phi-x)dx \right] d\phi \label{eq:N_splt} \\
\displaystyle \left. \frac{\partial N_{\rm tot}}{\partial t} \right|_{\rm cnc} 
&=& 2 \int_0^{\infty} \left[ n(\phi) \int_0^{\infty}n(x)m(\phi,x)dx \right] d\phi \label{eq:N_cnc}.
\end{eqnarray}

Further we want to deduce total change rates of flux amount by surface processes.
By multiplying Equation (\ref{eq:n_mrg})-(\ref{eq:n_emrg}) by $\phi$ and integrating on $\phi$ from $0$ to $\infty$, we obtain 
\begin{eqnarray}
\displaystyle \left. \frac{\partial \Phi_{\rm tot}}{\partial t} \right|_{\rm emrg} &=& 
2 \int_0^{\infty} \left[ \phi n(\phi) \int_0^{\infty}n(\phi+x)m(\phi,\phi+x)dx \right] d\phi \label{eq:phi_emrg} \\
\displaystyle \left. \frac{\partial \Phi_{\rm tot}}{\partial t} \right|_{\rm mrg} &=& 
\int_0^{\infty} \left[ \phi n(\phi) \int_0^{\infty}n(x)l(\phi,x)dx \right] d\phi \label{eq:phi_mrg} \\
\displaystyle \left. \frac{\partial \Phi_{\rm tot}}{\partial t} \right|_{\rm splt} &=&
\int_0^{\infty} \left[ \phi n(\phi) \int_0^{\phi}k(x,\phi-x)dx \right] d\phi \label{eq:phi_splt} \\
\displaystyle \left. \frac{\partial \Phi_{\rm tot}}{\partial t} \right|_{\rm cnc} &=&
2 \int_0^{\infty} \left[ \phi n(\phi) \int_0^{\infty}n(x)m(\phi,x)dx \right] d\phi \label{eq:phi_cnc}.
\end{eqnarray}

The time scale for each process are defined by dividing total flux amount, Equation (\ref{eq:flux}), by total change rate of flux amount 
, Equation (\ref{eq:phi_emrg})-(\ref{eq:phi_cnc}).
\begin{eqnarray}
\tau_{\rm emrg} &=& \Phi_{\rm tot} \left[ \left. \frac{\partial \Phi_{\rm tot}}{\partial t} \right|_{\rm emrg} \right]^{-1} \notag \\
\displaystyle &=& \dfrac{\displaystyle \int_0^{\infty}\phi \, n(\phi) d\phi}
{\displaystyle 2 \int_0^{\infty} \left[ \phi n(\phi) \int_0^{\infty}n(\phi+x)m(\phi,\phi+x)dx \right] d\phi} \label{eq:tau_emrg}
\end{eqnarray}
\begin{eqnarray}
\tau_{\rm mrg} &=& \Phi_{\rm tot} \left[ \left. \frac{\partial \Phi_{\rm tot}}{\partial t} \right|_{\rm mrg} \right]^{-1} \notag \\
\displaystyle &=& \dfrac{\displaystyle \int_0^{\infty}\phi \, n(\phi) d\phi} 
{\displaystyle \int_0^{\infty} \left[ \phi n(\phi) \int_0^{\infty}n(x)l(\phi,x)dx \right] d\phi} \label{eq:tau_mrg} \\
\tau_{\rm splt} &=& \Phi_{\rm tot} \left[ \left. \frac{\partial \Phi_{\rm tot}}{\partial t} \right|_{\rm splt} \right]^{-1} \notag \\
\displaystyle &=& \dfrac{\displaystyle \int_0^{\infty}\phi \, n(\phi) d\phi}
{\displaystyle \int_0^{\infty} \left[ \phi n(\phi) \int_0^{\phi}k(x,\phi-x)dx \right] d\phi} \label{eq:tau_splt} 
\end{eqnarray}
\begin{eqnarray}
\tau_{\rm cnc} &=& \Phi_{\rm tot} \left[ \left. \frac{\partial \Phi_{\rm tot}}{\partial t} \right|_{\rm cnc} \right]^{-1} \notag \\
\displaystyle &=& \dfrac{\displaystyle \int_0^{\infty}\phi \, n(\phi) d\phi}
{\displaystyle 2 \int_0^{\infty} \left[ \phi n(\phi) \int_0^{\infty}n(x)m(\phi,x)dx \right] d\phi} \label{eq:tau_cnc}.
\end{eqnarray}

At the last step, we deduce lifetime of patches.
Patches disappear at merging or canceling with a patch that has larger flux content.
The probability of disappearance is written from Equation (\ref{eq:p_mrg}) and Equation (\ref{eq:p_cnc}) as 
\begin{eqnarray}
\frac{\partial P_{\rm disapp}}{\partial t} (\phi) &=& \frac{1}{2} \frac{\partial P_{\rm tcnc}}{\partial t} + 
\frac{\partial P_{\rm tmrg}}{\partial t} \notag \\
&=&  \int_{\phi}^{\infty}n(x)m(\phi,x)dx + \int_{\phi}^{\infty}n(x)l(\phi,x)dx.
\end{eqnarray}

The lifetime is an inverse of probability of disappearance, namely
\begin{eqnarray}
\tau(\phi)&=&\left[ \frac{\partial P_{\rm disapp}}{\partial t} \right]^{-1} \notag \\
&=&\left[ \int_{\phi}^{\infty}n(x)m(\phi,x)dx + \int_{\phi}^{\infty}n(x)l(\phi,x)dx \right]^{-1}.
\end{eqnarray}

\subsection{Consideration of Detection Limit}
\label{sec:magch_lmt}

We want to introduce one idea, detection limit here.
As is explained in Chapter 2, there are many unipolar appearances and disappearances in the observational data.
One possible interpretation is surface process among involved patches below the observational limit as 
reported by \cite{lam2010}.
In a statistical analysis, it probably leads a misunderstanding when we investigate in the range where most patches are missed.
However, we cannot avoid missing the patches with smaller size than spatial resolution and weaker signal than observational limit.
We introduce an analytical limit of flux content, $\phi_{\rm th}$, to evaluate the effect quantitatively.
The term a detection limit is defined as a flux content above which we can pick up all patches.
We obtain it from the frequency distribution of flux content. 
We obtain $\phi_{\rm th}\thicksim 10^{19}$ for MDI FD data (Figure \ref{fig:dist_par}).
We cannot justify the assumption completely.
However, one justification of the interpretation is that frequency distribution by SOT/NFI, namely high-resolution data, 
belongs to the same power-law distribution although it is flattened in MDI FD line.

With a limit of statistical analysis, the forms for observables should be changed.
We can pick up processes where all involved patches have flux content larger than $\phi_{\rm th}$.
The number density of patches, flux density, total number of patches and total flux amount 
are obtained simple by changing the range of integration from [0,$\infty$] to [$\phi_{\rm th}$,$\infty$],
\begin{eqnarray}
\displaystyle N_{\rm tot}^{\rm APP}&=& A \int_{\phi_{\rm th}}^{\infty}n(\phi) d\phi \label{eq:num_app}\\
\displaystyle \Phi_{\rm tot}^{\rm APP}&=&A \int_{\phi_{\rm th}}^{\infty}\phi \, n(\phi) d\phi \label{eq:phi_app}.
\end{eqnarray}
where superscript ${\rm APP}$ means apparent observable taking into the detection limit into account.

On the other hand, apparent frequency distributions of surface processes become
\begin{equation}
\displaystyle \frac{\partial n_{\rm emrg}^{\rm APP}}{\partial t}(\phi) = 2 n(\phi) \int_0^{\infty}n(\phi+x)m(\phi,\phi+x)dx \label{eq:n_emrg_th}
\end{equation}
\begin{eqnarray}
\displaystyle \frac{\partial n_{\rm mrg}^{\rm APP}}{\partial t}(\phi) &=&  n(\phi) \int_{\phi_{\rm th}}^{\infty}n(x)l(\phi,x)dx \label{eq:n_mrg_th}\\
\displaystyle \frac{\partial n_{\rm splt}^{\rm APP}}{\partial t}(\phi) &=&  n(\phi)\int_{\phi_{\rm th}}^{\phi-\phi_{\rm th}}k(x,\phi-x)dx \label{eq:n_splt_th}\\
\displaystyle \frac{\partial n_{\rm cnc}^{\rm APP}}{\partial t}(\phi) &=& 2 n(\phi)\int_{\phi_{\rm th}}^{\infty}n(x)m(\phi,x)dx \label{eq:n_cnc_th}.
\end{eqnarray}
Note that apparent frequency distribution of emergence does not change from the actual frequency distribution of emergence 
because number of cancellation with a decrease of $\phi$ in the system does not change. 
In the same manner, apparent probability distributions, apparent total occurrence rates, total change rates of flux amount, and apparent time scales 
are evaluated as 
\begin{eqnarray}
\displaystyle \frac{\partial P_{\rm emrg}^{\rm APP}}{\partial t}(\phi) &=&  2 \int_0^{\infty}n(x)n(\phi+x)m(\phi,\phi+x)dx \label{eq:p_emrg_th}\\
\displaystyle \frac{\partial P_{\rm mrg}^{\rm APP}}{\partial t}(\phi) &=&  \int_{\phi_{\rm th}}^{\infty}n(x)l(\phi,x)dx \label{eq:p_mrg_th}\\
\displaystyle \frac{\partial P_{\rm splt}^{\rm APP}}{\partial t}(\phi) &=& \int_{\phi_{\rm th}}^{\phi-\phi_{\rm th}}k(x,\phi-x)dx \label{eq:p_splt_th}\\
\displaystyle \frac{\partial P_{\rm cnc}^{\rm APP}}{\partial t}(\phi) &=& 2 \int_{\phi_{\rm th}}^{\infty}n(x)m(\phi,x)dx \label{eq:p_cnc_th}.
\end{eqnarray}
\begin{eqnarray}
\displaystyle \left. \frac{\partial N_{\rm tot}^{\rm APP}}{\partial t} \right|_{\rm emrg} 
&=& 2 \int_{\phi_{\rm th}}^{\infty} \left[ n(\phi) \int_0^{\infty}n(\phi+x)m(\phi,\phi+x)dx \right] d\phi \label{eq:N_emrg_th} \\
\displaystyle \left. \frac{\partial N_{\rm tot}^{\rm APP}}{\partial t} \right|_{\rm mrg} 
&=& \int_{\phi_{\rm th}}^{\infty} \left[ n(\phi) \int_{\phi_{\rm th}}^{\infty}n(x)l(\phi,x)dx \right] d\phi \label{eq:N_mrg_th}\\
\displaystyle \left. \frac{\partial N_{\rm tot}^{\rm APP}}{\partial t} \right|_{\rm splt} 
&=& \int_{\phi_{\rm th}}^{\infty} \left[ n(\phi) \int_{\phi_{\rm th}}^{\phi-\phi_{\rm th}}k(x,\phi-x)dx \right] d\phi \label{eq:N_splt_th} \\
\displaystyle \left. \frac{\partial N_{\rm tot}^{\rm APP}}{\partial t} \right|_{\rm cnc} 
&=& 2 \int_{\phi_{\rm th}}^{\infty} \left[ n(\phi) \int_{\phi_{\rm th}}^{\infty}n(x)m(\phi,x)dx \right] d\phi \label{eq:N_cnc_th}.
\end{eqnarray}
\begin{eqnarray}
\displaystyle \left. \frac{\partial \Phi_{\rm tot}^{\rm APP}}{\partial t} \right|_{\rm emrg} &=& 
2 \int_{\phi_{\rm th}}^{\infty} \left[ \phi n(\phi) \int_0^{\infty}n(\phi+x)m(\phi,\phi+x)dx \right] d\phi \label{eq:phi_emrg_th}
\end{eqnarray}
\begin{eqnarray}
\displaystyle \left. \frac{\partial \Phi_{\rm tot}^{\rm APP}}{\partial t} \right|_{\rm mrg} &=& 
\int_{\phi_{\rm th}}^{\infty} \left[ \phi n(\phi) \int_{\phi_{\rm th}}^{\infty}n(x)l(\phi,x)dx \right] d\phi \label{eq:phi_mrg_th} \\
\displaystyle \left. \frac{\partial \Phi_{\rm tot}^{\rm APP}}{\partial t} \right|_{\rm splt} &=&
\int_{\phi_{\rm th}}^{\infty} \left[ \phi n(\phi) \int_{\phi_{\rm th}}^{\phi-\phi_{\rm th}}k(x,\phi-x)dx \right] d\phi \label{eq:phi_splt_th} \\
\displaystyle \left. \frac{\partial \Phi_{\rm tot}^{\rm APP}}{\partial t} \right|_{\rm cnc} &=&
2 \int_{\phi_{\rm th}}^{\infty} \left[ \phi n(\phi) \int_{\phi_{\rm th}}^{\infty}n(x)m(\phi,x)dx \right] d\phi \label{eq:phi_cnc_th}.
\end{eqnarray}
\begin{eqnarray}
\displaystyle \tau_{\rm emrg}^{\rm APP} &=& \frac{\displaystyle \int_{\phi_{\rm th}}^{\infty}\phi \, n(\phi) d\phi}
{\displaystyle 2 \int_{\phi_{\rm th}}^{\infty} \left[ \phi n(\phi) \int_0^{\infty}n(\phi+x)m(\phi,\phi+x)dx \right] d\phi} \label{eq:tau_emrg_th} \\
\displaystyle \tau_{\rm mrg}^{\rm APP} &=&  \frac{\displaystyle \int_{\phi_{\rm th}}^{\infty}\phi \, n(\phi) d\phi}
{\displaystyle \int_{\phi_{\rm th}}^{\infty} \left[ \phi n(\phi) \int_{\phi_{\rm th}}^{\infty}n(x)l(\phi,x)dx \right] d\phi} \label{eq:tau_mrg_th} \\
\displaystyle \tau_{\rm splt}^{\rm APP} &=& \frac{\displaystyle \int_{\phi_{\rm th}}^{\infty}\phi \, n(\phi) d\phi}
{\displaystyle \int_{\phi_{\rm th}}^{\infty} \left[ \phi n(\phi) \int_{\phi_{\rm th}}^{\phi-\phi_{\rm th}}k(x,\phi-x)dx \right] d\phi} \label{eq:tau_splt_th} \\
\displaystyle \tau_{\rm cnc}^{\rm APP} &=& \frac{\displaystyle \int_{\phi_{\rm th}}^{\infty}\phi \, n(\phi) d\phi}
{\displaystyle 2 \int_{\phi_{\rm th}}^{\infty} \left[ \phi n(\phi) \int_{\phi_{\rm th}}^{\infty}n(x)m(\phi,x)dx \right] d\phi} \label{eq:tau_cnc_th}.
\end{eqnarray}

\begin{figure}[p]
\centering
\includegraphics[bb=0 0 500 800,width=0.8\textwidth]{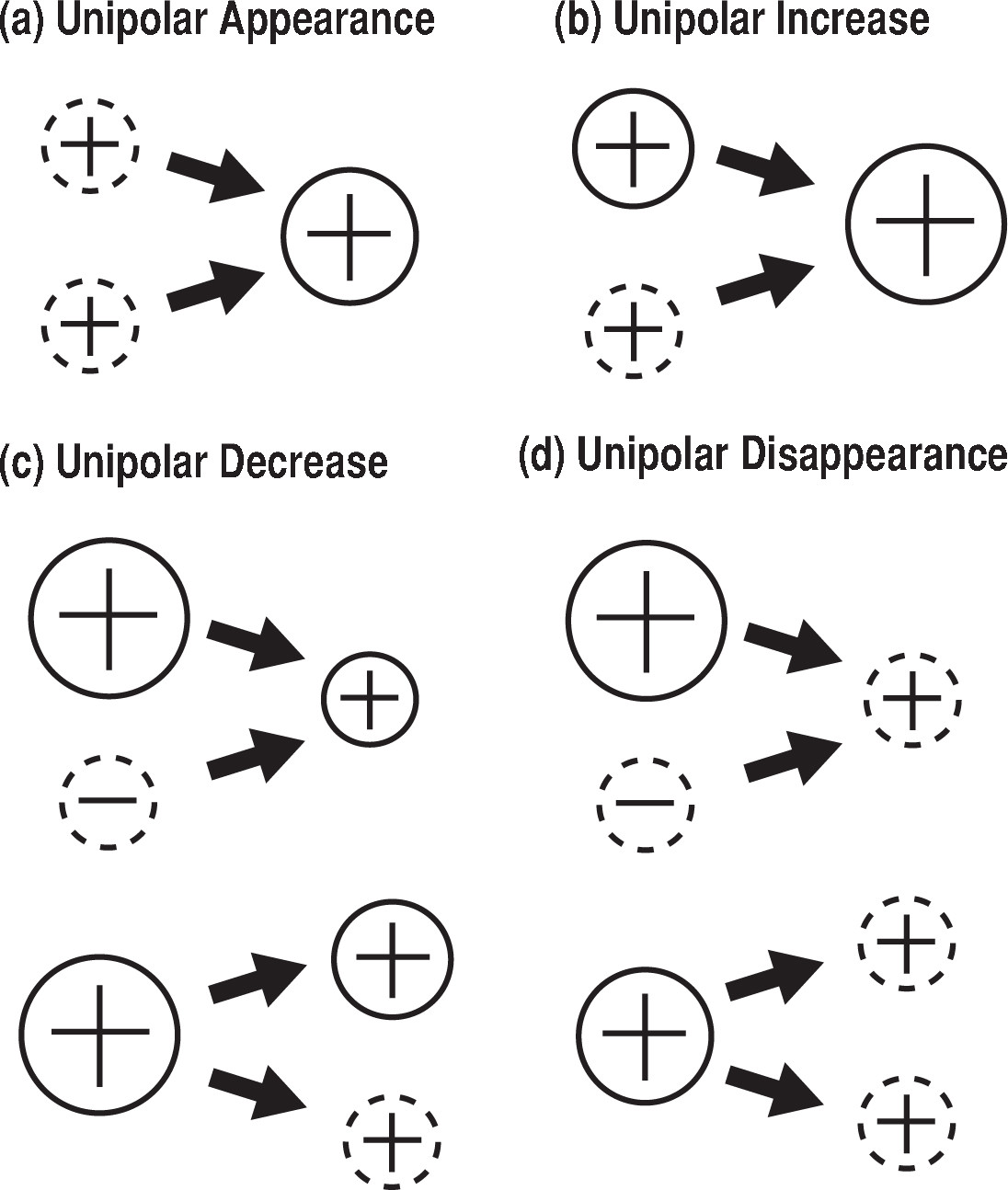}
\caption[Schematic pictures of unipolar flux change events.]{Schematic pictures of unipolar flux change events, 
namely (a) Unipolar appearance, (b) Unipolar increase, (c) Unipolar decrease, and (d) Unipolar disappearance.
}
\label{fig:pic_mono}
\end{figure}

Next we evaluate total occurrence rate of apparent unipolar processes and total change rates of flux amount by them.
There are four unipolar flux change events, namely unipolar appearance, unipolar increase, unipolar decrease, and unipolar disappearance.
Figure \ref{fig:pic_mono} summarizes schematic pictures of them.

Unipolar appearance corresponds to merging where flux contents of both merging patches are below $\phi_{\rm th}$ (Figure \ref{fig:pic_mono}a). 
Flux contents of both patches are below $\phi_{\rm th}$. 
Sum of flux content is larger than $\phi_{\rm th}$.
Noting these conditions, total occurrence rate and total change rate of flux amount are deduced to 
\begin{equation}
\left. \frac{\partial N_{\rm tot}}{\partial t} \right|_{\rm app}^{\rm APP} 
= \int_{\phi_{\rm th}}^{2 \phi_{\rm th}} \left[ \int_{\phi-\phi_{\rm th}}^{\phi_{\rm th}}n(x)n(\phi-x)l(x,\phi-x)d\phi \right] d\phi
\end{equation}
\begin{equation}
\left. \frac{\partial \Phi_{\rm tot}}{\partial t} \right|_{\rm app}^{\rm APP} 
= \int_{\phi_{\rm th}}^{2 \phi_{\rm th}} \phi \left[ \int_{\phi-\phi_{\rm th}}^{\phi_{\rm th}}n(x)n(\phi-x)l(x,\phi-x)d\phi \right] d\phi.
\end{equation}

Unipolar increase corresponds to merging where flux content of one merging patch is above $\phi_{\rm th}$ 
but that of the other is below $\phi_{\rm th}$ (Figure \ref{fig:pic_mono}b).
Total occurrence rate and total change rate of flux amount are deduced to in the same manner as
\begin{eqnarray}
\left. \frac{\partial N_{\rm tot}}{\partial t} \right|_{\rm inc}^{\rm APP} 
&=& \int_{\phi_{\rm th}}^{2 \phi_{\rm th}} \left[ \int_0^{\phi-\phi_{\rm th}}n(x)n(\phi-x)l(x,\phi-x)dx \right] d\phi \notag\\
&\ & + \int_{2 \phi_{\rm th}}^{\infty} \left[ \int_0^{\phi_{\rm th}}n(x)n(\phi-x)l(x,\phi-x)dx \right] d\phi\\
\left. \frac{\partial \Phi_{\rm tot}}{\partial t} \right|_{\rm inc}^{\rm APP} 
&=& \int_{\phi_{\rm th}}^{2 \phi_{\rm th}} \phi \left[ \int_0^{\phi-\phi_{\rm th}}n(x)n(\phi-x)l(x,\phi-x)dx \right] d\phi \notag\\
&\ & + \int_{2 \phi_{\rm th}}^{\infty} \phi \left[ \int_0^{\phi_{\rm th}}n(x)n(\phi-x)l(x,\phi-x)dx \right] d\phi.
\end{eqnarray}

Unipolar decrease corresponds to cancellation with a patch whose flux content is below $\phi_{\rm th}$ or splitting where
flux content of one produced patch is below $\phi_{\rm th}$ (Figure \ref{fig:pic_mono}c).
Total occurrence rate and total change rate of flux amount are deduced to in the same manner as
\begin{eqnarray}
\left. \frac{\partial N_{\rm tot}}{\partial t} \right|_{\rm dec}^{\rm APP} 
&=& \int_{\phi_{\rm th}}^{\infty} \left[ \int_0^{\phi_{\rm th}}n(x)n(\phi+x)m(x,\phi+x)dx \right] d\phi \notag\\
&\ & + 2\int_{\phi_{\rm th}}^{2 \phi_{\rm th}} n(\phi) \left[ \int_0^{\phi-\phi_{\rm th}} k(x,\phi-x) dx \right] d\phi \notag\\
&\ & + 2\int_{2 \phi_{\rm th}}^{\infty} n(\phi) \left[ \int_0^{\phi_{\rm th}} k(x,\phi-x) dx \right] d\phi \\
\left. \frac{\partial \Phi_{\rm tot}}{\partial t} \right|_{\rm dec}^{\rm APP} 
&=& \int_{\phi_{\rm th}}^{\infty} \phi \left[ \int_0^{\phi_{\rm th}}n(x)n(\phi+x)m(x,\phi+x)dx \right] d\phi \notag\\
&\ & + 2\int_{\phi_{\rm th}}^{2 \phi_{\rm th}} \phi n(\phi) \left[ \int_0^{\phi-\phi_{\rm th}} k(x,\phi-x) dx \right] d\phi \notag\\
&\ & + 2\int_{2 \phi_{\rm th}}^{\infty} \phi n(\phi) \left[ \int_0^{\phi_{\rm th}} k(x,\phi-x) dx \right] d\phi.
\end{eqnarray}

Unipolar disappearance corresponds to cancellation with a patch whose flux content is below $\phi_{\rm th}$ or splitting where
flux content of one produced patch is below $\phi_{\rm th}$ (Figure \ref{fig:pic_mono}d).
Total occurrence rate and total change rate of flux amount are deduced to in the same manner as
\begin{eqnarray}
\left. \frac{\partial N_{\rm tot}}{\partial t} \right|_{\rm mdisapp}^{\rm APP} 
&=& \int_{\phi_{\rm th}}^{2\phi_{\rm th}} \left[ \int_{\phi-\phi_{\rm th}}^{\phi_{\rm th}}n(x)n(\phi+x)m(x,\phi+x)dx \right] d\phi \notag\\
&\ & + \int_{\phi_{\rm th}}^{2 \phi_{\rm th}} n(\phi) \left[ \int_{\phi-\phi_{\rm th}}^{\phi_{\rm th}} k(x,\phi-x) dx \right] d\phi \\
\left. \frac{\partial \Phi_{\rm tot}}{\partial t} \right|_{\rm mdisapp}^{\rm APP} 
&=& \int_{\phi_{\rm th}}^{2\phi_{\rm th}} \phi \left[ \int_{\phi-\phi_{\rm th}}^{\phi_{\rm th}}n(x)n(\phi+x)m(x,\phi+x)dx \right] d\phi \notag\\
&\ & + \int_{\phi_{\rm th}}^{2 \phi_{\rm th}} \phi n(\phi) \left[ \int_{\phi-\phi_{\rm th}}^{\phi_{\rm th}} k(x,\phi-x) dx \right] d\phi.
\end{eqnarray}

We evaluate lifetime of patches. 
Patch disappears in the analysis range when it cancels with a larger patch, merges with a larger patch 
or disappears by unipolar disappearance.
Then disappearance probability, $\partial P_{\rm disapp}^{\rm APP} / \partial t (\phi)$, is deduced as 
\begin{eqnarray}
\displaystyle &\ &\frac{\partial P_{\rm disapp}^{\rm APP}}{\partial t}(\phi) \\
&=& \frac{1}{2}\frac{\partial P_{\rm tcnc}^{\rm APP}}{\partial t} + \frac{\partial P_{\rm tmrg}^{\rm APP}}{\partial t} 
+ \frac{\partial P_{\rm mdisapp}^{\rm APP}}{\partial t} \notag \\
&=&
\left\{ \begin{array}{ll}
& \displaystyle \int_{\phi}^{\infty} n(x)m(\phi,x)dx + \int_{\phi}^{\infty} n(x)l(\phi,x)dx + 
\int_{\phi-\phi_{\rm th}}^{\phi_{\rm th}}k(x,\phi-x)dx   \notag \\
& \ \ \ \ \ \
\displaystyle + \frac{1}{n(\phi)} \int_{\phi-\phi_{\rm th}}^{\phi_{\rm th}}n(x)n(\phi+x)m(x,\phi+x)dx \ \ \ \ (\phi_{\rm th} \le \phi < 2\phi_{\rm th})\\
& \displaystyle \int_{\phi}^{\infty} n(x)m(\phi,x)dx + \int_{\phi}^{\infty} n(x)l(\phi,x)dx \ \ \ \ \ \ \ \ \ \ \ \ (2 \phi_{\rm th} \le \phi).
\end{array} \right.
\end{eqnarray}

The lifetime is an inverse of disappearance probability, namely
\begin{eqnarray}
\displaystyle &\ &\tau^{\rm APP}(\phi) \label{eq:life_app} \\
&=& \left[ \frac{\partial P_{\rm disapp}^{\rm APP}}{\partial t}(\phi) \right]^{-1} \notag
\end{eqnarray}
\begin{equation}
=
\left\{ \begin{array}{ll}
\displaystyle \ & \displaystyle \left[ \int_{\phi}^{\infty} n(x)m(\phi,x)dx + \int_{\phi}^{\infty} n(x)l(\phi,x)dx
+ \int_{\phi-\phi_{\rm th}}^{\phi_{\rm th}}k(x,\phi-x)dx \right.  \notag \\
\displaystyle &\ \ \ \ \ \
\displaystyle  + \left. \frac{1}{n(\phi)} \int_{\phi-\phi_{\rm th}}^{\phi_{\rm th}}n(x)n(\phi+x)m(x,\phi+x)dx \right]^{-1} 
\ \ \ \ (\phi_{\rm th} \le \phi < 2\phi_{\rm th})\\
\displaystyle\ & \displaystyle \left[ \int_{\phi}^{\infty} n(x)m(\phi,x)dx + \int_{\phi}^{\infty} n(x)l(\phi,x)dx \right]^{-1} 
 \ \ \ \ \ \ \ \ \ \ \ \ (2 \phi_{\rm th} \le \phi).
\end{array} \right.
\end{equation}
There are two domains in the apparent lifetime.
One is the range below $2\phi_{\rm th}$, where there is a unipolar disappearance by merging and cancellation and the apparent lifetime is different
 from the actual lifetime.
Another is the range above $2\phi_{\rm th}$, where apparent lifetime is same as the actual lifetime.
\vspace{2.5cm}

\section{Discussions of Magnetic Patch Characters}
\label{sec:discuss_patch}

\subsection{Flux Injection and Convective Effect}

Data set 1 and data set 2 have a different character in the variability of total flux amount.
The total flux amount in data set 1 is almost constant during the observational period.
On the contrary, that in data set 2 changes by $\thicksim$ 50$\%$. 
This difference is caused by large flux emergences, which may be injected from below the photosphere.
Despite this difference, we find frequency distributions of flux content are the same power-law forms, 
which are also consistent with previous study by \cite{par2009}.
It means that the injected flux is rapidly fragmented, namely surface processes maintain a frequency distribution of flux content.

\vspace*{3.5cm}

\subsection{Lifetime of the Patches}

We obtained the averaged lifetime of patches as $17.3$ minutes in data set 1 and $23.2$ minutes in data set 2.
It is also found that lifetime has a power-law dependence as a number distribution on flux content with 
an index of -1.45 and -1.53, respectively (see Figure \ref{fig:dist_lt_d1} and Figure \ref{fig:dist_lt_d2}).
The power-law distribution continues from $\thicksim 70$ minutes down to $\thicksim 5$ minutes, 
which is near the time resolution of the data set, i.e. $3$ minutes.
The power-law distribution indicates that the averaged lifetime is not a typical one of the patches and that the 
value itself is not important. 
Further, we found that the corresponding power-law index of frequency distribution is 
$\thicksim -2.45$ and $-2.53$, respectively.
The power-law index smaller than $-2$ means that patches with shorter time scale determine the obtained averaged lifetime.
So the mechanism that determines the apparent lifetime is the process that is the most frequent one for 
their disappearance. From the results of Section 4.1, it is probably the unipolar decrease. See also Appendix B for the investigation
of the flux amount of it.

\section[Discussions of Surface Processes]{Discussions of  Surface Processes of Magnetic Patches}
\label{sec:discuss_obs}

We did a statistical investigation of frequencies and change rates of flux amount by surface processes through an auto-detection code.
The obtained results are as follows:\\
1) Merging and splitting are much frequent than cancellation and emergence.\\
2) Probability distribution of merging has only weak dependence on flux content, namely $0.28$ as a power-law index, and 
that of splitting has a similar tendency but rapid decrease around $\phi_{\rm th}$.\\
3) Frequency distribution of cancellation has a strong dependence on flux content, namely $-2.70$ as a power-law index.\\
We speculate physical picture of flux maintenance in quiet regions from these results.

\subsection{Comparison with Previous Studies of Bright Points}

Previous studies indicate that bright points in the photospheric and chromospheric lines 
(e.g. G-band, \ion{Ca}{II} H, and \ion{H}{$\alpha$} etc.) 
correspond to magnetic concentrations on the solar surface \citep{mul1975, mul1983, mul1984, berg1996, ish2007, abr2010b}.
The time scales of merging and splitting of bright points are reported \cite{berg1996, berg1998, berg1998b}.
They use high-resolution ground-based observation data taken by the Swedish Vacuum Solar Telescope on La Palma and investigate 534 bright points
by an auto-tracking method.
They report 320 and 404 seconds as the mean occurrence times for merging and splitting of bright points \citep{berg1998b}, which are significantly shorter
than our results, namely 2000 seconds.
We think that the difference of bright points may come from the fact that the bright point has a spatial scale of 200km \citep{utz2009}, 
which are shorter than our size threshold.
The change of bright points occurs from the granular motion because the timescale of merging and splitting are the same order of 
that of granulation \citep{berg1998b}.
On the contrary, our investigation may reflect the plasma motion larger than the granular motion
because we set the size threshold of patches as $1.0 \times 10^6$ km$^2$, which is the typical size of granulation.
The processes of bright points are controlled predominantly by the granular scale motion since the obtained time scales are similar 
(Berger et al. 1998a). 
On the other hand, those of our magnetic patches are controlled by the larger scale motion due to our setup in the size threshold.

\subsection{Probability Density Distributions of Merging and Splitting in M-C equation}
\label{sec:discuss_form}

We evaluate the probability density distributions of merging and splitting in 
a M-C equation from the observational results in this section.
Because the probability distributions, which we obtained in this study, are obtained by integrating them in the flux content once, 
we have to put at least one assumption to evaluate them.

The merging function $l(x,y)$ is given as follows. 
We obtain the form as a probability distribution.
By comparing Equation (\ref{eq:p_mrg_th}) and the result of Section \ref{sec:mrg}, we obtain  
\begin{equation}
\displaystyle \int_{\phi_{\rm th}}^{\infty}n(x)l(\phi,x)dx=p_{0,{\rm mrg}}\left( \frac{\phi}{\phi_0} \right)^{\beta_{\rm mrg}} 
\ \ \ (\phi \ge \phi_{\rm th}). \label{eq:rel_mrg}
\end{equation}
In the left-handside of this equation, the variable $\phi$ appears only in the $l(\phi,y)$.
We assume a simple form satisfying this relation namely,
\begin{equation}
l(x,y) \propto x^{\beta_{\rm mrg}}
\end{equation}
From the symmetry of $l(x,y)=l(y,x)$, this relation deduces 
\begin{equation}
l(x,y) = l_0 \left( \frac{x}{\phi_0} \right)^{\beta_{\rm mrg}} \left( \frac{y}{\phi_0} \right)^{\beta_{\rm mrg}}.
\end{equation}
Substituting it into Equation (\ref{eq:rel_mrg}) and Equation (\ref{eq:dist}) for $n(\phi)$, we obtain
\begin{equation}
\displaystyle l_0=\frac{\displaystyle p_{0,{\rm mrg}}}{\displaystyle n_0 \phi_0^{\gamma-\beta_{\rm mrg}} \int_{\phi_{\rm th}}^{\infty}x^{-\gamma+\beta_{\rm mrg}}dx}.
\end{equation}
The upper value of integration is limited on the actual Sun and we put it as $\phi_{\rm max}$.
We substitute the value obtained in the thesis
, namely $n_0=1.21\times10^{-36}$ Mx$^{-1}$ cm$^{-2}$, $\gamma=1.78$, $p_{0,{\rm mrg}}=2.56\times10^{-4}$ s$^{-1}$, 
$\beta_{\rm mrg}=0.28$, $\phi_{\rm max}=10^{19}$ Mx, $\phi_{\rm th}=10^{17.5}$ Mx, and $\phi_0=10^{18}$ Mx, and
 obtain $l_0=7.24 \times 10^{13}$ cm$^2$ s$^{-1}$.

Next we evaluate the probability density function of splitting.
The apparent frequency distribution of splitting has more complicated form than that of merging because splitting into patches below the 
detection threshold changes actual frequency distribution more than merging. 
It is difficult to evaluate the actual frequency distribution from the apparent one.
The strategy we choose here is that we assume the form of splitting function i.e., $k(x,y)$, and evaluate the apparent frequency distribution from that.
Then we can see the difference between the evaluation and the observational result. 
From our observations, the probability distribution of splitting events $\partial P^{\rm APP}_{\rm splt}(\phi) / \partial t$ is suggested to be independent
of the parent patch flux:
\begin{equation}
\frac{\partial P^{\rm APP}_{\rm splt}}{\partial t}(\phi)=k_0=\rm{constant} \ (\rm{for \ all} \ \it{\phi}).
\end{equation}
This claim is observationally supported at least in the range $\phi > \phi_{\rm{th}}$ (see Figure \ref{fig:distsplt_d1}).
The drop off below $\phi_{\rm{th}}$ is discussed immediately below.
If the splitting ratio between the daughter patches is randomly determined i.e.,
\begin{equation}
\frac{\partial}{\partial x} \left[ k(x,\phi-x) \right] = 0 \ (\rm{for} \ 0 < \it{x} < \phi),
\end{equation}
then we obtain,
\begin{equation}
k(x,y)=\frac{k_0}{x+y}.
\end{equation}
When the flux content of the daughter patch is below $\phi_{\rm{th}}$, such events are not recognized as a splitting event in our procedure. 
The probability distribution will be given as
\begin{equation}
\frac{\partial P^{\rm APP}_{\rm splt}}{\partial t}(\phi) = \int_{\phi_{\rm{th}}}^{\phi-\phi_{\rm{th}}}k(x,\phi-x)\, dx = k_0 (1-\frac{2 \phi_{\rm{th}}}{\phi}).
\end{equation}
The black and blue dashed curves in Figure \ref{fig:distsplt_d1} indicate analytical curves with 
$k_0=1.0\times10^{-3}$ s$^{-1}$ and $k_0=5.0\times10^{-4}$ s$^{-1}$, respectively.
This curve fits the drop of the observational line well, which supports the above assumptions.
We obtain $k(x,y)=k_0/(x+y)$ as a splitting function, where $k_0=5.0\times10^{-4}$ s$^{-1}$.

\subsection{The time Independent Solution of Splitting Process}
\label{sec:discuss_splt}

Since our observations show that merging and splitting are much more frequent than emergence and cancellation, 
it is suggested that the former two have much influence of the maintenance on the power-law distribution.
Though we have not find the solution of M-C equation with both merging and splitting, 
we show the time-independent solution only with splitting.
Numerical calculation of M-C equation may be useful to find the solution with splitting and merging in the future works.

The frequency of emergence, merging, splitting, and cancellation are represented by $S(\phi)$, $l(x,y)$, $k(x,y)$, and $m(x,y)$ respectively.
We use the equation only including the splitting terms by setting $S(\phi)=0$, $l(x,y)=0$, $m(x,y)=0$, namely 
\begin{equation}
\frac{\partial n(\phi)}{\partial t} = 2\int_{\phi}^{\infty}n(x)k(\phi,x-\phi)\, dx - \int_{0}^{\phi}n(\phi)k(x,\phi-x)\, dx. \label{eq:mc_splt}
\end{equation}
Here the splitting process, $k(x,y)$ is only taken into account. 
We obtained the function form of $k(x,y)$ in Section \ref{sec:discuss_form} and obtained $k(x,y)=k_0/(x+y)$.
After substituting it into Equation (\ref{eq:mc_splt}) and differentiating with $\phi$, we obtain
\begin{equation}
\frac{\partial^2n(\phi)}{\partial \phi \partial t} = - \frac{k_0}{\phi^2}\frac{\partial}{\partial \phi}\left[\phi^2n(\phi)\right].
\end{equation}
This equation has a time-independent solution $n(\phi) \propto \phi^{-2}$.
This power-law index of the flux content is in good agreement with the observational result, $\gamma=-1.8$.
It is possible that n($\phi$) has a solution with a power-law function by using both the splitting and merging processes 
in the above analysis though it is not simple task due to its mathermatical difficulty.
One further investigation is to investigate time evolution of the frequency distribution 
of the flux content during decay of active regions and frequencies of merging and splitting. 
If the merging and splitting make the power-law distribution of the flux content, we will observe the change of the power-law index to $-1.8$.

This scale-free distribution comes from constant dependencies of splitting, namely constancies on parent flux content 
and daughter flux content.
These constancies may come from splitting dominated by convection or flux tube instability.
In order to distinct these models, we propose two future observations: 
One is an investigation of convection flow field by using {\it Hinode}/FG data. 
The other is that of fine-scale structure in the flux tube and its dynamics by using {\it Hinode}/SP.

\subsection{Interpretation of Frequency Distribution of Cancellation and Emergence}
\label{sec:dsc_cncemg}

\begin{figure}[p]
\centering
\includegraphics[bb=0 0 1000 400,width=21cm,clip,angle=90]{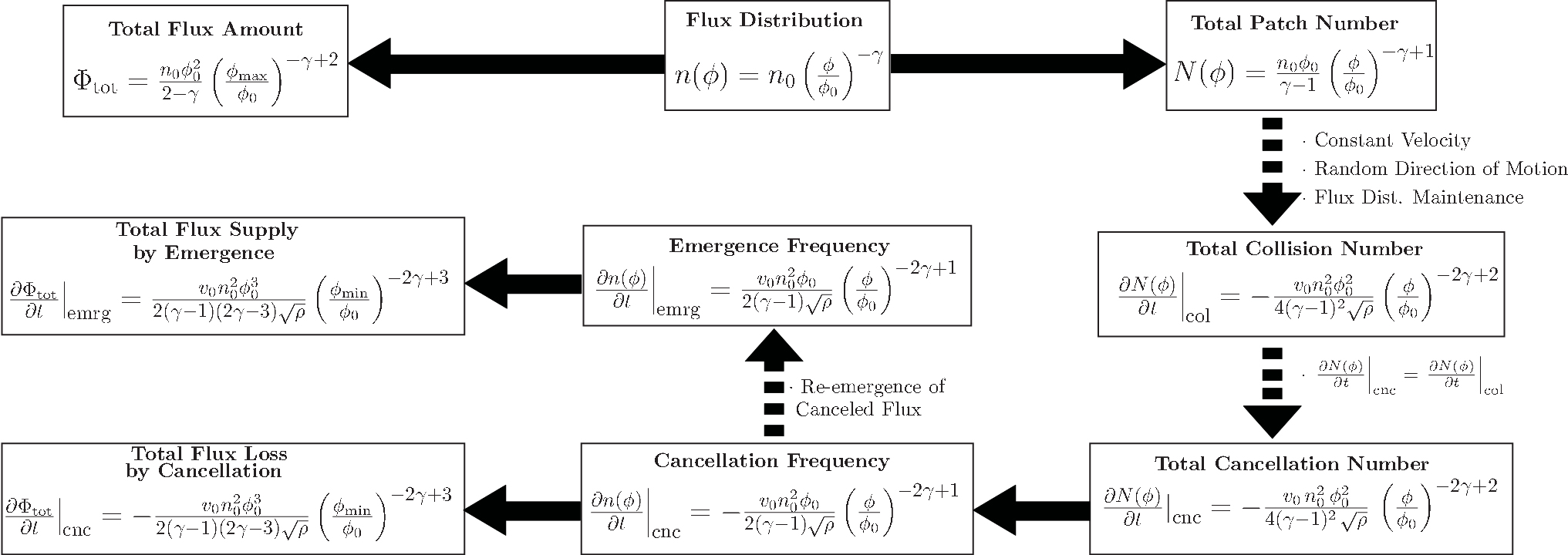}
\caption[Relationship model among the frequency distribution of the flux content, cancellation, and emergence.]{
Relationship model among the frequency distributions of the flux content, cancellation, and emergence with $1.5<\gamma<2$.
The solid arrows indicate mathematical relations, while the dashed arrows indicate relationships with some assumptions.
}
\label{fig:model_cnc}
\end{figure}

We discuss why the slope of frequency distribution of cancellation is so steep compared to those of merging and splitting 
in this section.
We suggest a relationship model among the frequency distributions of the flux content, cancellation, and emergence.
The constant velocity and randomness of direction of motion are assumed here.
Figure \ref{fig:model_cnc} shows a schematic view of this model.
We put a power-law distribution of flux content as 
\begin{equation}
n(\phi)=n_0 \left( \frac{\phi}{\phi_0} \right)^{-\gamma}. \label{eq:model_dist}
\end{equation}
The power-law index, $\gamma$, is derived as $1.5<\gamma<2$ by our observation and the previous studies \citep{par2009,zha2010}.
The maximum flux content in the system ($\phi_{\rm{max}}$) is assumed to be much larger than the minimum ($\phi_{\rm{min}}$) in the following 
discussion.
We calculate $\hat{N}(\phi)$, a total patch number with a flux content larger than $\phi$, by integrating from $\phi$ to $\phi_{\rm{max}}$ as
\begin{equation}
\hat{N}(\phi)=\int_{\phi}^{\phi_{\rm{max}}}n(\phi) \, d\phi \approx \frac{n_0 \phi_0}{\gamma-1} \left( \frac{\phi}{\phi_0} \right)^{-\gamma+1}.
\end{equation}
\cite{sch1997b} evaluated a collision rate of opposite patches from a total patch number with assumptions of a constant velocity 
and a randomness of patch motion along a network.
They obtained the collision frequency, $\nu$, as
\begin{equation}
\nu = \frac{v_0}{4 \sqrt{\rho}}N_t^2
\end{equation}
where $v_0$, $\rho$, and $N_t$ mean a typical velocity of patches, a number density of network cell, and a number density of patches, 
respectively. We multiply $1/2$ taking the double counting into account.
We obtained that the frequency of merging and splitting are much larger than that of cancellation.
It can be deduced that the frequency distribution of flux content is maintained rapidly by merging and splitting compared to the time scale of cancellation.
This enables us to treat that the number density of patches is time independent in the evaluation of cancellation and apply the same analogy 
to the number density expanded in the dimension of the flux content.
The collision frequency of opposite polarities, $\left. \partial \hat{N}(\phi)/\partial t \right|_{\rm{col}}$, is evaluated as
\begin{equation}
\left. \frac{\partial \hat{N}(\phi)}{\partial t} \right|_{\rm col}= - \frac{v_0 n_0^2 \phi_0^2}{4(\gamma -1)^2 \sqrt{\rho}} 
\left( \frac{\phi}{\phi_0} \right)^{-2\gamma+2}.
\end{equation}
Note that this total collision number becomes time-independent with an assumption of maintenance of a power-law distribution of flux content.
We assume that the total number of cancellation, 
$\partial \hat{N}_{\rm cnc}(\phi)/\partial t$, equals the total number of collision events of the opposite polarities, 
$\partial \hat{N}_{\rm col}(\phi)/\partial t$, namely,
\begin{equation}
\frac{\partial \hat{N}_{\rm{cnc}}(\phi)}{\partial t} =  \frac{\partial \hat{N}_{\rm{col}}(\phi)}{\partial t}.
\end{equation}
The frequency distribution of cancellation, $\partial n_{\rm cnc}(\phi)/\partial t$ is given by differentiating equation  with $\phi$, namely
\begin{equation}
\frac{\partial n_{\rm cnc}(\phi)}{\partial t}= 
\frac{\partial ^2 \hat{N}_{\rm cnc} (\phi)}{\partial \phi \partial t} = 
- \frac{v_0 n_0^2 \phi_0}{2(\gamma-1) \sqrt{\rho}} \left( \frac{\phi}{\phi_0} \right)^{-2\gamma+1} . \label{eq:model_distcnc}
\end{equation}
This assumption means that there are no scripting patches in the collision of opposite polarities once they collide, which is difficult 
to check and we justify this assumption from the comparison of the 
obtained frequency distribution of cancellation in this model and that in the observation. 
Based on the above discussion our estimation predicts a frequency distribution of cancellation as 
$\partial n_{\rm cnc} / \partial t \sim - 9.4 \times 10^{-42} \times \left( \phi / 1.0 \times 10^{18} \, \rm{Mx} \right)^{-2.86\pm0.14}$ 
Mx$^{-1}$ cm$^{-2}$ s$^{-1}$ where we adopt $\rho = 1.0 \times 10^{-19} \ \rm{cm}^{-2}$ \citep{hag1997}, $v_0 = 1.2 \ \rm{km \  s^{-1}}$ 
(Section 3.4), $n_0=6.8\times10^{-37}$ Mx$^{-1}$ cm$^{-2}$ (Section 3.3), and $\gamma=1.93$ from data set 2.
The observed frequency distribution of cancellation for data set 2 is $\partial n_{\rm cnc}/\partial t = - 1.29 \times 10^{-41} \times
\left( \phi / 1.0 \times 10^{18} \rm{Mx} \right)^{-2.48\pm0.24}$.
The absolute values of the distributions are in consistent within a factor of 2 and 
the power-law index is in consistent within the $1\sigma$-error.
The difference of absolute value may be explained by the fact that the patches have the finite size and it increase the rate of the collisions.

We also evaluate a frequency distribution of emergence with assumptions of small amount of flux supply from the outside of the system
 and re-emergences of canceled fluxes.
These assumptions lead to the relationship in which a flux distribution of emergence nearly equates that of cancellation,
\begin{equation}
\frac{\partial n_{\rm emrg}(\phi)}{\partial t}  \approx
- \frac{\partial n_{\rm cnc}(\phi)}{\partial t} =
\frac{v_0 n_0^2 \phi_0}{2(\gamma-1)\sqrt{\rho}} \left( \frac{\phi}{\phi_0} \right)^{-2\gamma+1}. \label{eq:model_distemrg}
\end{equation}
\cite{tho2011} reports a power-law frequency of emergence as 
$\partial n_{\rm emrg} / \partial t \sim 1.5 \times 10^{-40} \times \left( \phi / 
1.0 \times 10^{18} \, \rm{Mx} \right)^{-2.7}$ Mx$^{-1}$ cm$^{-2}$ s$^{-1}$.
The steepness of the estimated power-law distribution, -2.48$\pm0.24$, is in good agreement with the observational results, -2.7.
It supports the idea of recycling of flux content in a quiet region. 
The discrepancy in absolute value between our estimation and \cite{tho2011} may be due to the difference in patch 
detection methods. 
It is reported \citep{def2007} that there is a two-orders-of-magnitude enhancement in the number of detected patches by the clumping method 
(ours) over the downhill one \citep{tho2011}.
To investigate this difference, we will investigate the frequency distributions of emergence and cancellation in the same data set with 
the same detection method as a future work.
As for the direct investigation, the flux recycling is difficult to be determined because we cannot observe the magnetic 
structure below the solar surface.
One possible investigation is to make pairs of emergence and cancellation by the correlation time and the distance between them.

Further we calculate the time scales of cancellation and emergence to see their dependences on the detection limit, 
which is reported from the observation as that they become much shorter with the higher resolution.
The total flux amount in the system is calculated from Equation (\ref{eq:model_dist}) as
\begin{equation}
\Phi_{\rm tot} = \int_{\phi_{\rm min}}^{\phi_{\rm max}} \phi \, n(\phi) \, d\phi \approx \frac{n_0 \phi_0^2}{2-\gamma} 
\left( \frac{\phi_{\rm max }}{\phi_0} \right)^{-\gamma+2}
\end{equation}
Since $\gamma < 2$, this result shows that the total flux is dominated by patches with larger flux content.
On the other hand,  we obtain a total flux loss amount by cancellation, $\left. \partial \Phi_{\rm tot }(\phi)/\partial t \right|_{\rm cnc}$, 
from Equation (\ref{eq:model_distcnc}) as 
\begin{equation}
\left. \frac{\partial \Phi_{\rm tot }}{\partial t} \right|_{\rm{cnc}} = 
\int_{\phi_{\rm{min}}}^{\phi_{\rm{max}}}\phi \left. \frac{\partial n(\phi)}{\partial t} \right|_{\rm{cnc}} \, d\phi
\approx - \frac{v_0 n_0^2 \phi_0^3}{2 ( \gamma - 1)(2 \gamma - 3) \sqrt{\rho}} \left( \frac{\phi_{\rm{min}}}{\phi_0} \right)^{-2\gamma+3}.
\end{equation}
The total flux supply by recycled emergence is also evaluated in the same manner from Equation (\ref{eq:model_distemrg}) as 
\begin{equation}
\left. \frac{\partial \Phi_{\rm tot }}{\partial t} \right|_{\rm{emrg}}
=\int_{\phi_{\rm{min}}}^{\phi_{\rm{max}}}\phi \left. \frac{\partial n(\phi)}{\partial t} \right|_{\rm{emrg}} \, d\phi
\approx \frac{v_0 n_0^2 \phi_0^3}{2 (\gamma - 1)(2 \gamma - 3) \sqrt{\rho}} \left( \frac{\phi_{\rm{min}}}{\phi_0} \right)^{-2\gamma+3}.
\end{equation}
Then time scales of cancellation and emergence are evaluated as
\begin{eqnarray}
\tau_{\rm cnc}=\tau_{\rm emrg}
& = &\Phi_{\rm tot}/ ( \left. \partial \Phi_{\rm tot}/\partial t \right|_{\rm cnc})
=\Phi_{\rm tot}/ ( \left. \partial \Phi_{\rm tot}/\partial t \right|_{\rm emrg}) \nonumber \\
& = &\frac{2 ( \gamma - 1 ) (2 \gamma - 3 )}{( 2 - \gamma ) v_0 n_0 \phi_0 \sqrt{\rho}} \left( \frac{\phi_{\rm{max}}}{\phi_0} \right)^{-\gamma+2}
\left( \frac{\phi_{\rm{min}}}{\phi_0} \right)^{2 \gamma-3}.
\end{eqnarray}
The actual value of $\gamma$ was obtained as 1.8 $\--$ 1.9 in this study, which means that $2 \gamma -3 = 0.6$ 
$\--$ 0.8. The positive dependence on the detection limit, which is represented by $\phi_{\rm min}$ here, means that the flux replacement 
time scales becomes shorter with the smaller detection limit.
This result is qualitatively consistent with the observational fact that 
the flux replacement time scales by emergence and cancellation becomes shorter with higher resolution in the previous 
studies \citep{mar1985, sch1998, hag2001}.

\section[Flux Maintenance in Quiet Regions]{The Interpretation of Flux Maintenance in Quiet Regions}
\label{sec:phys_pict}
\begin{figure}[tp]
\centering
\includegraphics[bb=0 0 800 800,width=1.2\textwidth]{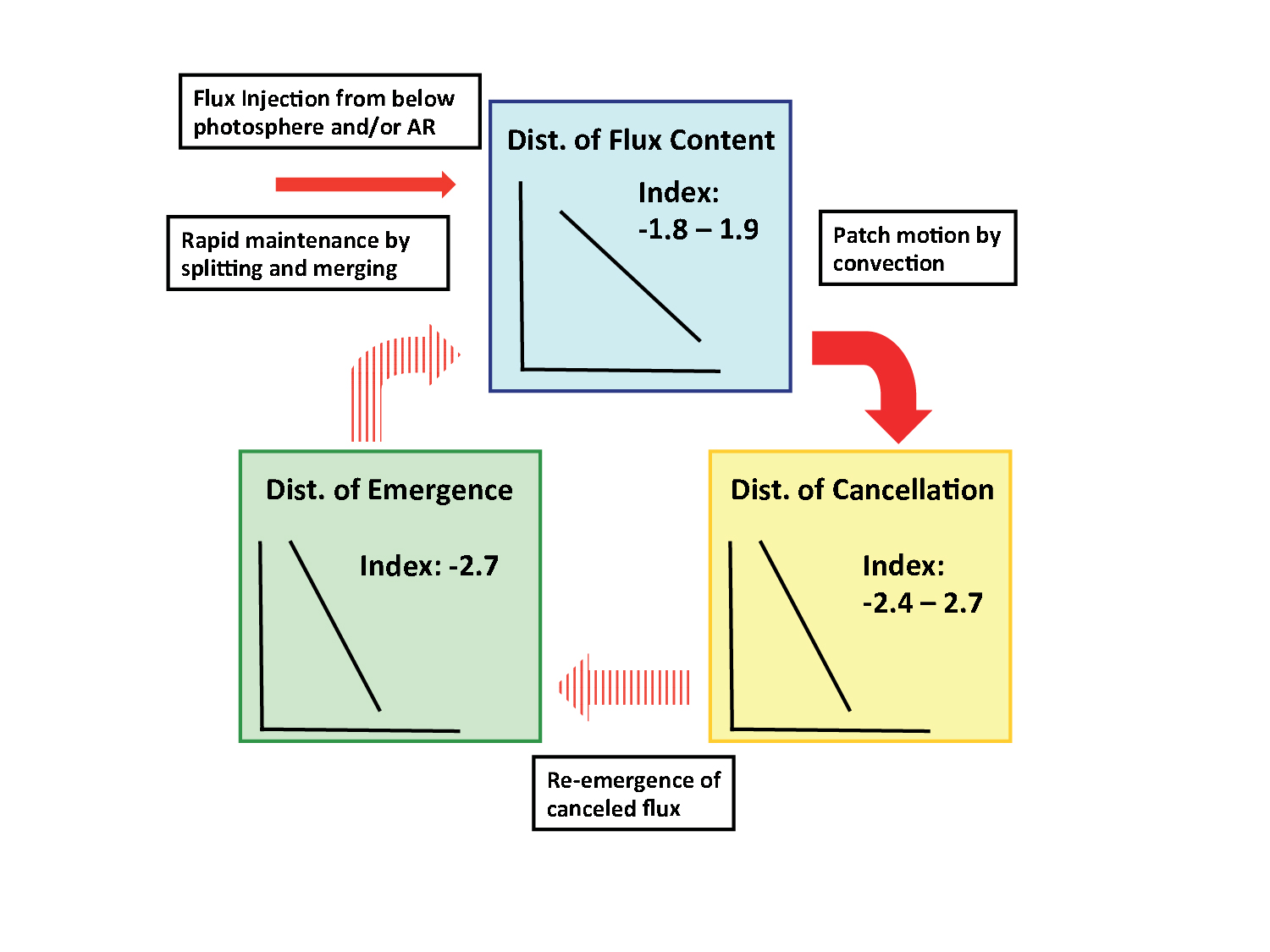}
\caption[Schematic picture of our speculation of flux maintenance in quiet regions.]{
Schematic picture of our speculation of flux maintenance in quiet regions.
}
\label{fig:model_whole}
\end{figure}

We summarize our interpretation from the discussion here.
Figure \ref{fig:model_whole} shows a schematic picture of the interpretation.
First, a power-law frequency distribution of flux content is rapidly maintained by merging and splitting, which is supported 
by the result of a comparison of change rate of flux amount by each activity.
In addition to this, we have found that a power-law frequency distribution with an index of $-2$ is a time-independent solution of splitting.
Cancellation is caused by convective motion and frequency distribution of cancellation should naturally become a steep power-law distribution. 
Emergence is interpreted as a re-emergence of submerged flux by cancellation, which is consistent with a steep frequency distribution of cancellation in 
our study and emergence reported by \cite{tho2011}.
One of the possibilities but the important point of this model is that the apparent flux transport by emergence and cancellation 
becomes much more drastic with higher resolution even when there is no transport.

\ifodd \arabic{page}
\else
  \thispagestyle{fancy}
  \mbox{}
  \newpage
  \clearpage
\fi

\chapter{Conclusion and Future Works}
\thispagestyle{fancy}

\section{Conclusion}
\pagestyle{fancy}
Our conclusion is that frequency distribution of magnetic flux in quiet regions is maintained by surface magnetic processes.
The frequency distribution of the flux content has a powe-law dependence with an idex of 
$-1.8 \- -1.9$ which continues to $10^{17.5}$ Mx at least.
The patches with this distribution move randomly and with a typical velocity, which may be caused by 
convective motion on the solar surface.
The frequency distribution of cancellation has a steep power-law distribution with an index of $-2.48\pm0.24$.
This steep distribution is explained by a random convective motion and collisions of patches.
We speculate that the patches submerge through cancellation and re-emerge to solar surface as an emergence, which is 
supported by comparing the obtained frequency of cancellation and that of emergence reported by \cite{tho2011}.
Splitting and merging are caused with an timescale of $\thicksim 33$ minutes on the solar surface.
These timescales have only a weak dependence on flux content, $-0.28$ as a power-law index for merging and 
no significant dependence for splitting.
These weak dependence may cause a power-law distribution with an index of $\thicksim -2$.

In this thesis, we develop an auto-detection code of surface processes and apply it to two different data sets of 
magnetograms obtained by {\it Hinode} spacecraft.
In Chapter 4 of our thesis, we investigate relationships between flux content and size, proper velocity, and lifetime of patches statistically.
The proper velocity is found to have only weak dependence, $-0.28$ as a power-law index.
We treat the proper velocity of patches as constant in the later discussion from this result.
It is found that small patches determine the averaged life.
It suggests a possibility that the apparent fragmentation processes to undetected scales, namely cancellation 
and splitting involving undetected patches, are much frequent than actual disappearances.
On the other hand, there is an upper limit of lifetime, $\thicksim 60$ minutes.
These results indicates the dissociative nature of patches in the analyzed flux range.

The measurements of probability distributions of surface processes are done in Chapter 5.
It is found that splitting and merging are much more frequenct than cancellation and emergence.
We further investigate frequencies of merging, splitting, and cancellation.
Probability distribution of merging has only weak dependence on flux content with an index of $-0.28$.
That of splitting is also almost constant but there is rapid decrease around the detection limit.
Frequency distribution of cancellation has a steep power-law distribution with an index of $-2.48\pm0.24$. 
The obtained results are converted to probability density distributions.
Analysis of M-C equation shows that splitting has a time-independent solution of power-law distribution with an index of $-2$.
On the other hand, the steep power-law of frequency distribution of cancellation can be explained with a constant proper velocity evaluation, 
which we interpret they are advected by convective motion.

\vspace{5.5cm}

\section{Future Works}

\subsection{Extending Studies of Magnetic Structures on the Solar Surface}

In this thesis, we estimate each term in M-C equation, i.e. $l(x,y)$, $k(x,y)$, and $m(x,y)$, in an {\it indirect} manner under 
an assumption of the symmetry in one of two dependent variables (See Section 5.3.2 for detail). 
The main reason is a lack in numbers of events to obtain enough statistics. 
In order to do a {\it direct} measurement, two orders of magnitude larger number of events are necessary. 
At this moment, such data is unavailable in the {\it Hinode} database. 
The full Sun observation data of magnetic field by SDO is what we should work out in the near future.

The discussion in this thesis is based on the M-C equation that only has the time and magnetic flux as the independent variables, 
while the actual processes on the solar surface are much more complex with many degrees of freedom. 
For example, the filling-factor of the magnetic field is an important factor that controls the time scale and the rate of 
flux re-distribution in a splitting event. If the filling-factor is smaller in larger patches, then the convective flow can 
split them more easily. 
This may explain our result of non-dependence of splitting frequency  on the flux content. Convective collapse 
\citep{parke1978,naga2008,fis2009} 
in larger patches may explain this relative difference in the filling factor. 
Using the {\it Hinode}/SP observations, the detailed information in each splitting event would become available and 
will make it possible to evaluate this model.

The collaborative investigation with numerical calculations is useful.
We suggest the qualitative picture in this study but we need to confirm whether it is possible in quantitative sense.
It is achieved by solving the M-C equation numerically by using our measurements.
\cite{mey2011} report their two-dimensional numerical simulations based on the modeling \citep{par2002}. 
They investigate equilibrium state by injecting fluxes and controlling the splitting frequency as a free parameter. 
A power-law distribution of flux content is maintained with an input of a steep power-law frequency of emergence.
The induced frequency of cancellation becomes a steep power-law one, which is consistent with our model.

Recent progresses of numerical simulation enable more realistic simulations on the solar surface, 
namely radiative magnetohydrodynamics simulation.
Although it is still difficult to set the numerical domain large enough for a statistical investigation, 
such calculation would be very helpful to determine the physical interpretation of 
frequencies of surface processes of magnetic patches.

\subsection{Contributions to Studies of Other Solar Phenomena}

We think there are two solar phenomena that are related to the results of this thesis, namely 
the surface flux transport and the nature of X-ray bright point.

The flux transport to the solar poles is an essential ingredient of the flux-transport dynamo model \citep{wan1991}. 
The major part of the flux in this model is provided by the diffusion of active regions. 
So it is necessary to extend to cover the active regions beyond our current analysis limitted to quiet regions. 
It may also be necessary to cover much smaller scale including the intranetwork field.

\begin{figure}[tp]
\centering
\includegraphics[bb=1400 50 3000 1000,width=0.8\textwidth]{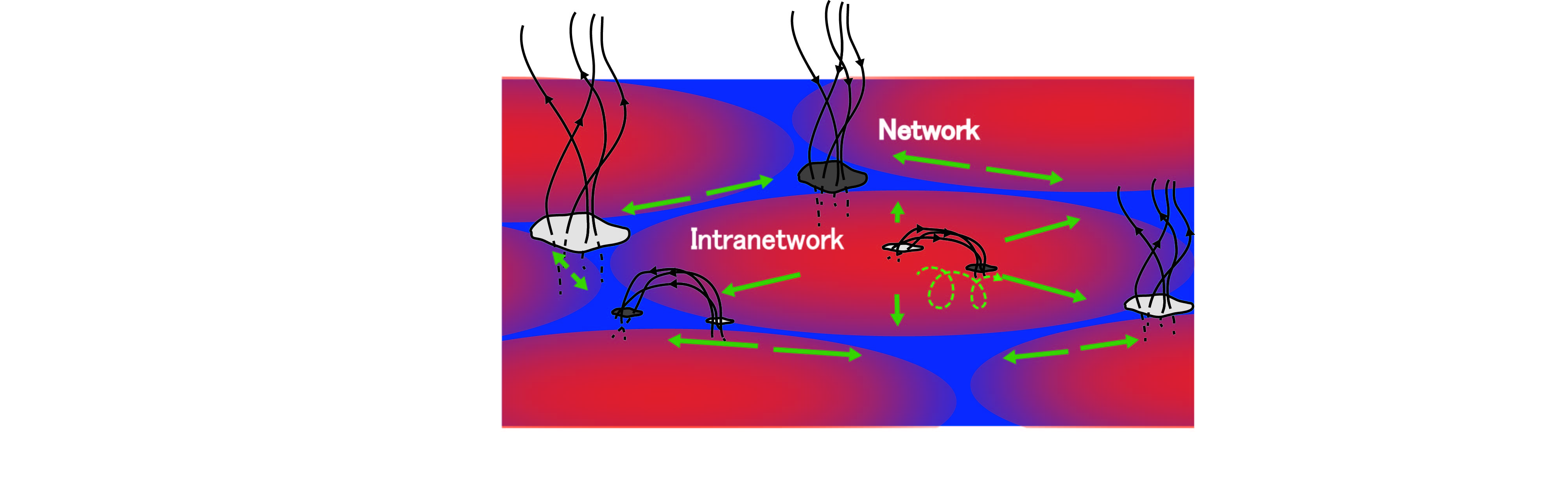}
\caption[A schematic picture of network field and internetwork field.]{
A schematic picture of network and internetwork field in quiet regions.
The blue (red) color in background shows upflow (downflow) motion in supergranule time scale.
The green arrows indicate the horizontal convective motion. 
Internetwork field is located in the supergranular cell. 
It is thought to be disturbed by smaller convective motion, namely granular cell and turbulence in the cell.
The network field is a magnetic field swept to the edge of supergranular cells.
The one at the conjunction point has a larger size and is more stable.}
\label{fig:pic_net}
\end{figure}

Transient Horizontal Magnetic Field (THMF) \citep[Figure \ref{fig:pic_net}; ][]{ish2009,ish2010}.
is the horizontal field with a short lifetime of $1\-- 10$ minutes and a flux content smaller than $10^{17.5}$ Mx.
\cite{ish2010b} reports a three-dimensional view of THMF, that THMF is an emerging flux tube within the internetwork.
THMF can be the recycled magnetic flux, which is supported by the fact that the flux amount of THMF, 
1.8 $\times$ 10$^{25}$ Mx day$^{-1}$ for the entire surface \citep{ish2010b}, 
is as large as the small emerging bi-pole detected in \citep{tho2011}, 3 $\times$ 10$^{25}$ Mx day$^{-1}$ for the entire surface.
Of course we need to investigate the spatial and temporal relationship between cancellations and THMF as the future works
to give a conclusion to this phenomena.

For the future investigation of these small field in the internetwork, the next high-resolution spacecraft mission, 
{\it Solar-C}, will be a great help.
It is planned to have a higher resolution by one-order of magnitude in magnetic flux content. 
We may be able to set the detection limit as $10^{16.5}$ Mx, which corresponds to one-order of magnitude in internetwork field domain.
It will be possible to do the statistical survey of such internetwork field based on the auto-detection of patches in {\it Solar-C} era.

The other extension of our analysis is to investigate the latitudinal difference of occurrence of cancellations.
We need to have the patches submerge or at least disappear at the high latitude for the polar reversal in solar cycles, 
both of which can be done only through cancellations.
So there is a possibility that there is a latitude dependence of cancellations.
Recent paper, Shiota et al. (accepted), reports another interesting result that the decrease in the net flux is caused by a decrease 
in the number and size of the large flux concentrations ($> 10^{18}$ Mx) as well as the appearance of patches with opposite polarity 
at lower latitudes. 
Our method of analysis will be also helpful to decide whether these large patches are convected as a small patches and maintained by mergings 
in the polar regions or convected as the large patches through the quiet regions.

The energetic events on the solar surface are related to magnetic activities as mentioned in Chapter 1.
X-ray Bright Points (XBPs) are the brightenings seen in the X-ray images \cite{gol1976}.
One model of XBPs is a magnetic reconnection in the higher solar atmosphere between different flux systems (Figure \ref{fig:pic_magac}).
This model is suggested from the fact that most XBPs correspond to cancellations \citep[80$\%$; ][]{har1997}, 
which are thought as submerging loops after reconnection \citep{iid2010}. 
We obtain an energy distribution of BPs from Equation (\ref{eq:model_distcnc}) and these relationships as
\begin{equation}
\displaystyle \frac{\partial n_{\rm BP}}{\partial t}(E) = 
\frac{v_0 n_0^2 \phi_0^2}{(\gamma-1)\alpha_{\rm BP} \sqrt{\rho} E_0} \left( \frac{E}{E_0} \right)^{\frac{-2 \gamma - \alpha_{\rm BP} +2}{\alpha_{\rm BP}}}.
\end{equation}
We set two assumptions here.
One is that all cancellations coincide magnetic reconnection in higher atmosphere, which is supported by the observational results 
in the {\it Yohkoh} era.
The other is that there is one-to-one relationship of a power-law distribution between canceling flux content and released energy, 
namely $\partial n_{\rm BP} / \partial t = E_0 \left( \phi /\phi_0 \right)^{\alpha_{\rm BP}} $.
Rough estimation, $\gamma=1.8$ and $\alpha_{\rm BP}=1.5$, deduces a power-law index of BP energy as $-2.1$. 
The power-law index steeper than -2 means that the smaller XBPs has a significant effect for the total energy release.
We want to note that the scale-free nature of energy distribution of BP comes from scaleless of flux content in this model.
There are some papers reporting the power-law index of the energy release of XBPs 
\cite{cro1993, shi1997, kru1998, par2000, ben2002, asc2002}. 
Some show that steeper than -2 but most show the power-law of $-1.5 \-- -1.7$, which are flatter than our estimation.
However, it should be noted that their target is the XBPs in the loop structure, namely not canceling XBPs.
We speculate that the appropriate model for the XBPs in the loop structure is the entangled flux tube suggested by \cite{parke1972},
 where the energy is released by the component magnetic reconnections of nearly paralleled field.
Thus we expect the different distribution of released energy as XBPs from our estimation.
The investigation distinguishing canceling XBPs and XBPs in the loop structure as the future works.

\ifodd \arabic{page}
\else
  \thispagestyle{empty}
  \mbox{}
  \newpage
  \clearpage
\fi

\appendix
\chapter[Flux Imbalance in Merging and Splitting]{Flux Imbalance in Merging and Splitting: Does It Affect Frequency Distribution of Cancellation?}
\thispagestyle{fancy}

Because merging and splitting are much more frequent than cancellation in our result, there is a possibility that flux imbalances 
during merging and splitting have significant effects in the evaluation of the frequency distributions of cancellation.
We examine this possibility by comparing the frequency distribution of flux imbalances during these events.

\begin{figure}[p]
\centering
\includegraphics[bb=0 0 700 700,width=0.99\textwidth,clip]{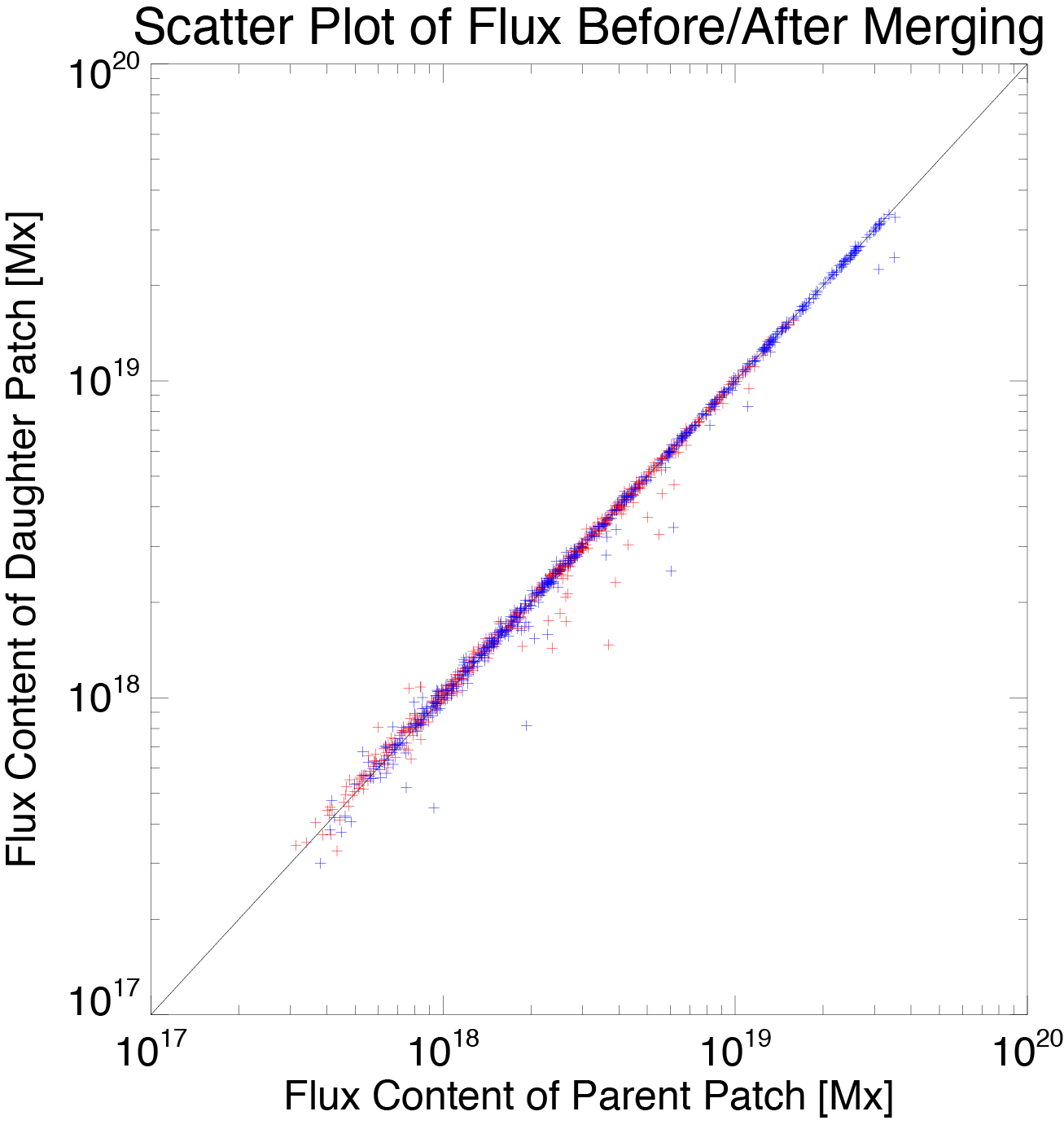}
\caption[A scatter plot of flux content before and after merging process.]{
A scatter plot of the flux content before and after merging processes in data set 1.}
\label{fig:scat_mrg}
\end{figure}

\begin{figure}[p]
\centering
\includegraphics[bb=0 0 700 700,width=0.99\textwidth,clip]{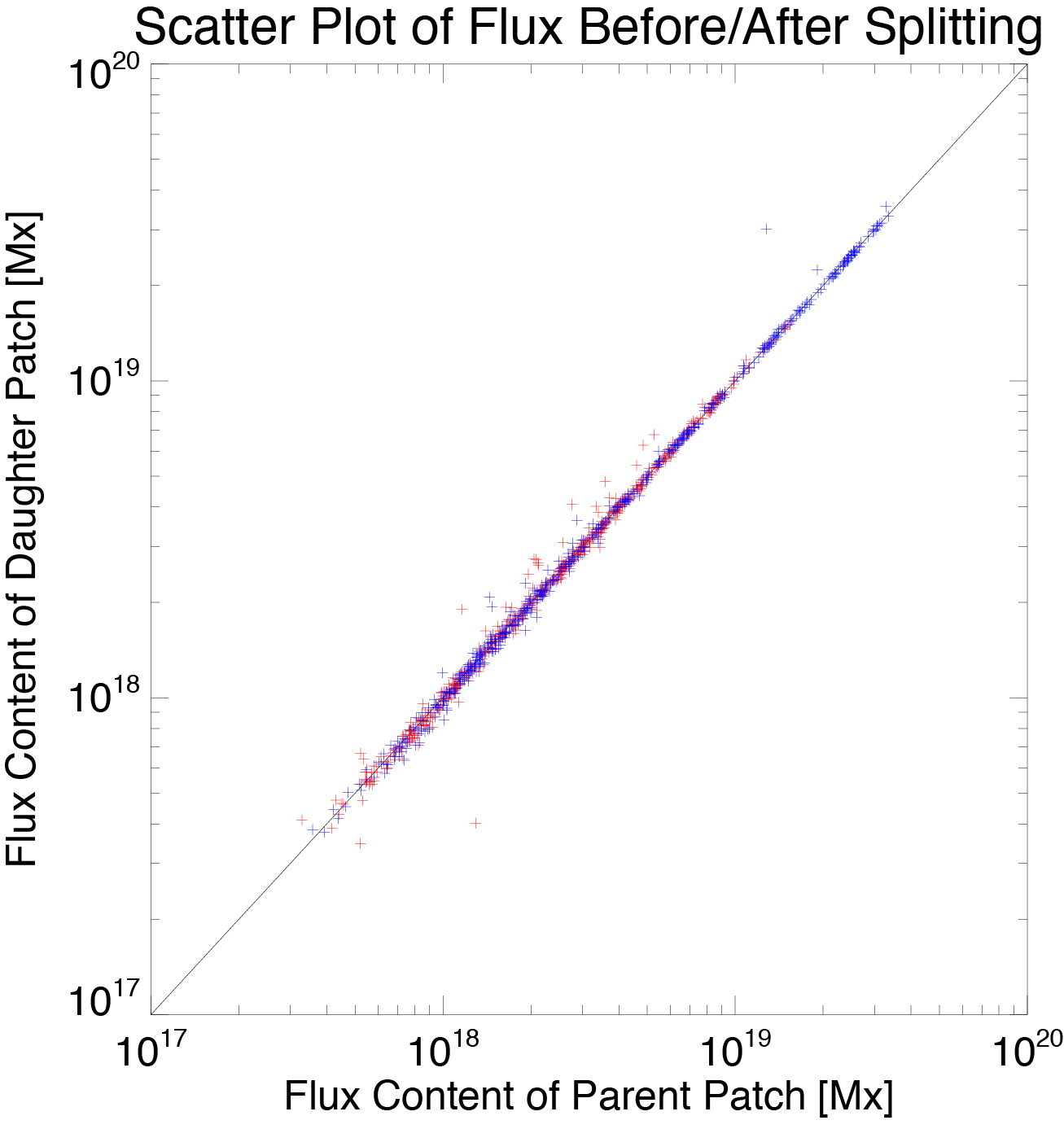}
\caption[A scatter plot of the flux content before and after splitting process.]{
A scatter plot of the flux content before and after splitting processes in data set 1.}
\label{fig:scat_splt}
\end{figure}

First we investigate flux imbalances during merging and splitting in data set 1.
Figure \ref{fig:scat_mrg} and Figure \ref{fig:scat_splt} shows scatter plots of flux content in parent patch and daughter patches.
We see that flux amount seems to be conserved well in the both processes from these figures.

\begin{figure}[p]
\centering
\includegraphics[bb=0 0 700 700,width=0.99\textwidth,clip]{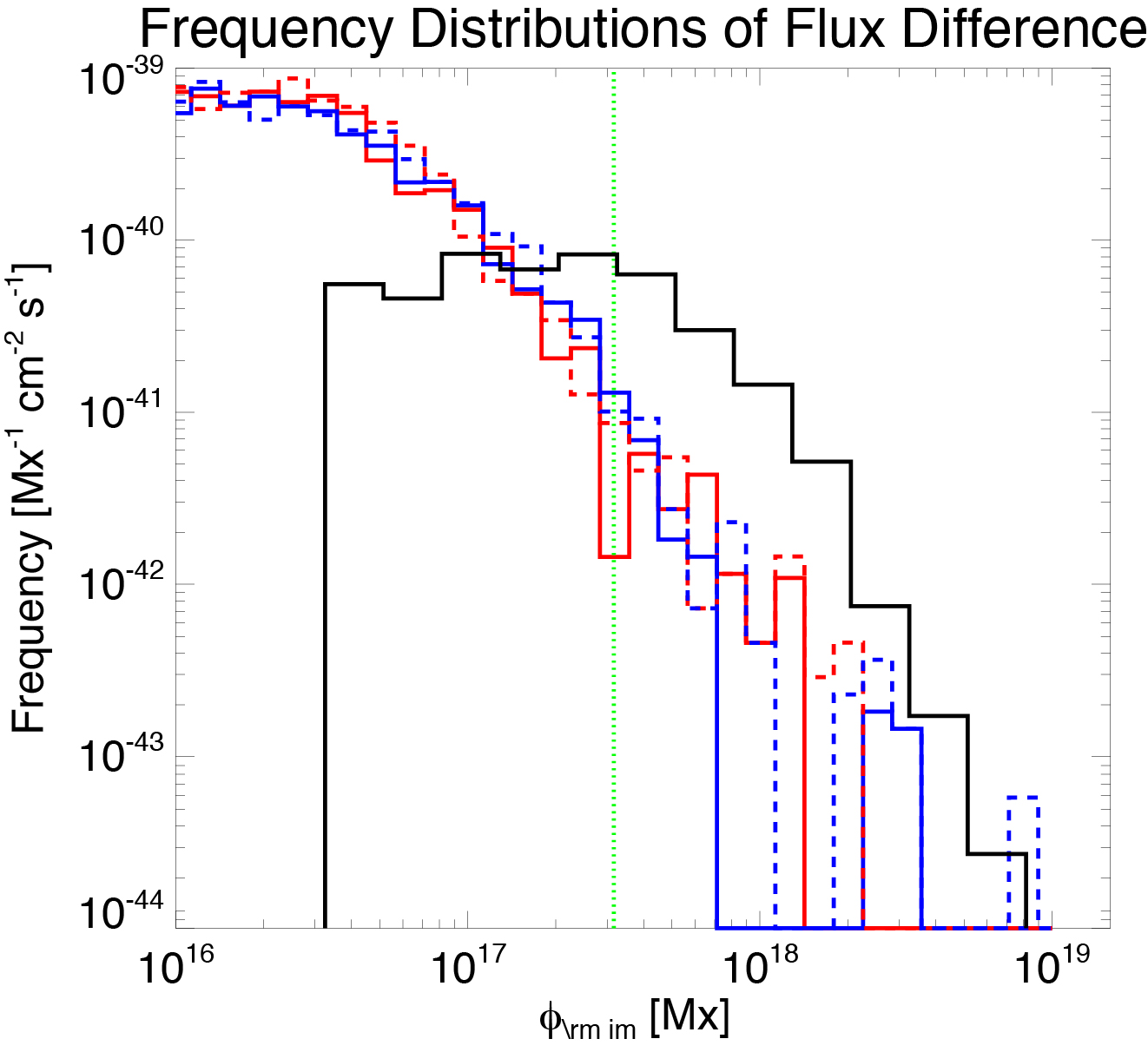}
\caption[Frequency distribution of flux imbalance in merging and splitting processes.]{
Frequency distribution of flux imbalance in merging and splitting processes. 
The red and blue solid histograms indicate that for merging of positive and negative polarity and the dashed ones for splitting.
The black solid histogram shows the frequency distribution of cancellation in data set 2.}
\label{fig:dist_imbalance}
\end{figure}

Next we investigate frequency distributions of flux imbalance of these processes.
The flux imbalance is defined as difference between sum of flux amount in parent patches and that in daughter patches, namely
\begin{equation}
\phi_{\rm im}=|\sum_{i=0}^{n_{\rm parent}} \phi_{i} - \sum_{j=0}^{n_{\rm daughter}} \phi_{j}|.
\end{equation}
Figure \ref{fig:dist_imbalance} shows the result. 
The solid histograms indicate that for merging and the dashed ones for splitting.
A green vertical line shows the detection limit of this study.
First, we see that flux imbalance is less than the detection limit.
It shows well-balanced flux content through merging and splitting.
Second, magnitude of frequency distribution in the range above detection limit is larger than that of cancellation, which is shown by 
black histogram, by one $\--$ two orders of magnitude in the analysis range.
We therefore conclude that flux imbalances in merging and splitting do not have a significant influence on frequency distribution of flux content.

\ifodd \arabic{page}
\else
  \thispagestyle{fancy}
  \mbox{}
  \newpage
  \clearpage
\fi

\chapter{Unipolar Processes Observed in This Study}
\thispagestyle{fancy}

We found many unipolar events, namely unipolar increase, appearance, decrease, and disappearance in our data sets as shown in Table 4.1.
Here we show examples of unipolar processes observed in this study to support the assumption that these events are interactions with 
patches below the detection limit, which is put throughout the thesis.

Although it is often mentioned that plenty of unipolar events are observed in magnetograms, there are few studies about these unipolar events.
\cite{lam2010} shows one example of unipolar appearances as a coalescence of patches below a resolution limit by 
using magnetograms with different spatial resolutions.

\begin{figure}[ph]
\centering
\includegraphics[bb=0 0 1100 700,width=18.5cm,clip,angle=270]{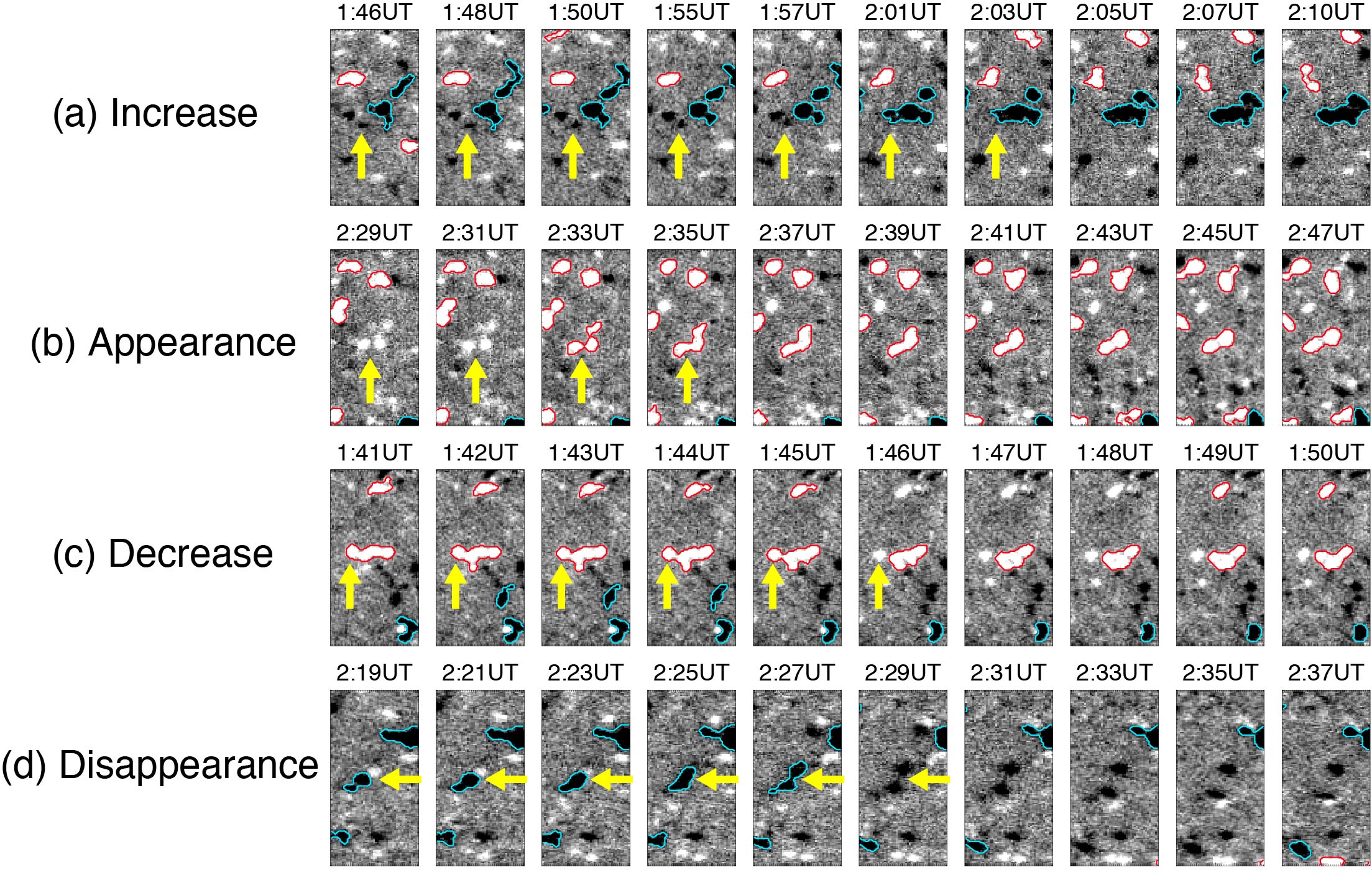}
\caption[Examples of apparent unipole events.]{
Examples of apparent unipolar events, namely, (a) unipolar increase, (b) unipolar appearance, 
(c) unipolar decrease, and (d) unipolar disappearance. 
The background shows the magnetic flux density obtained by ${\it Hinode}$/NFI.
The red (blue) contours indicate positive (negative) patches detected with our threshold.
The field of view is $9.6'' \times 19.2''$ for all images.}
\label{fig:example_uni}
\end{figure}

We interpret the unipolar events as interaction with patches below the detection limit.
Figure \ref{fig:example_uni} shows examples of unipolar processes detected in data set 1.
The background shows the magnetic flux density and the red (blue) contours indicate the positive (negative) patches with 
the detection condition. 
Figure \ref{fig:example_uni}(a) shows a unipolar increase.
There is a negative patch at 1:46UT indicated by the blue contour with the red arrow (patch A).
There are two small patches without contours (patches B $\&$ C) in the east-side of the detected patch.
They are gradually absorbed to the patch A and they are recognized as one large patch at 2:01UT.
The patch stays stable and recognized as one magnetic massif after the coalescence.
Because the patches B $\&$ C are below our detection limit, this event is recognized simply as an increase of the flux content in the patch A,
 i.e., a unipolar increase event.
Figure \ref{fig:example_uni}(b) shows a unipolar appearance of positive patch.
In this case, there are three small patches below the detection limit at 2:29-2:31UT.
They converge and coalesce to one large patch at 2:33UT.
It is difficult to see the inner structure of the patch even with the high spatial resolution of {\it Hinode} spacecraft.
Figure \ref{fig:example_uni}(c) shows a unipolar decrease. 
One detected positive patch is near the center of view at 1:41UT.
The east part of the patch changes its shape and one patchy structure can be seen there.
It gradually grows during 1:42$\--$1:45UT and is finally separated at 1:49UT.
The separated patch is small and it is below the detection limit.
Figure \ref{fig:example_uni}(d) shows a unipolar disappearance of the negative patch.
One negative patch is detected at 2:19UT.
However, a dipped shape is suddenly formed in the outline of the patch at 2:27UT.
The patch is divided into two small negative patches below the detection limit at 2:29UT.
These four events are explained as interactions with patches below the detection limit, which is the same result as \citep{lam2010}.
Through our interpretation, the unipolar events should be counted for splitting events. 
This leads to the underestimation of the number of splitting events and is the cause of an apparent drop in the probability distribution (Figure 4.4).
Such influence of the detection limit is taken into account in our analysis (Section 5.1.3).

\begin{figure}[p]
\centering
\includegraphics[bb=0 0 800 800,width=0.90\textwidth,clip]{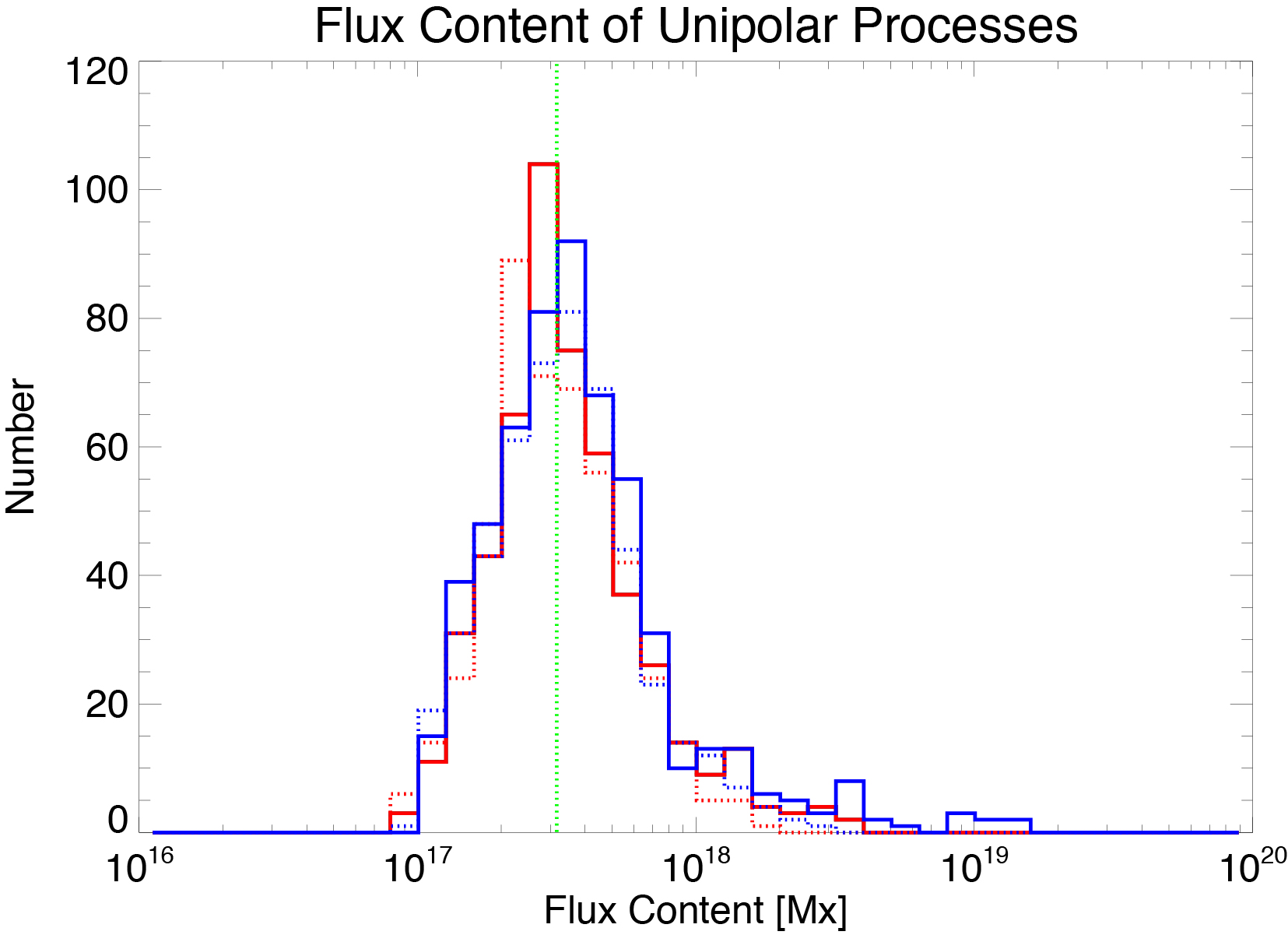}
\caption[Histogram of flux amount changed by unipolar processes in data set 1.]{
Histogram of flux amount changed by unipolar processes in data set 1.
The solid lines indicate increase processes and the dotted lines indicate decrease processes.
Colors indicate the polarity, namely red for positive and blue for negative.}
\label{fig:hist_uni_d1}
\end{figure}

Further, we investigate the flux amount changed by the unipolar processes in data set 1. 
It should concentrate around the detection limit based on our interpretation.
Figure \ref{fig:hist_uni_d1} shows the histogram of the total flux amount changed by unipolar processes.
We can see that all four histograms concentrate around the detection limit, namely the green vertical line.
This result supports the interpretation that the unipolar processes are the interaction with the patches below the detection limit. 
Because we take the influence of the detection limit, the power-law index induced in the thesis may not be affected by these unipolar events.

The further quantitative investigation of unipolar processes is needed to give a conclusion for the discussion here.
We believe that the auto-detection with the different conditions will be useful at this point and it should be done 
in the future works.

\ifodd \arabic{page}
\else
  \thispagestyle{plain}
  \mbox{}
  \newpage
  \clearpage
\fi

\addcontentsline{toc}{chapter}{Bibliography}
\bibliographystyle{apj}
\bibliography{iida}

\end{document}